\documentclass{elsarticle}
\usepackage{amsmath}
\usepackage{amssymb}
\usepackage{subcaption}
\usepackage{soul}
\sethlcolor{green}
\usepackage{ulem}
\usepackage{tikz}
\usepackage{bm}
\renewcommand{\Re}{\ensuremath{Re}}

\newcommand {\Wi}{W\!i}

\begin{document}

\begin{frontmatter}

\title{Understanding viscoelastic flow instabilities: Oldroyd-B and beyond\tnoteref{mytitlenote}
}
\tnotetext[mytitlenote]{Published in the Oldroyd-100 special issue of the Journal of non-Newtonian Fluid Mechanics}


%

\author{Hugo A. Castillo S\'{a}nchez \fnref{myfootnoteA}}   
\address{Department of Applied Mathematics and Statistics, Institute of Mathematics and Computational Sciences, University of Sao Paulo, Brazil}
\ead{hugo\_acs@icmc.usp.br}

\author{Mihailo R. Jovanovi\'{c}}
\address{Ming Hsieh Department of Electrical and Computer Engineering, University of Southern
California, Los Angeles, California 90089, USA}
\ead{mihailo@usc.edu}
\fntext[myfootnoteA]{The authors are listed alphabetically and contributed in their areas of expertise to this review, whose preparation was coordinated by V.~Shankar and G.~Subramanian.}

\author{Satish Kumar} 
\address{Department of Chemical Engineering and Materials Science, University of Minnesota, Minneapolis, MN 55455, USA}
\ead{kumar030@umn.edu}

\author{Alexander Morozov} 
\address{SUPA, School of Physics and Astronomy, The University of Edinburgh, Edinburgh EH9 3FD, UK}
\ead{Alexander.Morozov@ed.ac.uk}

\author{V.~Shankar} 
\address{Department of Chemical Engineering, Indian Institute of Technology, Kanpur 208016, India}
\ead{vshankar@iitk.ac.in}

\author{Ganesh Subramanian} 
\address{Engineering Mechanics Unit, Jawaharlal Nehru Center for Advanced Scientific Research, Bangalore 560064, India}
\ead{sganesh@jncasr.ac.in}

\author{Helen J. Wilson} 
\address{Mathematics Department, University College London, Gower Street, London WC1E 6BT, UK}
\ead{helen.wilson@ucl.ac.uk}


%
%

\begin{abstract}
The Oldroyd-B model has been used extensively to predict a host of instabilities in shearing flows of viscoelastic fluids, often realized experimentally using polymer solutions. The present review, written on the occasion of the birth centenary of James Oldroyd, provides an overview of instabilities found
across major classes of shearing flows. These comprise (i) the canonical rectilinear shearing flows including plane Couette, plane and pipe Poiseuille flows; (ii) viscometric shearing flows with curved streamlines such as those in the Taylor-Couette, cone-and-plate and parallel-plate geometries; (iii) non-viscometric shearing flows with an underlying extensional flow topology such as the flow in a cross-slot device; and (iv) multilayer shearing flows. While the underlying focus in all these cases is on results obtained using the Oldroyd-B model, we also discuss their relation to the actual instability, and as to how the shortcomings of the Oldroyd-B model may be overcome by the use of more realistic constitutive models.
All the three  commonly used tools of stability analysis, viz., modal linear stability, nonmodal stability, and weakly nonlinear stability  analyses are discussed,  with supporting evidence from  experiments and numerical simulations as appropriate. Despite only accounting  for a shear-rate-independent viscosity and first normal stress coefficient, the Oldroyd-B model is able to qualitatively predict the majority of instabilities in the aforementioned shearing flows. The review also highlights, where appropriate, open questions in the area of viscoelastic stability.\\
\end{abstract}


\end{frontmatter}


\section{Introduction}
\label{sec:Intro}
Compelling differences between Newtonian and viscoelastic flow phenomena in the same geometry have been well highlighted in textbooks \citep{birdvol1}, and this contrast also applies to instabilities occurring in the same base-flow configuration. Viscoelastic flows are prone to novel instabilities that arise due to elasticity alone, or due to a combination of elastic and inertial effects, and such instabilities  evidently have no Newtonian counterparts. Initial interest in the understanding of these instabilities was driven by the need to  prevent their occurrence during polymer processing operations
\citep{Petrie_Denn1976,larson1992}, and thereby circumvent the associated restrictions on processing rates; fluid inertia is usually negligible in these processes, and the focus is therefore on purely elastic instabilities. For instance, during extrusion of highly viscous entangled polymer melts, the extrudate often exhibits a spiral or wavy distortion, a phenomenon referred to as `melt fracture' \citep{Denn2004},
and thought to occur via a hydrodynamic instability; although, physicochemical effects such as wall slip likely play a role \citep{Denn2001}.
%
A second, more recent, motivation for studying viscoelastic flow instabilities arose from their discovery in the standard rheometric geometries \,(e.g. the Couette, cone-and-plate and parallel-plate set ups); for instance, see \citep{Shaqfeh1996,muller_review}. The occurrence of purely elastic instabilities, in dilute polymer solutions subject to simple curvilinear shearing flows in rheometric devices, hampered their use for purposes of rheological characterization; an understanding of these instabilities is clearly essential for defining the operating range of these devices. The elastic instabilities above have their origins in normal stress differences present in viscoelastic shear flows. For the curvilinear shearing flows in particular, the first normal stress difference leads to a hoop stress, on account of streamline curvature, which drives an instability.

Viscoelastic flow instabilities can also be beneficial depending on the scenario. For instance, with regard to the rheometric example above, increasing the shear rate causes the initial elastic instability to eventually saturate in a complex disorderly nonlinear flow state termed `elastic turbulence'\,(ET)   
\citep{Groisman2000,Groisman2001,Steinberg2021}. There have been reports of similar ET-like states in rectilinear flows of dilute polymer solutions through micro-channels, especially when the flow is perturbed by finite-amplitude obstacles at the channel inlet \citep{Pan_2013_PRL,qin2017elastic,qin2019elastic,jha2020universal}. The inaccessibility of Newtonian turbulence on microfluidic scales (owing to the modest Reynolds numbers) implies that the ET state, in either of the two cases above, can instead be exploited towards increasing mixing efficiencies, as has indeed been demonstrated in earlier efforts \citep{Groisman2001}.

While the above instances probed the low Reynolds number ($Re$) regime where fluid inertial effects are negligible,
there have also been many reports of `early turbulence' in pipe flow of polymer solutions at higher Re, albeit still lower than the oft-quoted Newtonian threshold of $Re \sim 2000$ \citep{ram_etal_1964,goldstein_etal_1969,hansen_etal_1973,hansen_little_1974,hoyt_1977,zakin_etal_1977,draad_etal_1998};
 these early studies were, however, not systematically corroborated in the subsequent literature. 
In contrast, the recent experiments of Hof and coworkers \citep{Samanta2013,choueiri2021experimental}, involving pipe flow of polymer solutions at concentrations below or close to the overlap value, have unambiguously demonstrated  transition from the laminar state at $Re \sim 1000$,
thereby confirming the aforementioned observations. The ensuing flow was found to be neither laminar, nor to bear a resemblance to Newtonian turbulence, and was therefore christened `elasto-inertial turbulence' (EIT) to emphasize the importance of both fluid elasticity and inertia, in contrast to the ET states discussed above. This EIT state was further shown to be linked \citep{Choueiri2018} to the asymptotic maximum drag reduction (MDR) regime, a universal state that arises with the progressive addition of polymers to turbulent Newtonian pipe flow \citep{Virk}. This link is an important one. The phenomenon of turbulent drag reduction \citep{Tom1977} is undoubtedly one of the most spectacular manifestations of viscoelasticity, and while there exists a vast body literature in this regard \citep{white_mungal_2008,Graham2014,Xi2019DRreview}, the prevailing viewpoint regards the aforementioned MDR regime as a drag-reduced state accessible only from the Newtonian turbulent state; even relatively recent dynamical-systems-based interpretations have attempted to understand MDR in terms of the existence of essentially Newtonian coherent structures, modified by elasticity \citep{stone_graham2002,stone_graham2003}. As a consequence, advances in viscoelastic stability and drag reduction have occurred largely independently with very little cross-pollination of theoretical viewpoints. However, as we discuss later in this review, the aforementioned  pipe flow experiments, together with recent theoretical work \citep{Piyush_2018}, show that the MDR regime, at least for moderate $Re$, can be viewed as a `drag-enhanced' state arising from an elastoinertial instability of the laminar state. The above efforts for pipe flow, and other recent efforts that include theoretical \citep{roy2021}, computational \citep{BISTAGNINO2007} and experimental \citep{muller_review} investigations of other rectilinear  and curvilinear shearing flow configurations, reflect an increasing interest in the origin and dynamics of elastoinertial instabilities.
 
%
%

The present review article, written on the occasion of the birth centenary of James Oldroyd who proposed the now-eponymous constitutive equation  \citep{Oldroyd1950}, attempts to provide a state-of-the-art summary of the understanding of flow instabilities using the Oldroyd-B model. That said, however, where appropriate, we also discuss the use of more refined constitutive models, with physics outside of the Oldroyd-B framework, that are often crucial to obtaining better agreement with experimental observations. Indeed, from a fundamental standpoint, an accurate prediction of viscoelastic flow instabilities may also be regarded a rigorous test that aids in the eventual development of physically sensible constitutive equations \citep{muller_review,Wilson2006}.
While there have been many earlier review articles on the subject of viscoelastic flow instabilities \citep{Petrie_Denn1976,larson1992,Shaqfeh1996,muller_review}, the present review focuses on the developments over the last two decades. Further, in contrast to some of the review articles above, which have focused almost exclusively on the hoop-stress-driven elastic instabilities that arise in the curvilinear rheometric geometries, this review covers instabilities in both the canonical rectilinear and curvilinear shearing flows, with the latter encompassing both viscometric and non-viscometric flow configurations. In fact, the experimental observations mentioned in the preceding paragraph, pertaining to the moderate-to-high $Re$ regimes in flow through pipes and channels, have spawned renewed interest in viscoelastic instabilities that occur in rectilinear shearing flows, and the present review lays a greater emphasis on these more recent findings. The review by Renardy and Thomases \cite{Renardy2021} in this special issue presents a different perspective, by focusing on open mathematical challenges related to the Oldroyd-B model, and the write-up of Hinch and Harlen \citep{Hinch_Harlen2021} provides a conceptual account of Oldroyd's formulation of upper- and lower-convected derivatives.
While the review by Shaqfeh and Khomami \citep{Shaqfeh_Khomami_2021}, also a part of this special issue, addresses instabilities in curvilinear viscoelastic flows based on the Oldroyd-B model, as already mentioned, we also place an emphasis  (in Sec.~\ref{sec:curvilinear} below) on how the use of genuinely nonlinear constitutive equations that account for shear-thinning effects plays an important role in more accurate predictions of experimental observations.
Another recent multi-author review article \citep{datta2021perspectives}, based on the virtual workshop on viscoelastic flow instabilities and elastic turbulence organized by the Princeton Center for Theoretical Sciences, also provides a state-of-the-art summary of the various challenges in this broad area. 

It is worth recalling that Oldroyd, in his seminal 1950 paper \citep{Oldroyd1950}, proposed a constitutive model for viscoelastic flows purely from a continuum viewpoint, by requiring the model to satisfy the principle of material frame indifference. As discussed by Hinch and Harlen  \citep{Hinch_Harlen2021}, Oldroyd argued that the frame indifference implied that the time derivative of the stress tensor be computed in a  reference frame that undergoes local translation, rotation, and deformation with the material.
The usual material derivative in fluid mechanics denotes the instantaneous rate of change of a fluid property (e.g. velocity, temperature) in a reference frame that translates with a given fluid element, and is thereby  appropriate only for a point microstructure. In contrast, the upper convected derivative introduced by Oldroyd denotes the rate of change in a  reference frame that, in addition, deforms\,(affinely) with the fluid motion, thereby accounting\,(implicitly) for an underlying orientable microstructure.

Interestingly, the Oldroyd-B constitutive equation can also be derived from a coarse-grained, mesoscopic model wherein a dilute solution of polymer molecules is modeled as a suspension of non-interacting Hookean `dumbbells' in a Newtonian solvent, each
polymer molecule being idealized as a dumbbell comprising an infinitely extensible Hookean spring connecting two beads which account for the drag experienced by the polymer molecule  \citep{birdvol1,larson1988constitutive}.  
Any additional physical effects entering this mesoscopic picture, for instance, a nonlinear spring force, a configuration-dependent bead drag, or hydrodynamic interaction between the beads, invariably lead to a closure problem, and derivation of the equation governing the polymeric stress, to be used in the continuum mechanical formulation, requires appropriate pre-averaging approximations.
The stress tensor in the Oldroyd-B constitutive equation is conveniently divided into two parts that correspond to the polymeric and Newtonian solvent contributions. Thus, the stress is written as $\boldsymbol{\tau} = 
\boldsymbol{\tau}_p + \boldsymbol{\tau}_s$, where the solvent contribution $\boldsymbol{\tau}_s = \eta_s (\nabla \boldsymbol{v} + \nabla \boldsymbol{v}^T)$, $\eta_s$ being the solvent viscosity, and the polymeric contribution $\boldsymbol{\tau}_p$ is given by the `upper-convected Maxwell' (UCM) constitutive equation:
\begin{equation}
\lambda \left(\frac{\partial \boldsymbol{\tau}_p}{\partial t}+ \boldsymbol{v} \cdot \nabla  \boldsymbol{\tau}_p-\left(\nabla \boldsymbol{v} \right)^{T} \cdot\boldsymbol{\tau}_p-\boldsymbol{\tau}_p \cdot \left(\nabla \boldsymbol{v}\right) \right) +\boldsymbol{\tau}_p =  \eta_p \left(\nabla \boldsymbol{v} +\nabla \boldsymbol{v}^{T} \right) \, .
\label{eq:UCM}
\end{equation}
%
The terms within parentheses on the left-side of the above equation represent the upper-convected derivative of $\boldsymbol{\tau}_p$ described above, $\lambda$ is the relaxation time of the dumbbell, being related to the longest relaxation time of the polymer molecule, and $\eta_p$ is the polymeric contribution to the steady-shear viscosity.  The  nonlinear coupling between the velocity and the stress in the upper-convected derivative suggests that, even in the absence of fluid inertia, one may anticipate bifurcations from a given base flow, leading to qualitatively new instabilities - such bifurcations are indeed the basis of the purely elastic instabilities mentioned earlier. In the limit of zero solvent viscosity, $\eta_s = 0$, the Oldroyd-B model reduces to the UCM model.

%

For steady simple shear flow, the Oldroyd-B model predicts a shear-rate-independent viscosity and first normal stress coefficient $\Psi_1 = N_1/\dot{\gamma^2}$ (where $N_1$ is the first normal stress difference and $\dot{\gamma}$ is the shear rate), and yields a zero second-normal stress difference\,($N_2=0$).  The shear-rate-independence is due to the infinite extensibility of the Hookean dumbbells in the mesoscopic picture underlying the Oldroyd-B model. The Hookean response is valid only for small deformations from the equilibrium conformation of a polymer molecule. 
Consequently, the Oldroyd-B model is not applicable for strongly shear-thinning systems such as polymer melts or water-based dilute polymer solutions. The Oldroyd-B model also cannot describe phenomena ascribable exclusively to an $N_2$ which, in the present context, include spanwise instabilities in a rectilinear shearing flow \citep{ramachandran_leighton_2008,Hinchyoutube}. While the model correctly predicts a thickening behavior in extensional flows, it also predicts an unbounded growth of the extensional viscosity beyond a threshold extension rate, this growth again arising from the infinite extensibility of the Hookean dumbbells. Thus, the Oldroyd-B model is expected to be more relevant to shear-dominated flows.

One way to rectify the aforementioned shortcomings of the Oldroyd-B model is to go beyond the Hookean dumbbell assumption, by incorporating a nonlinear spring with the spring force diverging at maximum extension; the most commonly used form of the force law leads to the so-called finitely extensible nonlinear elastic (`FENE') springs \citep{Bird1980}. The nonlinearity of FENE springs does not allow for an analytical solution of the Smoluchowski equation in the underlying kinetic theory framework \citep{Herrchen_Ottinger_1997}. Consequently, preaveraging approximations are needed to obtain a closed-form relation between the stress and the strain rate, and the constitutive equation thus obtained is referred to as the `FENE-P' model (`P' being Peterlin, who proposed this approximation). The FENE-P model is characterized by an additional dimensionless parameter, $L$, which is the ratio of the fully stretched length to the equilibrium coil dimension; $L \rightarrow \infty$ recovers the Oldroyd-B model. The nonlinear stiffening of the spring, and the resulting decrease in the relaxation time implies that the FENE-P model predicts shear thinning of both the viscosity and the first normal stress coefficient at high shear rates.
The inclusion of a nonlinear spring force also removes the divergence of  the extensional viscosity, causing it to saturate at a large but finite value, in accord with experimental observations \citep{larson1988constitutive}.
A closely related nonlinear constitutive equation that also incorporates a finitely extensibile spring is the FENE-CR model\,(proposed by Chilcot and Rallison \cite{chilcott1988creeping}), which predicts a constant shear viscosity, while allowing a for a shear-rate dependence of the first normal stress coefficient, and is thereby especially suited for the so-called Boger fluids; see footnote~5 in Sec.~\ref{subsec:purelyelasticTC}.

A second way of addressing the deficiencies of the Oldroyd-B model, one appropriate for concentrated polymer solutions, is to recognize the anisotropy of a given dumbbell's environment, this anisotropy arising from the the stretched dumbbells in its neighborhood; this may be incorporated via an anisotropic tensorial correction to the relaxation term. Giesekus \citep{Giesekus1982} postulated that the tensor characterizing the anisotropy is proportional to the  (deviatoric)\,stress itself, leading to the Giesekus constitutive equation. The proportionality constant, $\alpha$, measures the amplitude of anisotropy, with $\alpha = 1$ denoting maximum anisotropy, and $\alpha = 0$ denoting the original isotropic relaxation in the UCM model \citep{larson1988constitutive}. For any non-zero $\alpha$, the Giesekus model includes an additional term quadratic in the stress tensor which becomes important in the nonlinear regime. Similar to the FENE-P model, the Giesekus model predicts shear thinning of both the viscosity and first normal stress coefficient, and does not exhibit any singularity in the extensional viscosity.


%

Even for the simplest shearing flows driven by the motion of rigid boundaries, and that are characterized by a single length\,($H$) and velocity\,($V$) scale, the stability of an Oldroyd-B fluid is governed by three dimensionless parameters: the Reynolds number $Re = H V \rho/\eta$, the Weissenberg number $Wi = \lambda V/H$ which is the product of the polymer relaxation time and a typical shear rate, and the ratio of solvent to solution viscosity $\beta = \eta_s/(\eta_s + \eta_p)$; here, $\rho$ and $\eta = \eta_p + \eta_s$ are, respectively, the density and total  viscosity of the polymer solution. For purely elastic instabilities, $Wi$ and $\beta$ are the relevant parameters; the elasticity number $E = Wi/Re = \lambda \eta/(\rho H^2)$, that is independent of the flow, being the ratio of the polymer relaxation and the momentum diffusion timescales, may also be used (instead of either $Re$ or $Wi$) when describing elastoinertial instabilities. We note in passing that the capillary number, denoting the ratio of viscous to surface tension forces, will become relevant for shearing flows with a free surface. For viscoelastic flows in particular, the `elastocapillary' number, which is the ratio of Weissenberg and capillary numbers, and measures the relative importance of elastic and capillary forces, may be used \citep{McKinley2005}. In flow configurations involving multiple length scales and/or in the presence of an unsteady shearing, the Deborah number ($De$), defined as the ratio of the relaxation time to a characteristic flow time $T$, is also used; here, $T$ can be either the residence time of a fluid element in the region of interest or  the time period of an oscillatory shear flow. The Deborah and Weissenberg numbers are often  interchangeably used for steady shearing flows. For the curvilinear viscometric flows examined in  Sec.~\ref{sec:curvilinear}, $De$ and $Wi$ will be seen to be related to each other by the aspect ratio of the particular flow configuration \citep{poole2012}.

In light of the above, the subject of transition in viscoelastic shearing flows, unlike their Newtonian counterparts, is not a `single problem'. For a given viscoelastic shearing flow, one could be in an inertially dominant regime with weak elasticity\,($Wi \ll 1, Re \gg 1$), a strongly elastic near-Stokesian regime\,($Wi \sim O(1)$, $Re \ll 1$), or an elasto-inertial regime with inertia and elasticity being of comparable importance\,($Wi, Re > 1, E \sim O(1)$); the nature of transition would depend sensitively on the particular asymptotic regime. In addition, the solvent viscosity ratio $\beta$, which is a proxy for polymer concentration, allowing one to span the regimes from ultra-dilute polymer solutions ($\beta \to 1$) to polymer melts ($\beta \rightarrow 0$), is also expected to influence the transition. Thus, transition from the steady laminar state, to states with nontrivial spatiotemporal dynamics, can occur via multiple pathways in $Re$-$Wi$-$\beta$ space; we return to the aspect of multiple transition scenarios, in the context of rectilinear flows, in Sec.~\ref{subsec:finiteRe}.

The first step in analyzing the stability of a laminar flow is to consider its response to infinitesimal disturbances, which allows for the linearization of the governing equations about the laminar base state. Within this linear stability framework, there are two different approaches. The classical approach is modal stability, and involves expressing the perturbation fields in the normal mode form with an exponential dependence in time, in turn leading to an eigenvalue problem for the growth rate as a function of the wavenumber and other relevant dimensionless parameters. According to the convention usually followed, a change in sign of the imaginary part of the eigenvalue\,(the growth rate), from negative to positive, corresponds to the onset of instability. For sheared base states in particular, the non-normality of the differential operator governing linear stability implies that the aforementioned modal stability analysis only pertains to the asymptotic behavior for long times; when unstable, this long-time evolution is dominated by exponentially growing modes \citep{Drazinreid}. Even in the absence of such unstable modes, however, small amplitude disturbances can grow algebraically for shorter times. The machinery for a detailed analysis of this so-called transient growth is now well developed, and has been extensively applied in the Newtonian context \citep{Schimid-Henningson}.

Going beyond linear stability, finite amplitude disturbances are often considered within the framework of an amplitude expansion, an approach that originated in the efforts of Stuart \citep{stuart_1960} and Watson \citep{Watson_1960} in the Newtonian context. In fact, the nonmodal and nonlinear stability approaches have acquired prominence owing to the failure of the classical modal approach to explain transitions in any of the canonical Newtonian shearing flows (plane Couette, plane Poiseuille and pipe flows).  All the three approaches above are covered in this review within the context of the Oldroyd-B model and its refinements. It is important to note that, in recent times, direct numerical simulations of viscoelastic flows complement the aforementioned approaches, providing detailed structural information in the nonlinear regime \citep{dubief_white_terrapon_shaqfeh_moin_lele_2004,Dubief2013,page2020exact,Shekar2019,Shekar_2020,song_wan_liu_lu_khomami_2021,song_lin_liu_lu_khomami_2021}. The computational expense, however, implies that the parameter space explored by such simulations is necessarily restricted, a limitation that is amplified by the high-dimensional nature of the parameter space characterizing viscoelastic shearing flows.

The remainder of this review is organized as follows. We  begin with instabilities in simple rectilinear flows in Sec.~\ref{sec:rectilinear}. After a brief summary of the principal features of the Newtonian spectrum in Sec.~\ref{Newt:spectrum}, we discuss the nature of the viscoelastic spectrum in the inertialess limit in Sec.~\ref{subsec:zeroRe}. Herein, we point out that rectilinear shearing flows are generally linearly stable in the $Wi$-$\beta$ space; although, plane Poiseuille flow has recently emerged as an exception in this regard, becoming unstable for $Wi \gg 1$ and $\beta \to 1$. In Sec.~\ref{subsec:finiteRe}, the  elasto-inertial spectrum for the canonical rectilinear shearing flows is discussed, and it is shown that while plane Couette flow is always stable in the $Re$--$Wi$--$\beta$ space, both plane- and pipe-Poiseuille flows are unstable in significant domains of this space. The nature of instabilities in these flows is discussed briefly, and based on our current knowledge, we provide an overview of various possible transition scenarios in the $Re$-$Wi$-$\beta$ space.
In Sec.~\ref{subsec:Interfacial}, we discuss instabilities in two-layer flows of viscoelastic fluids wherein a jump in the first normal stress difference leads to a novel instability absent for Newtonian two-layer flows. This section also includes a brief summary of instabilities in shear-banded flows; while shear banding itself is outside the purview of the Oldroyd-B model, owing to the absence of nonmonotonicity in the constitutive curve, the instabilities in the banded state can nevertheless be usefully interpreted using the Oldroyd-B model.


Instabilities in curvilinear viscometric shearing flows are surveyed in Sec.~\ref{sec:curvilinear}. 
Here, we first discuss (Sec.~\ref{subsec:VEonNewtonianTC}) the role of elasticity on the Newtonian (centrifugal) instability in the Taylor-Couette geometry, before moving on to a discussion of the purely elastic instability in the same geometry (Sec.~\ref{subsec:purelyelasticTC}). The effects of finite-gap widths (relative to the radius of the inner cylinder) and nonaxisymmetric disturbances are summarized in Sec.~\ref{subsec:finitegap}. The role of inertia in leading to additional instabilities, and the nature of the resulting dominant mode in the $Wi-Re$ plane, is discussed in Sec.~\ref{subsec:inertialeffects}. Elastic instabilities in other curvilinear viscometric flows, including the those in the cone-and-plate and parallel-plate geometries are discussed in Sec.~\ref{subsec:coneplate}. The issues that bedevil the comparison of experimental observations and theoretical predictions of purely elastic instabilities are then discussed in Sec.~\ref{subsec:exptsnonisothermal}. The role of rheological features beyond the scope of the Oldroyd-B model (such as a nonzero second normal-stress difference) is analyzed in Sec.~\ref{subsec:beyondOldB}; for the rectilinear shearing flows surveyed in Sec.~\ref{sec:rectilinear}, the relatively nascent state of  understanding, of instabilities that drive transition, has meant that the consequences of refinements to the Oldroyd-B model are only beginning to be explored.
We end Sec.~\ref{sec:curvilinear} with a discussion of the Pakdel-McKinley criterion for instabilities in shearing flows with curved streamlines, and in particular, its use in understanding the role of shear thinning in limiting the elastically unstable domain in the relevant parameter plane\,(Sec.~\ref{subsec:PakdelMcKinley}). Section~\ref{sec:nonviscometric} examines instabilities in non-viscometric settings such as the cross-slot geometry\,(Sec.~\ref{sec:CrossSlotInstabilties}), contraction-expansion flows\,(Sec.~\ref{sec:contraction-expansion}), and flow past a circular cylinder\,(Sec.~\ref{subsec:flowcylinder}). While the approach used in the sections above is the classical modal one, Sec.~\ref{sec:nonmodal} provides an account of recent efforts, within the non-modal transient growth framework, applied to viscoelastic flows. The role of finite-amplitude disturbances, within a nonlinear stability framework, is discussed in Sec.~\ref{sec:nonlinear}, the emphasis being on the purely elastic scenario. Finally, in Sec.~\ref{sec:Concl} we end with a brief discussion on some of the outstanding issues in this field.

\section{Rectilinear shearing flows: Results from modal analyses in the $Re$-$Wi$-$\beta$ space}
\label{sec:rectilinear}

In this section, we examine the stability of canonical rectilinear shearing flows, comprising plane Couette, plane Poiseuille  and pipe Poiseuille flows, from the modal perspective. Further results from the non-modal viewpoint are presented later in Sec.~\ref{sec:nonmodal}. A prerequisite to understanding the non-trivial structure of the full elastoinertial spectrum, and associated instabilities, is an understanding of the spectra arising from inertial and elastic forces acting separately. Thus, we begin with a discussion of the Newtonian spectrum below, and follow it up with a discussion of the inertialess elastic spectrum associated with an Oldroyd-B fluid. This discussion also highlights the recent, and unexpected, discovery of a purely elastic instability in plane Poiseuille flow.

\begin{figure}
  \centering
    \includegraphics[width=0.5\textwidth]{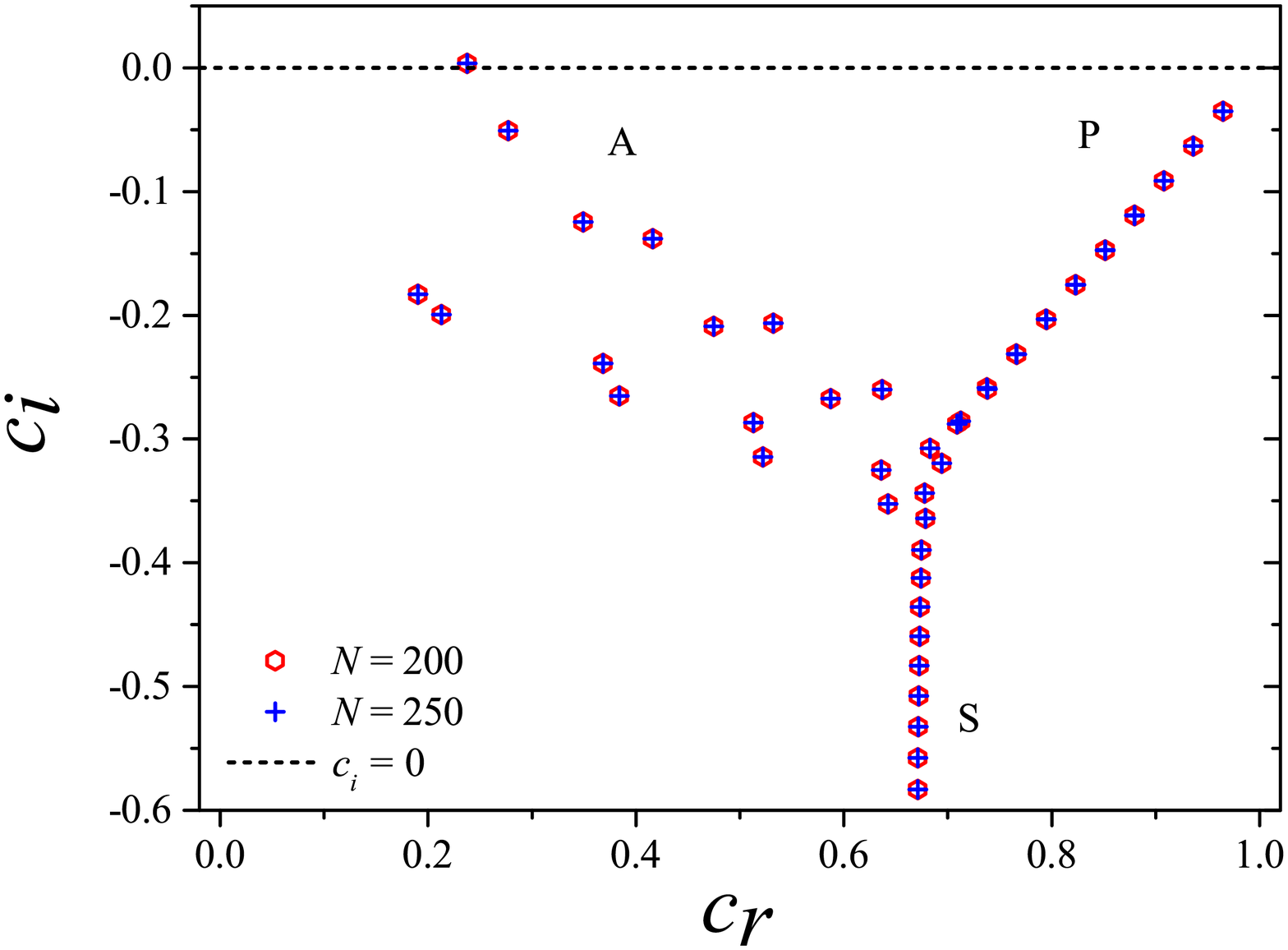}
    \caption{The eigenspectrum for Newtonian plane Poiseuille flow  at $Re = 10^4$, $k = 1$ illustrating the A, P, and S branches characteristic of the Newtonian spectrum. One of the modes belonging to the A branch is unstable for the chosen parameters. The spectrum is obtained for two different values of $N$ (the number of collocation points in the spectral collocation method), and the convergence of the eigenvalues with $N$ demonstrates that the eigenvalues are physically genuine.}
\label{fig:Newtonianspectrum}
\end{figure}

\subsection{The Newtonian Spectrum}
\label{Newt:spectrum}

It is useful to first recall features of the $Re$-dependent Newtonian eigenspectrum for the canonical rectilinear shearing flows \citep{Schimid-Henningson}. By way of illustration, consider plane Poiseuille flow in the $x$-direction with velocity profile $U(y) \propto [1-(y/H)^2]$; here, $2H$ is the separation between the walls with the spanwise base-state vorticity pointing along the $z$-direction. In the linear stability analysis, and within the modal framework, the perturbed velocity field is assumed to be of the form $[U(y) + v_x'(x,y,z,t),v_y'(x,y,z,t),v_z'(x,y,z,t)]$, where the primes denote the perturbation components which are taken to be Fourier modes $v_{x,y,z}'(x,y,z,t) = \tilde{v}_{x,y,z}(y) \exp[ikx + ilz - ikct]$; here, $k$ and $l$ are the wavenumbers in the $x$ and $z$-directions, $\tilde{v}_{x,y,z}(y)$ are the shapes (eigenfunctions) of the perturbations in the $y$ direction, and the complex wavespeed $c = c_r + i c_i$ is the (unknown)\,eigenvalue. If $c_i> 0$, the flow is temporally unstable, and if $c_i < 0$, the flow is asymptotically stable in that perturbations decay away exponentially for sufficiently long times. On account of Squire's theorem, which remains valid for both Newtonian \citep{Drazinreid} and Oldroyd-B \citep{BISTAGNINO2007} fluids, it is sufficient to restrict attention to two-dimensional perturbations\,($l=0$). Note, however, that the theorem is applicable only within the normal-mode ansatz, and it is possible to have a larger nonmodal growth of three-dimensional perturbations, as will indeed be seen in Sec.~\ref{sec:nonmodal}.

Figure~\ref{fig:Newtonianspectrum} shows the Newtonian spectrum at $Re = 10^4$ and $k = 1$,  which has a characteristic `Y-shape' structure in the $c_r$-$c_i$ plane. It consists of (i) the A branch, corresponding to `wall modes' with phase speeds approaching zero with decreasing decay rates, (ii) the P branch corresponding to `center modes' with phase speeds approaching the base-state maximum, again with decreasing decay rates, and (iii) the vertical S branch corresponding to modes having a common phase speed intermediate between the wall velocity and the base-state maximum, and with decay rates extending to infinity. The aforementioned Y-shaped structure only emerges above a threshold $Re$. Below this threshold, whose value is dependent on the particular base-state profile, the Newtonian spectrum comprises only of the S-modes.  For sufficiently large $Re$, the Y-locus itself becomes invariant, with the density of modes along each of the three (invariant)\,branches increasing with increasing $Re$. As evident from Figure~\ref{fig:Newtonianspectrum}, the first mode belonging to the A branch is unstable for the chosen parameters, and corresponds to the well known Tollmien-Schlichting\,(TS) instability. The Newtonian pipe flow spectrum is qualitatively similar to that of plane Poiseuille flow, with the characteristic A, P, and S branches, albeit being stable at all $Re$ \citep{meseguer_trefethen_2003}. In addition, Newtonian plane Couette flow has also been found to be linearly stable. However, unlike plane and pipe Poiseuille flow, on account of an exact antisymmetry about the centerline, plane Couette flow does not possess a P branch; instead, both arms of the Y correspond to A branches with the corresponding eigenfunctions exhibiting a mirror symmetry about the centerline.

\subsection{The purely elastic spectrum and the elastic centermode instability\,($\Re = 0, Wi \neq 0$)}
\label{subsec:zeroRe}

In the absence of inertia, the governing equations in the Newtonian case reduce to the Stokes equations. For Stokes flows driven by the motion of rigid boundaries, the quasi-steady nature of the governing equations and boundary conditions implies there can be no associated spectrum. With reference to the preceding subsection, the $S$-modes in the finite-$Re$ Newtonian spectrum recede down to negative infinity in the Stokes limit.  In contrast, the stress relaxation term in the Oldroyd-B equation provides for an intrinsic time scale, and as discussed below, gives rise to a non-trivial spectrum even in the absence of inertia. We discuss below the nature of this inertialess spectrum whose structure is a function of $Wi$ and $\beta$. Unstable modes in this spectrum correspond to {\it purely elastic} instabilities.

The simplest flow is, of course, plane Couette flow. The elastic plane Couette eigenspectrum was first examined in the UCM limit by Gorodtsov \& Leonov~\cite{Gorodtsov1967}, who showed, analytically, that there is a continuous spectrum (abbreviated as `CS' henceforth) along with two discrete modes, all of which are stable. We refer to the two stable discrete modes as the zero-Reynolds number Gorodtsov-Leonov (`ZRGL') modes. The elastic continuous spectrum is a generic presence, and owes its origin to the spatially local evolution of the polymeric stress\,(in accordance with the simple fluid paradigm, which stipulates a local relation between stress and deformation in a fluid \citep{larson1988constitutive}); the CS eigenfunctions decay exponentially on the scale of the polymer relaxation time. The above picture was generalised to the Oldroyd-B fluid by Wilson, Renardy \& Renardy in 1999~\cite{Wilson1999}. While the flow continues to remain stable, the spectrum becomes considerably more complicated, with new discrete modes appearing for any non-zero $\beta$, thereby pointing to the singular nature of the UCM elastic spectrum. The continuous spectrum associated with the UCM fluid is qualitatively unchanged, as are the two ZRGL modes. But, there is an additional stable continuous spectrum which moves in from $c_i = -\infty$ as $\beta$ increases from zero. Further, unlike the UCM-continuous spectrum, this new so-called viscous continuous spectrum is associated with a branch cut, and discrete eigenvalues can emerge from, or disappear into, the viscous continuous spectrum with varying $\beta$. The number of discrete modes increases with decreasing $\beta$, with there existing an infinite sequence of discrete modes in the limit $\beta \rightarrow 0$; for moderate $\beta$, all of these discrete modes are all more stable than the viscous continuous spectrum modes.

In the aforementioned effort, the authors also analyzed the spectrum of plane Poiseuille flow. In the UCM limit, the authors showed that the equivalent of the Gorodtsov--Leonov spectrum has six discrete modes (instead of the two for plane Couette flow above); numerical computations showed that the discrete modes continued to remain stable. The addition of a solvent viscosity, leading to the  Oldroyd-B model, again resulted in a spectrum similar to plane Couette flow; thus, a second viscous continuous spectrum arose for any non-zero $\beta$, along with a large family of stable discrete modes that disappear into this spectrum with increasing $\beta$.

\begin{figure}
\begin{center}
\includegraphics[scale=0.3]{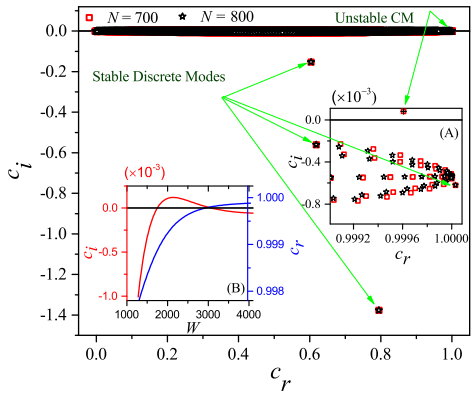}
\end{center}
\caption{\label{fig:spectrum} The  inertialess plane-Poiseuille spectrum of an Oldroyd-B fluid shows five discrete modes for $\beta = 0.997$, $k = 0.75$, $Wi = 2500$, with three in the main figure and the remaining two visible in inset~A; $N$ is the number of Chebyshev polynomials in the spectral expansion.  Inset~A shows an enlarged view of the region near the unstable center mode (`unstable CM'). Inset~B shows the variation of $c_r$ and $c_i$ with $Wi$. Figure reproduced with permission from Khalid \textit{et al.} \citep{khalid2021centermode}.}
\end{figure}

\begin{figure}
\includegraphics[scale=0.26]{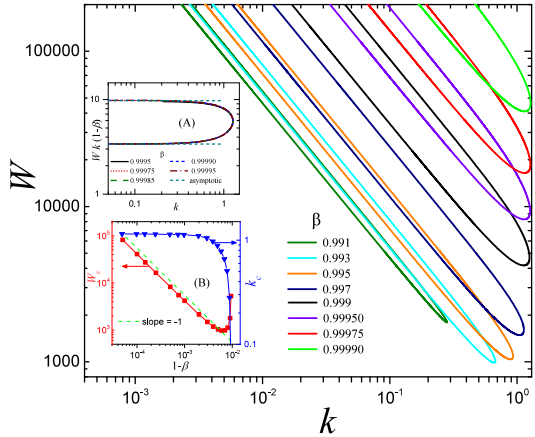}
\caption{\label{neutralcurves} Neutral curves in the $Wi$--$k$ plane for different $\beta$'s in the creeping-flow limit; inset~A shows the collapse for $\beta \rightarrow 1$ when plotted as $Wi\,k(1-\beta)$ vs $k$, and the results obtained from the reduced equations in that limit.
Inset~B shows the variation of the critical Weissenberg number $Wi_c$ and wavenumber $k_c$ with $(1-\beta)$. Figure reproduced with permission from Khalid \textit{et al.} \citep{khalid2021centermode}.}
\end{figure}

The preceding two paragraphs had, until very recently, represented our understanding of the elastic stability characteristics of rectilinear shearing flows. Thus, although never proven, such shearing flow configurations have nonetheless been thought to be linearly stable\,(this is the scenario even for Newtonian pipe flow, although in this case observations clearly point to nonlinear mechanisms). As a consequence, purely elastic linear instabilities are synonymous with curvilinear flow configurations \citep{Shaqfeh1996}, with the analog of such instabilities in rectilinear flows thought to have a nonlinear character\,(see section~\ref{morozov_subsection_A}); in either case, streamline curvature is regarded as a necessary prerequisite for instability \citep{Pakdel_McKinley}. However, recent work by Khalid, Shankar, and Subramanian~\cite{khalid_creepingflow_2021} has demonstrated that inertialess plane Poiseuille flow of an Oldroyd-B fluid is, in fact, linearly unstable at sufficiently high $Wi$\,(of $O(1000)$), and for $\beta > 0.99$. Figure~\ref{fig:spectrum} shows the structure of the elastic spectrum at such high $Wi$'s, and Fig.~\ref{neutralcurves} shows neutral curves, which are in the form of the unstable tongues in the $Wi-k$ plane; the instability appears to arise due to a critical-layer mechanism\footnote{The term `critical-layer'  refers to the location where the base-flow velocity equals the phase speed of the eigenmode. The instability arises due to stretched polymers being rotated away from flow-alignment by the perturbation shear, as they are swept past by the base-state parabolic flow. The differential rate of convection becomes small near the critical layer, owing to the phase speed of the eigenmode approaching the base-flow velocity.
As a result, the time available for the perturbation-shear-induced rotation (of the stretched polymers) increases,
and the resulting accumulation of perturbation elastic
shear stress drives a reinforcing flow, leading to exponential growth. Close to neutrality, this mechanism leads to stress eigenfunctions that exhibit singular features in the neighborhood of the critical layer.} \citep{khalid_creepingflow_2021}, in contrast to the hoop-stress-based mechanism that is operative in curvilinear shearing flows. The work of Buza, Page, and Kerswell~\citep{buza_page_kerswell_2021}, using the FENE-P model, has confirmed that the aforementioned instability continues to exist with the incorporation of finite extensibility. The said authors have also carried out a weakly nonlinear stability analysis to show that the instability in the creeping-flow limit is subcritical, pointing to a potentially larger unstable region in the $Wi-\beta$ plane. At present, it is not yet clear whether
this instability is directly relevant to recent experimental observations from the Paulo Arratia \citep{qin2017elastic,qin2019elastic}  and 
Victor Steinberg groups~\cite{jha2020universal} which clearly indicate an ET-like state for $\Re \ll 1$ even in rectilinear shearing flows, albeit for smaller $\beta \sim$ $0.5$--$0.7$.

\subsection{The elasto-inertial spectrum\,($\Re, Wi \neq 0)$: the wall- and center-mode instabilities instabilities at finite $Re$}
\label{subsec:finiteRe}

The work of Gorodtsov and Leonov\cite{Gorodtsov1967}, referred to in the section above in the context of the inertialess elastic spectrum, also analyzed plane Couette flow of a UCM fluid for small but finite $\Re$. In addition to the aforementioned pair of stable ZRGL modes, the authors found a new class of modes, corresponding to damped shear waves in a viscoelastic fluid with phase speeds of $O(\sqrt{G/\rho})$, $G = \eta/\lambda$ being the shear modulus. We refer to this family of modes as the high-frequency-Gorodtsov-Leonov (`HFGL') modes since, in units of the base-state velocity scale, the phase speed\,(frequency) of the HFGL modes is $O(Wi \Re)^{-\frac{1}{2}}$, and therefore these modes recede to infinity\,(parallel to the $c_r$-axis) in the inertialess limit. Although Gorodtsov and Leonov \cite{Gorodtsov1967} predicted an instability due to the HFGL modes in the limit $k Wi \gg 1$, this was later shown to be incorrect \cite{renardy1986linear}; the HFGL modes remain damped for any finite $Wi$, with $c_i \rightarrow - 1/2kWi$ for $Re \rightarrow 0$. Note that the original elastic continuous spectrum continues to be present at finite $Re$, with the CS-modes having phase speeds in the base-state range of velocities, with decay rates of $c_i = -1/kWi$\,(this corresponds to the dimensional decay rate equaling the inverse relaxation time, as mentioned in section \ref{subsec:zeroRe}). Thus, the elastoinertial spectrum of plane Couette flow of a UCM fluid has been shown \citep{Kumar2005,renardy1986linear} to consist of the finite-$Re$ continuation of the ZRGL modes, the elastic continuous spectrum, and the HFGL modes. Although a rigorous proof does not exist, plane Couette flow does appear to be stable in the $\Re$--$Wi$ plane for $\beta = 0$; the conclusion remains unchanged on consideration of an Oldroyd-B fluid. \citep{lee_finlayson_1986,renardy1986linear,chokshi_kumaran2009,chaudhary_etal_2019}. The stability of viscoelastic plane Couette flow therefore mirrors that of its Newtonian counterpart \citep{Drazinreid}, although there exists a rigorous proof in the latter case  \citep{romanov1973}. In summary, there appears no evidence of a linear instability of plane Couette flow in the $Re$--$Wi$--$\beta$ space.

\begin{figure}
  \centering
  \begin{subfigure}[htp]{0.35\textwidth}
    \includegraphics[width=\textwidth]{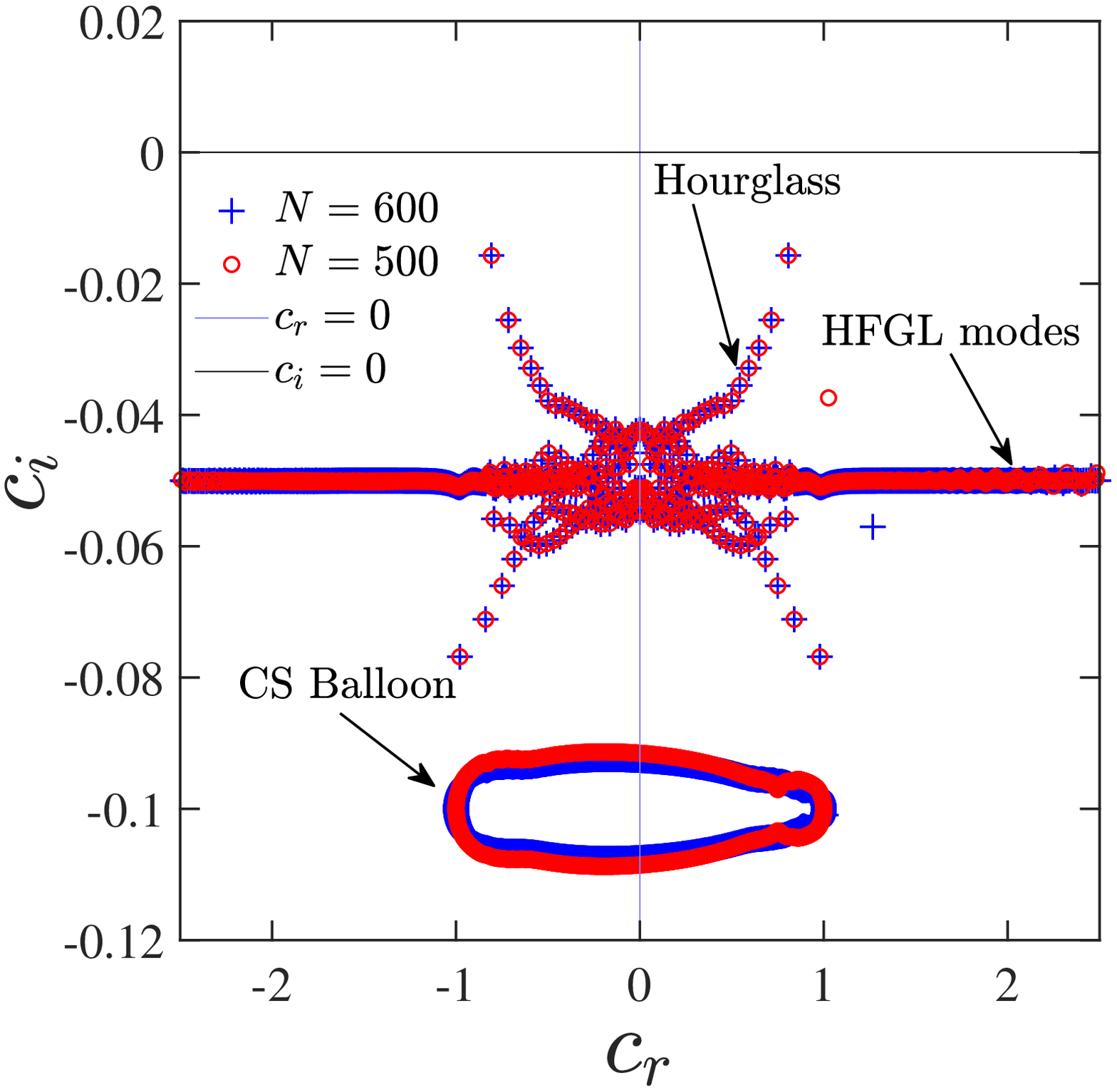}
    \caption{Plane Couette flow}
    \label{fig:uespcf_E0pt001_k1pt0Re10000N600500}
  \end{subfigure}
  \quad\quad
  \begin{subfigure}[htp]{0.35\textwidth}
    \includegraphics[width=\textwidth]{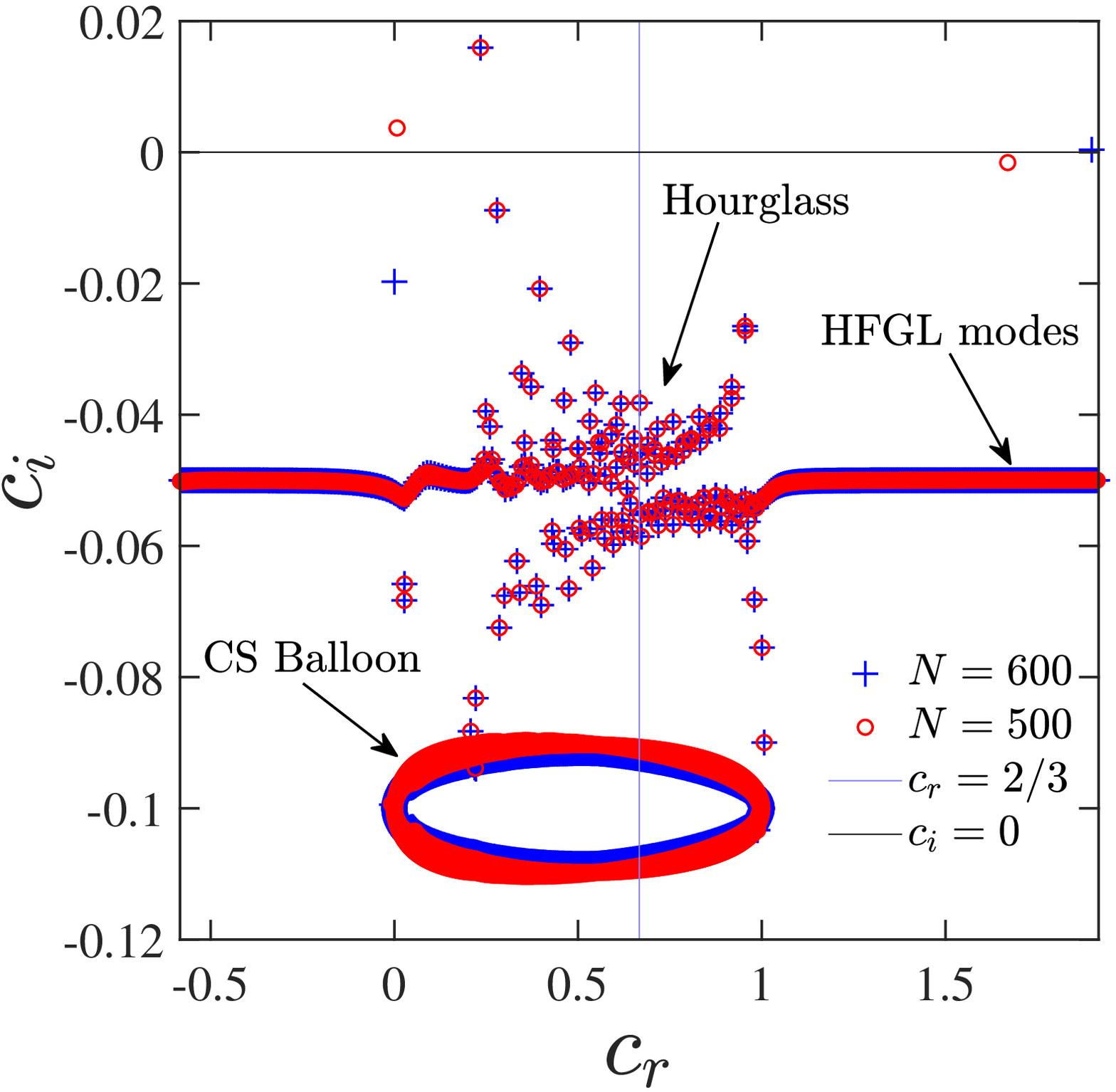}
    \caption{Plane Poiseuille flow}
    \label{fig:uesppf_E0pt001_k1pt0Re10000N600500}
  \end{subfigure}
  \caption{Eigenspectra for (a) plane Couette and (b) plane
    Poiseuille flows of a UCM fluid ($\beta = 0$) for $E = 0.001, k = 1.0, \Re = 10^4$ with $N = $
    $500$ and $600$. The distinctive features are labelled within the
    spectra.  Reproduced with permission from Ref~\cite{chaudhary_etal_2019}.}
  \label{fig:PCFPPF}
\end{figure}

\begin{figure}
  \centering
    \includegraphics[width=0.35\textwidth]{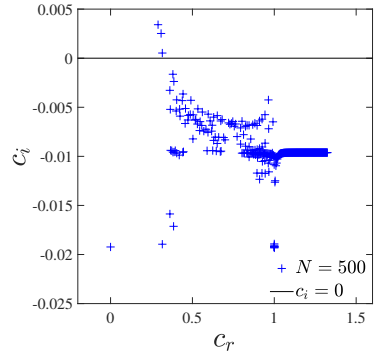}
    \caption{Eigenspectrum for plane Poiseuille flow of a UCM fluid showing three unstable modes for
    $E = 0.005, k = 1.3, \Re = 8000$. Only the converged eigenvalues are shown, and thus the CS balloons are absent in this filtered spectrum. Reproduced with permission from Ref.~\cite{chaudhary_etal_2019}.}
\label{fig:planePoiseuilleUCM}
\end{figure}

In contrast, plane Poiseuille flow of a Newtonian fluid ($Wi = 0$) becomes
susceptible to the TS instability \citep{Drazinreid} at $Re_c \approx 5772$. As already shown in section \ref{Newt:spectrum}, the unstable TS eigenvalue belongs to the A-branch, and is therefore a wall mode. A continuation of this instability is expected for small $Wi$ regardless of $\beta$, including for the case of a UCM fluid. The key question is whether there are new unstable modes in plane Poiseuille flow of a UCM fluid that have an essentially elastic origin, and are therefore absent in the Newtonian limit. This question was first addressed by Porteus and Denn \citep{porteous1972linear}, who found three unstable modes for sufficiently high $\Re\,(> 2000)$, only one of which was a continuation of the TS mode; the other two unstable modes are absent in the Newtonian limit. 
 The authors showed that, for the TS mode, increasing elasticity in the range $0 < E < 10^{-2}$ resulted in a decrease in $Re_c$ from its Newtonian value to $Re_c \sim 2000$. Elasticity was also shown to have a destabilizing effect on one of the other two unstable modes, albeit with limited data.
On the other hand, plane Poiseuille flow of a UCM fluid was found to be stable at low $Re$ \citep{Hodenn1977,lee_finlayson_1986}. Sureshkumar and Beris \citep{sureshkumar1995linear} found two different unstable families, of which one was a continuation of the Newtonian TS mode. The critical Reynolds number $Re_c$ showed a nonmonotonic behavior, showing an initial decrease for very small $E$'s and an eventual decrease at higher $E$. Both the modes analyzed in Ref.\citep{sureshkumar1995linear}, and the initial decrease found in $Re_c$ with $E$, are consistent with the earlier results of Porteus and Denn \citep{porteous1972linear}.

\begin{figure}
\begin{center}
    \includegraphics[width=0.32\textwidth]{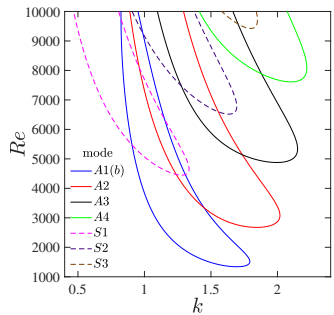}
\end{center}
  \caption{Neutral stability curves in the $Re$--$k$ plane corresponding to the different wall modes  for $E = 0.005$. The modes labeled A1(b), A2, A3, and A4 are antisymmetric, while S1, S2 and S3 are symmetric. Reproduced with permission from Ref.~\cite{chaudhary_etal_2019}.}
  \label{fig:wallmodeneutralloops}
\end{figure}

The recent study of Chaudhary \textit{et al.} \citep{chaudhary_etal_2019} presented a more comprehensive picture of the elasto-inertial spectrum of a UCM fluid, emphasizing features common to both plane Couette and Poiseuille flows. As shown in Fig.~\ref{fig:PCFPPF}, for $Re \sim 1000$ and higher, the elastoinertial spectrum for both flows contains: (i) a ballooned manifestation of the horizontal line\,($c_i = -1/k\,Wi$; $c_r\,\epsilon\,[-1,1]$ for plane Couette, and $c_r\,\epsilon\,[0,1]$ for plane Poiseuille) corresponding to the elastic continuous spectrum, (ii) a horizontal string of eigenvalues corresponding to the aforementioned HFGL modes, and (iii) a roughly `hourglass' shaped structure that extends above and below the HFGL line; note that the length of the HFGL sequence obtained is a function of the numerical resolution of the spectral method, and is smaller for plane Poiseuille flow due to the lower $N$. Despite both spectra conforming to a common template, all modes remain stable for plane Couette flow, as mentioned above, while some of the eigenvalues belonging to the small-$c_r$ `arm' of the hourglass  become unstable at sufficiently high $\Re$ and $E$, for plane Poiseuille flow; see Fig.~\ref{fig:planePoiseuilleUCM}.

%

Chaudhary \textit{et al.} \citep{chaudhary_etal_2019} further showed that plane Poiseuille flow of a UCM fluid is susceptible to an apparently infinite hierarchy of elasto-inertial wall mode instabilities.  In contrast to the antisymmetric Newtonian TS mode, these unstable elastoinertial modes can have either symmetry\,(symmetry, here, is based on the variation of the streamwise velocity eigenfunction, about the centerline, in the wall-normal direction). The multiple unstable tongues in the $Re-k$ plane, for both the antisymmetric and symmetric wall-mode instabilities, are shown in Fig.~\ref{fig:wallmodeneutralloops} for $E = 3.5 \times 10^{-3}$.  The lowest critical Reynolds number was found to be $Re_c \approx 1210.9$ for $E = 0.0066$; $Re_c$ was found to diverge in the limit $E \ll 1$, although the scalings differed for the symmetric\,($Re_c \propto E^{-1}$) and antisymmetric\,($Re_c \propto E^{-\frac{5}{4}}$) modes. Both the unstable wall modes above, that are part of the hour-glass structure, and the HFGL modes, are found to be strongly stabilized on introduction of a solvent viscosity component\,(non-zero $\beta$) \citep{khalid_solvent, khalid2021centermode}. Thus, although relevant to the UCM limit, the wall-mode instabilities are not relevant to the dilute solutions on which most experiments have been performed.

%

 \begin{figure}[htp]
 \centering
                \includegraphics[width=0.32\textwidth]{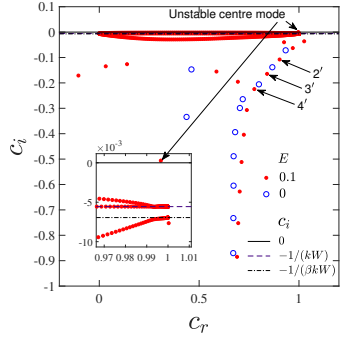}
                \caption{Eigenspectrum for pipe flow of an Oldroyd-B fluid for $E = 0.1$, $\beta = 0.8$, $\Re = 600$, and $k = 3$, and $N = 200$ (red filled circles). The inset shows the unstable center mode. The modes labelled $2'$, $3'$ and $4'$ are discrete center modes that emerge out of the CS as $E$ is increased. The vertical locations of the CS lines and the pipe-flow Newtonian spectrum (open blue circles) at the same $Re$ and $k$ are shown for reference. Reproduced with permission from Ref.~\citep{chaudharyetal_2021}.}
                \label{fig:pipecentermode}
        \end{figure}

%
\begin{figure}
  \centering
  \begin{subfigure}[htp]{0.26\textwidth}
    \includegraphics[width=\textwidth]{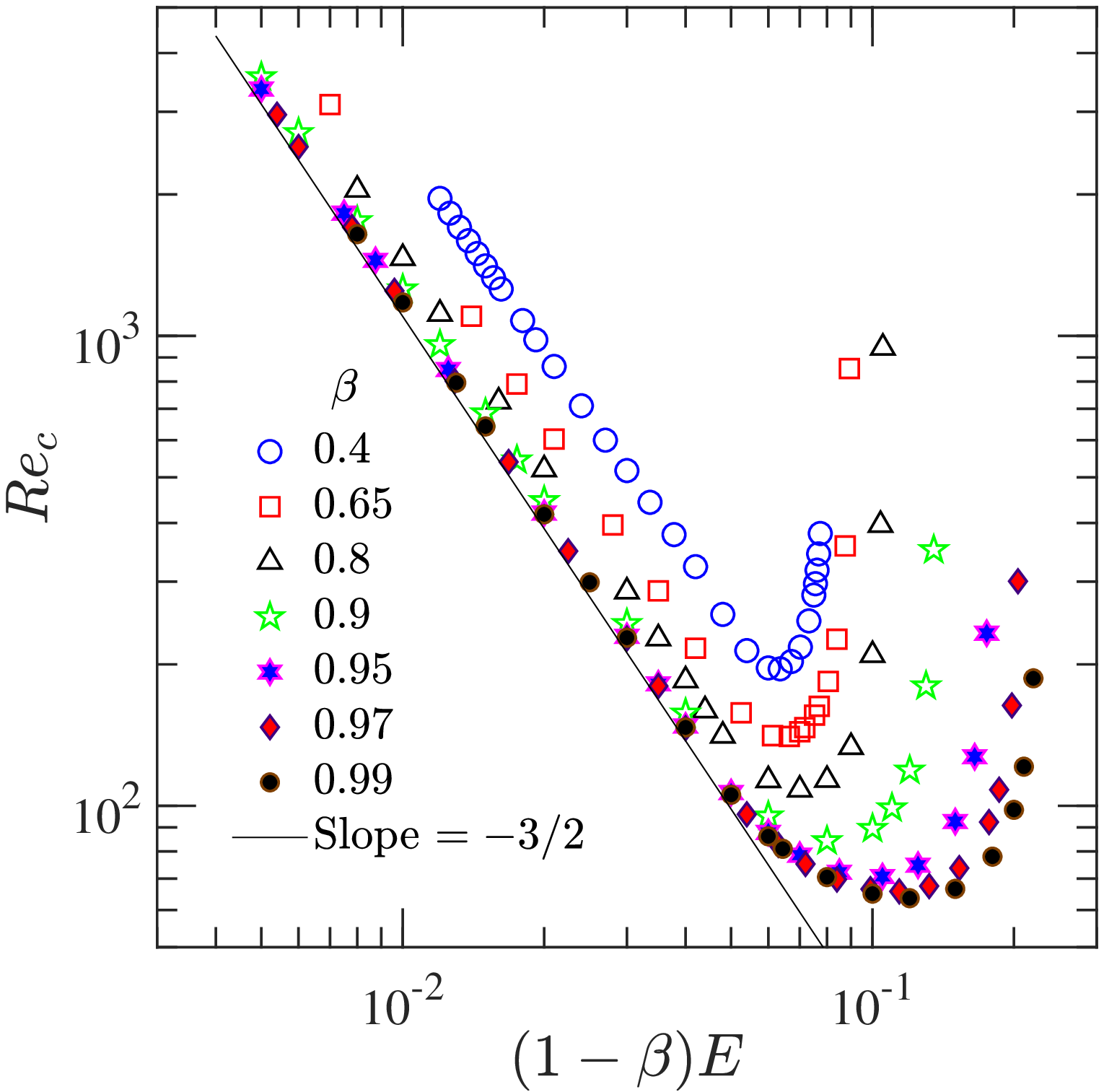}
    \caption{Pipe-Poiseuille flow}
    \label{fig:pipeRecvsE}
  \end{subfigure}
  \quad
  \begin{subfigure}[htp]{0.31\textwidth}
    \includegraphics[width=\textwidth]{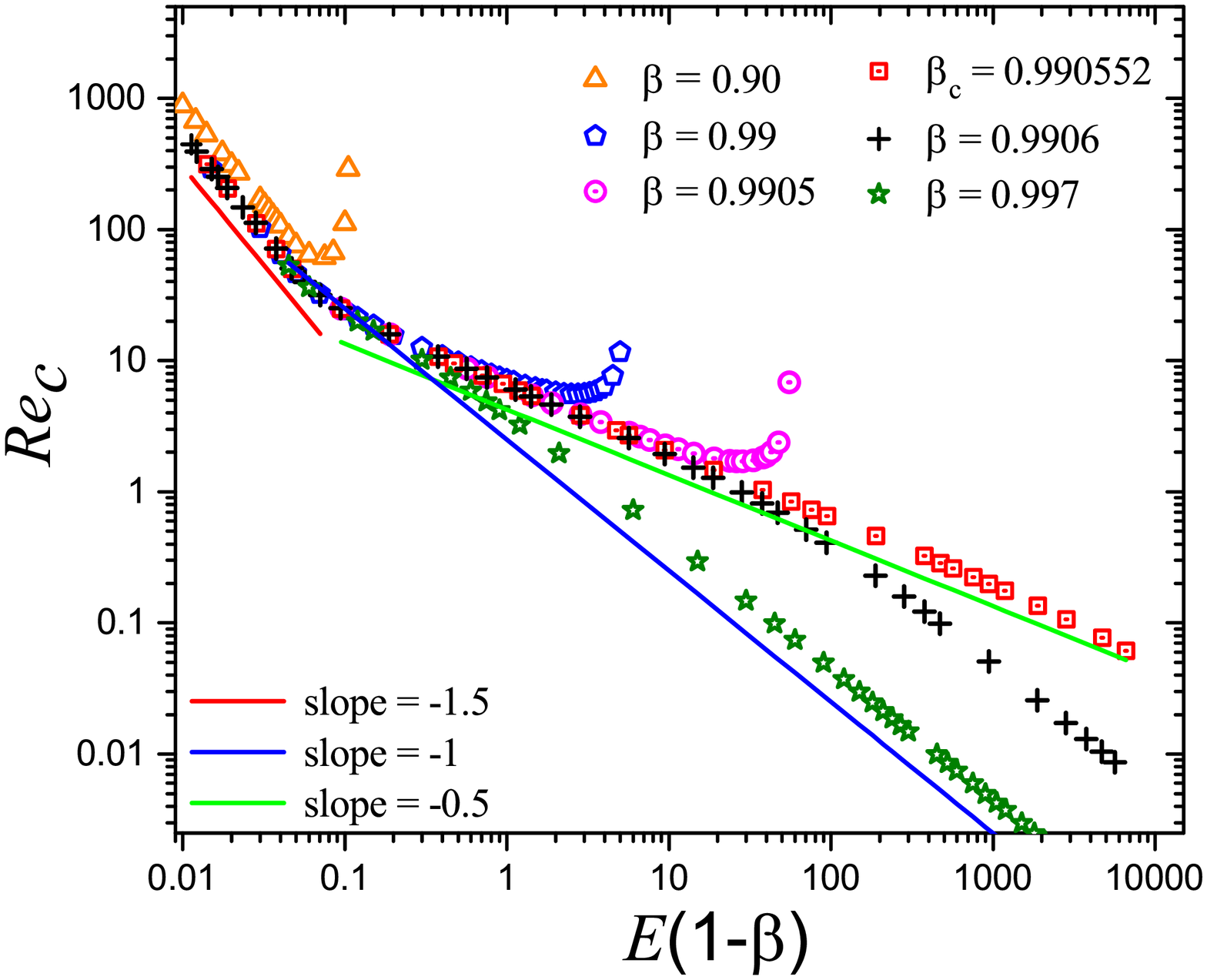}
    \caption{Plane-Poiseuille flow}
    \label{fig:channelRecvsE}
  \end{subfigure}
  \caption{Variation of the critical Reynolds number $Re_c$ with $E(1-\beta)$ for different $\beta$ for both pipe and plane-Poiseuille flows. Figures from Ref.~\citep{chaudharyetal_2021} and Ref.~\citep{khalid2021centermode} reproduced with permission.}
  \label{fig:pipechannelWRe}
\end{figure}

In contrast to the many studies that have focused on the stability of plane Poiseuille flow of an Oldroyd-B fluid, rather surprisingly, there had not been a single study, until recently (see \citep{Piyush_2018, chaudharyetal_2021}),  analyzing the stability of pipe flow of an Oldroyd-B fluid. The only stability analysis in the literature by Hansen \citep{hansen1973,hansen_etal_1973} had neglected the crucial convected nonlinearities in the Oldroyd-B model. The lack of emphasis on pipe flow could perhaps be attributed to the linear stability of Newtonian pipe flow for all $\Re$ \citep{meseguer_trefethen_2003}, in turn leading to the assumption of viscoelastic pipe flow  also being linearly stable in $Re$--$Wi$--$\beta$ space; an assumption that has often found an explicit mention in the literature \citep{bertola2003experimental,Morozov2005,Pan_2013_PRL,Sid_2018_PRF}. This is despite the absence of a systematic exploration of the larger (three-dimensional) parameter space, and inspite of the recent pipe flow experiments of Samanta \textit{et al.}\,\cite{Samanta2013} showing the existence of an perturbation-amplitude-independent threshold $Re$ for transition from the laminar state in sufficiently elastic polyacrylamide solutions, the amplitude independence being a  clear signature of an underlying linear instability.

It is worth pointing out here that two protocols were adopted by Samanta \textit{et al.}\cite{Samanta2013} in the experiments above: in the first protocol, the flow was forced by radial fluid injection near the inlet, resulting in the oft-quoted threshold $Re \approx 2000$ for the Newtonian case. The second protocol did not involve any external forcing, corresponding therefore to a `natural' transition, and occurred at $Re \approx 8000$ in the Newtonian limit. With increase in the polyacrylamide concentration, the threshold $Re$ for the natural transition decreased, while that for the forced transition is increased, and for concentrations greater than $300$ppm, the threshold $Re$ became independent of the  protocol. For the $500$ppm solution in particular, the threshold $Re$ was found to be as low as $800$, and the transition was bereft of signatures such as turbulent puffs that accompany the onset of Newtonian turbulence. As mentioned in the Introduction, the flow state that resulted after the non-hysteretic transition was referred to as elasto-inertial turbulence, to distinguish it from both purely-elastic turbulence and inertial Newtonian turbulence; this reduction in the transition $Re$ has been corroborated by later experiments in pipes of much smaller radii \citep{Bidhan2018,chandra_shankar_das_2020}. The subsequent experimental study of Choueiri \textit{et al.} \citep{Choueiri2018} showed that, at a fixed $Re < 3600$, as the polymer concentration is increased, the frictional drag decreased and approached the maximum-drag-reduction asymptote, in accordance with the well-established paradigm of turbulent drag reduction. However, in a significant departure from this scenario, further increase in polymer concentration resulted in the drag reduction exceeding the MDR asymptote, with the flow relaminarizing completely for a range of polymer concentrations. This laminar state becomes unstable when polymer concentration is increased further, eventually again approaching the MDR asymptote.  As alluded to in Ref.~\cite{Piyush_2018},  the MDR regime could thus be viewed as a `drag-enhanced' state directly accessible via an instability of the laminar state, rather than as a drag-reduced state accessible from Newtonian turbulence.

%

Motivated by the above experiments, and in a significant departure from the Newtonian scenario\,(see end of section \ref{Newt:spectrum}), the recent studies of Garg \textit{et al.}\, \citep{Piyush_2018} and Chaudhary \textit{et al.}\,\citep{chaudharyetal_2021} have shown that pipe Poiseuille flow of an Oldroyd-B fluid is indeed linearly unstable to an axisymmetric center-mode, consistent with the aforementioned experimental observations.  An analogous 2D center-mode instability is predicted for plane Poiseuille flow \citep{Piyush_2018,khalid2021centermode}. Figure\,\ref{fig:pipecentermode} highlights the existence of an unstable center-mode in the pipe elastoinertial spectrum for $\Re =600, E = 0.1 $ and $\beta = 0.8$. The relevance of exponentially growing axisymmetric/2D disturbances is consistent with the signatures seen in recent\footnote{It is worth pointing out that these recent DNS studies differ from the earlier ones on drag reduction \citep[for instance, Refs.][]{sureshkumar_1997,xigrahamPRL2010} in not using an artificially high stress diffusivity that might have led to the absence of spanwise EIT structures in those earlier efforts.} DNS studies \citep{Sid_2018_PRF,lopez_choueiri_hof_2019} of viscoelastic pipe and channel flows. In both cases, the characteristic structures in the elastoinertial turbulent regime are found to be spanwise oriented rolls, in contrast to the streamwise oriented spanwise varying streaks and counter-rotating vortices which are known to underlie the sub-critical Newtonian transition \citep{kerswell_2005}.

Unlike the wall-mode instabilities described above, the center-mode instability is not restricted to small $\beta$ for either pipe or plane Poiseuille flow. In fact, for both flows, the instability appears to require a combination of fluid inertia, elasticity and solvent viscous effects. This may be seen in the limit $Re \gg 1$ when the threshold Reynolds number  $Re_c \propto E^{-3/2}$ for both these flows, a scaling that can only be obtained by balancing fluid inertia, elasticity and solvent viscous effects in a thin layer near the pipe centerline/channel midplane \citep{chaudharyetal_2021}; see Figs~\ref{fig:pipeRecvsE}  and \ref{fig:channelRecvsE}. Thus, in sharp contrast to the known irrelevance of linear (modal)\,stability theory vis-a-vis the Newtonian transition in the canonical shearing flows, a common modal mechanism is predicted to underlie the transition to EIT, in plane- and pipe-Poiseuille flows of an Oldroyd-B fluid, over a significant domain of the $Re$-$Wi$-$\beta$ space, with supporting evidence from both simulations   
 \citep{Sid_2018_PRF,lopez_choueiri_hof_2019} and experiments \citep{choueiri2021experimental}.

While the center-mode instabilities for pipe- and plane-Poiseuille flows share many similarities, there are
crucial differences. The instability ceases to exist for $\beta < 0.5$ in channel flow \citep{khalid2021centermode}, while continuing down to $\beta \sim 10^{-3}$ for pipe flow \citep{Piyush_2018}. Recent work by Wan \textit{et al.} \citep{wan2021subcritical} has found that the center-mode instability is present even in the UCM limit\, ($\beta = 0$) for some isolated regions in the $Re$-$Wi$ space.
The more interesting limit is that corresponding to dilute solutions with $\beta \to 1$. As shown in Fig.~\ref{fig:pipeRecvsE}, for pipe flow, in the aforesaid limit, $Re_c \approx 63$ for $E \rightarrow \infty$, this being the minimum Reynolds number for instability onset. In contrast, as shown in Fig.~\ref{fig:channelRecvsE}, $Re_c$ for plane Poiseuille flow behaves in a similar manner only for $\beta < \beta_c \approx 0.990552$; for $\beta > \beta_c$, the center-mode instability continues to down arbitrarily low $\Re$, with $Re_c \propto E^{-1}$\,(corresponding to a threshold $Wi$) for $E \rightarrow \infty$; in the process, the elastoinertial center-mode morphs into the purely elastic center-mode instability described in section \ref{subsec:zeroRe} \citep{khalid_creepingflow_2021}. While the predictions for $Re_c$ in Figs.~\ref{fig:pipeRecvsE} and \ref{fig:channelRecvsE} correspond to an Oldroyd-B fluid, the use of a nonlinear constitutive equation, such as the FENE-P model, is expected to lead to curves that remain similar in the vicinity of the minimum Reynolds number, but that close up beyond a second larger critical $Re$, owing to shear-thinning-induced stabilization; this feature will again be seen, in the context of the curvilinear instabilities, in Sec.~\ref{subsec:PakdelMcKinley}.


\begin{figure}[htp]
 \centering
                \includegraphics[width=0.4\textwidth]{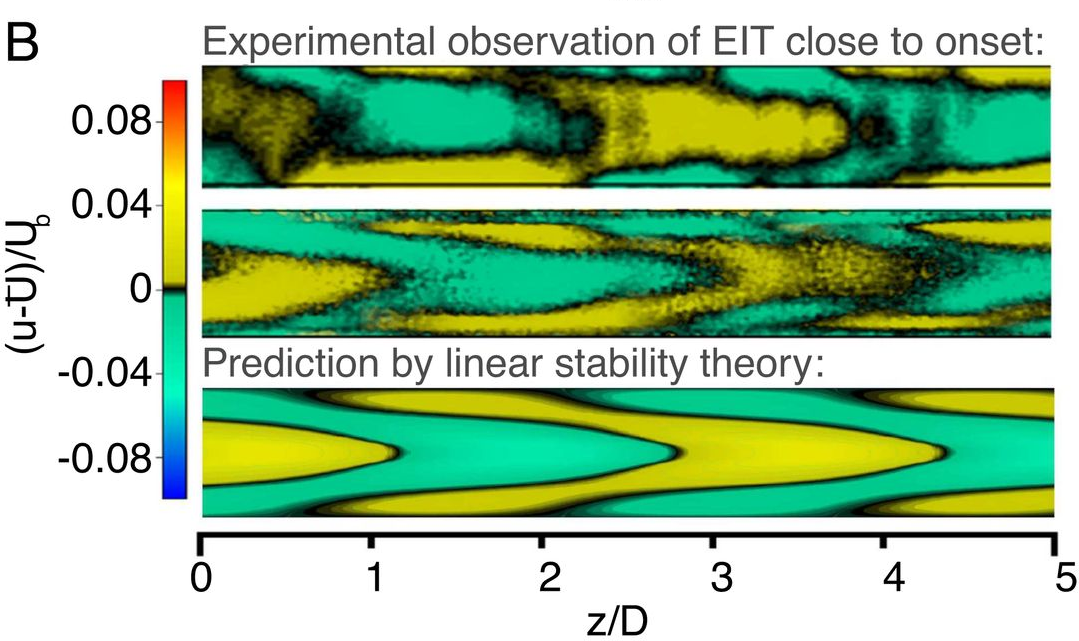}
                \caption{Comparison of experimental flow structures with the eigenfunction for the center mode. Top and Middle show streamwise velocity fluctuations obtained from PIV measurements in a longitudinal cross-section. Top shows flow structures at $Re = 5$ and corresponds to an experiment with $\beta=0.57,E=20.8, Wi=104$, whereas Middle corresponds to $Re=100$ with $\beta=0.5,E =3.4, Wi=304$. Lower shows the most unstable mode in the linear stability analysis; the solution plotted is intended for qualitative comparison only and was computed for different flow parameters: $Re = 100, E=0.6, Wi  =  60, \beta=0.9, n = 0$, and $k = 1$; here, $n$ and $k$ are the azimuthal and axial wave numbers, respectively. Flow direction is from right to left. Reproduced with permission from Fig.~1b of Choueiri \textit{et al.}~\citep[][]{choueiri2021experimental}}
                \label{fig:choueirifig}
        \end{figure}

Finally, getting down to actual numbers, linear stability theory \citep{Piyush_2018,chaudharyetal_2021} predicts a threshold $\Re \approx 800$ for transition, similar to the observations of Samanta \textit{et al.}\,\citep{Samanta2013}, albeit at much higher $E$'s. The theoretical predictions are in better agreement with the pipe flow experiments of Chandra \textit{et al.}\,\citep{Bidhan2018}, an aspect that might have to do with the differing methods used to determine the polymer relaxation times in the two efforts. The recent pipe-flow experiments of Choueiri \textit{et al.}\,\citep{choueiri2021experimental}, however, show excellent agreement between their observations, and theoretical predictions \citep{Piyush_2018,chaudharyetal_2021} for the threshold $Re$, for $E \leq 0.1$.  Further, the experiments  demonstrate a remarkable match (see Fig.~\ref{fig:choueirifig})  between the structures seen immediately after transition and the linear center-mode eigenfunction, while also pointing to a secondary transition to a wall mode.  For $E > 0.1$, the experiments reveal a monotonic decrease of the threshold $Re$ with $E$, while the theoretical predictions \citep{Piyush_2018,chaudharyetal_2021} predict a sharp upturn in $Re_c$\,(see Fig.~\ref{fig:pipeRecvsE}). The experimental threshold appears to indicate a transition from an elastoinertial instability to an elastic one, similar to plane Poiseuille flow, although the elastic branch\,(corresponding to the higher values of $E$) must then correspond to a subcritical (nonlinear) transition. For viscoelastic plane Poiseuille flow, the predictions \citep{khalid2021centermode} are in good agreement with the limited experimental data of  Srinivas and Kumaran \citep{Srinivas-Kumaran2017} for channels with a cross-sectional aspect ratio of $10$:$1$. 

\begin{figure*}
 \centering
 \includegraphics[width=0.7\linewidth]{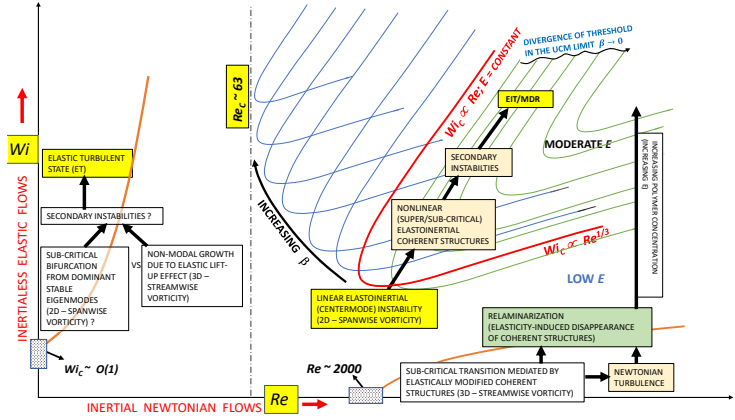}
                \caption{Schematic representation of various transition scenarios in viscoelastic pipe flow in the $Wi$--$Re$ plane.}
                \label{fig:schematicpipe}
\end{figure*}

\subsection{Transition scenarios in rectilinear viscoelastic shearing flows}
\label{subsubsec:scenarios}

\begin{figure*}
  \centering
  \begin{subfigure}[htp]{0.7\textwidth}
 \includegraphics[scale=0.5]{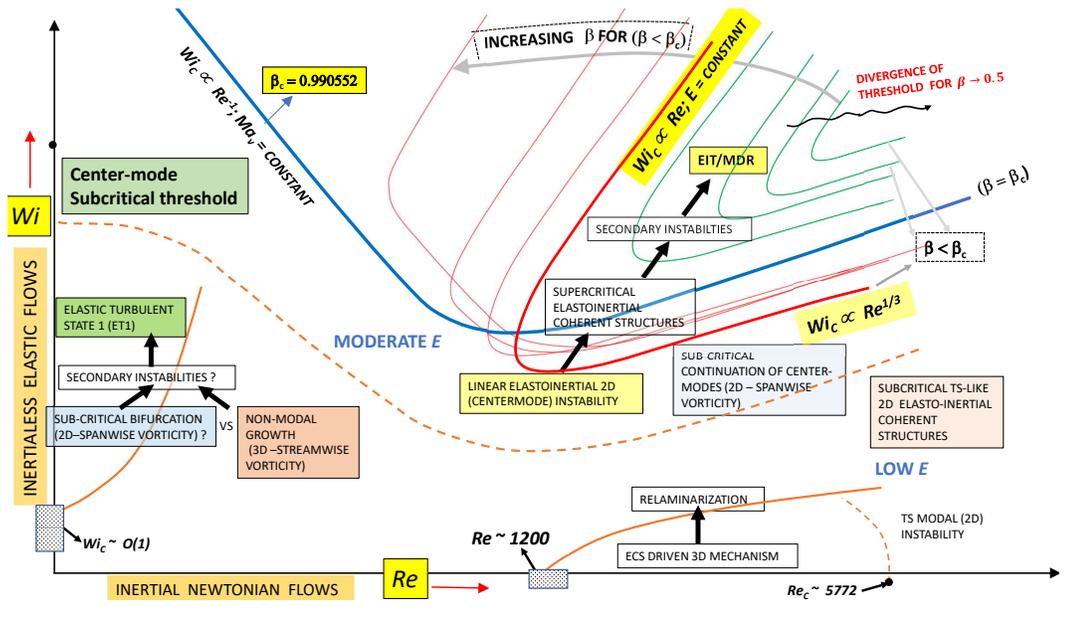}
    \caption{$\beta \leq \beta_c$}
    \label{fig:lowbeta}
  \end{subfigure}
  \quad\quad
  \begin{subfigure}[htp]{0.7\textwidth}
    \includegraphics[scale=0.5]{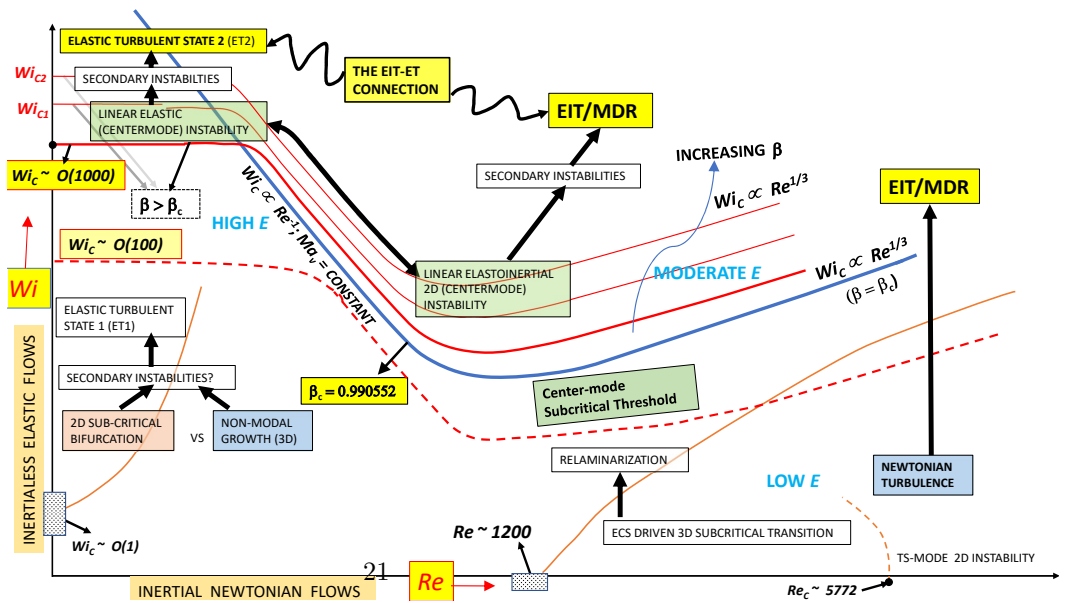}
    \caption{$\beta > \beta_c$}
    \label{fig:highbeta}
  \end{subfigure}
  \caption{Schematic representation of various transition scenarios in viscoelastic channel flow in the $Wi$--$Re$ plane.}
  \label{fig:channelscenarios}
\end{figure*}

Figures~\ref{fig:schematicpipe} and \ref{fig:channelscenarios} illustrate the various possible transition scenarios in the $Wi$-$Re$ plane, for different fixed $\beta$, for pipe and plane Poiseuille flows. In these schematic illustrations, we bring together ideas based on the section above, on the centermode instability, and other hypotheses based on earlier nonmodal and nonlinear analyses\,(see sections \ref{sec:nonmodal} and \ref{sec:nonlinear} respectively for a detailed discussion); we also comment briefly on a recent independent line of work by Graham and coworkers that proposes a new subcritical route to EIT based on elastoinertial TS-wave analogs \citep{Shekar2019,Shekar_2020,Shekar_2021}. In the aforementioned figures, the linearly unstable regions in the interior of the $Wi$-$Re$ plane correspond to the domain of the elastoinertial centermode instability, and are depicted using colored lines for different $\beta$. Regions adjacent to the $Re$ and $Wi$ axes correspond to the onset of predominantly inertial and elastic instabilities, respectively, with the former underlying the sub-critical Newtonian transition. Recall from section \ref{Newt:spectrum} that Newtonian pipe flow is believed to be linearly stable at all $Re$ in sharp contrast to the observed transition at $Re \approx 2000$. Likewise, the presence of the classical TS instability in Newtonian channel flow, at $Re \approx 5772$\,(see section \ref{Newt:spectrum}), is now known to be irrelevant to the observed transition at $Re \approx 1000$. Thus, the inertial Newtonian transition for either flow configuration has a nonlinear subcritical character. Indeed, transition in these flows is now understood to be a complex process triggered by the emergence, via saddle-node bifurcations, of novel three-dimensional solutions called `exact coherent states' (ECS) \citep{waleffe_2001,wedin_kerswell_2004,eckhardt_etal_2007}, with a sufficiently large number of such unstable solutions forming the scaffold of the turbulent attractor in an appropriate phase space.  

We begin with a brief discussion of the features common to the center-mode instability in both pipe and channel flows, before going on to describe those unique to channel flow in Figs.~\ref{fig:lowbeta} and \ref{fig:highbeta}. It has been shown that elasticity suppresses the 3D ECS solutions \citep{stone_graham2002,stone_graham2003,stone_graham2004,li_etal_2006,Graham2007}, making the nonlinear Newtonian-ECS-based mechanism irrelevant for weakly elastic flows. Although this suppression has been demonstrated specifically for plane Poiseuille flow, the prediction should be valid for pipe flow as well, on account of the similarity of the Newtonian ECSs across the different rectilinear shearing flows   
 \citep{Fabian_PRL_1998,waleffe_2001,wedin_kerswell_2004}. The elasticity-induced suppression of the ECS has been proposed to underlie delayed transition and eventual disappearance of the Newtonian turbulent state in the flow of polymer solutions. As a result, in the said figures, the Newtonian turbulent-like state is confined to a region between the $Re$-axis and a curve that corresponds to a critical $Re$-dependent $Wi$. 

For smaller $\beta$, as shown in Fig~\ref{fig:schematicpipe}, the aforementioned Newtonian turbulent state likely gives way to a laminar one with increasing elasticity. Indeed, the vertical arrow shown on the extreme right in Fig.~\ref{fig:schematicpipe} corresponds to the experimental path of Choueiri \textit{et al.} \citep{Choueiri2018} who, starting from Newtonian turbulence, first accessed an intermediate laminar state, and then the MDR regime, with increasing $Wi$, as discussed above in Sec.~\ref{subsec:finiteRe}. On the other hand, for very dilute solutions, as shown in Fig~\ref{fig:highbeta}, the intervening laminar state gives way to overlapping Newtonian and elastoinertial turbulent regions at higher $Re$. The vertical arrow shown in the figure, again on the extreme right, now corresponds to a `reverse' transition where the Newtonian turbulent state exhibits an increasing degree of spatiotemporal intermittency with increasing $Wi$, before giving way to EIT; this was observed to be the pathway, at higher $Re$, in Ref.~\citep{Choueiri2018}.

For sufficiently high elasticities, the linear center-mode instability, discussed in section \ref{subsec:finiteRe} above, becomes relevant. Although the extent of the linearly unstable region depends sensitively on flow-type and $\beta$, the unstable regions for both pipe and channel flows exhibit qualitative similarities for $0.5 < \beta < 0.98$, with $Wi \propto Re^{1/3}$ along the lower branch of the unstable region\,(this scaling corresponds to the $Re \propto E^{-3/2}$ regime in Figs.~\ref{fig:pipeRecvsE} and \ref{fig:channelRecvsE}), and $Wi \propto Re$ along the upper one\,(this corresponds to the near-vertical divergence of $Re_c$ in Figs~\ref{fig:pipeRecvsE} and \ref{fig:channelRecvsE}). Note that $Wi \propto Re$, in corresponding to a constant $E$, also represents an experimental path of increasing  flow rate for a given flow geometry and polymer solution. Thus, as shown in Figs \ref{fig:schematicpipe}, \ref{fig:lowbeta} and \ref{fig:highbeta}, for both the plane and pipe Poiseuille geometries, the centermode eigenfunction is likely to lead to supercritical nonlinear structures that, either directly, or through secondary instabilities, might underlie the dynamics of the EIT state. The centermode instability, for both pipe and channel flows, therefore provides a continuous pathway from the laminar state to the EIT/MDR regime, a prediction that now has been confirmed in experiments \citep{choueiri2021experimental}.

Beyond the aforementioned range of $\beta$, as mentioned above in section \ref{subsec:finiteRe}, there exist significant differences between the pipe and plane Poiseuille cases. Specifically, in the limit $\beta \rightarrow 1$, while the center-mode instability
appears to be restricted to $Re > 63$ for pipe flow (Figs.~\ref{fig:pipeRecvsE} and \ref{fig:schematicpipe}), it morphs into a purely elastic instability for channel flow, continuing to arbitrarily small $Re$ for $\beta > \beta_c \approx 0.990552$ (Fig.~\ref{fig:channelRecvsE}). Correspondingly, in Fig~\ref{fig:highbeta}, the lower boundary of the linearly unstable envelope (with $Wi \propto Re^{\frac{1}{3}}$) opens out into a plateau with decreasing $Re$, approaching a threshold $Wi$ for $Re \rightarrow 0$. This purely elastic instability might in turn lead to an ET state, and the implied continuous (modal) pathway between the EIT and ET states is shown schematically in Fig~\ref{fig:highbeta}. Note that the blue curve in Figs~\ref{fig:lowbeta} and \ref{fig:highbeta} corresponds to the neutral boundary for $\beta = \beta_c$ that demarcates, within a linearized framework, the pure-EIT regime, and the one that exhibits the EIT-ET connection.

In regions of the $Re$-$Wi$-$\beta$ space where the centermode is linearly stable, and the originally Newtonian ECS are stabilized by elasticity, novel subcritical mechanisms are expected to dominate the transition process. For plane Poiseuille flow, at moderate $Re$, recent work \citep{Shekar2019,Shekar_2020,Shekar_2021} has identified a nonlinear mechanism based on elastoinertial wall modes closely related to the stable Newtonian TS mode (although still disconnected from it in phase space until a $Re$ of $10^4$). Such a pathway could be especially relevant in a direct transition between the Newtonian and elastoinertial turbulent states (as in Fig~\ref{fig:highbeta}), with the near-wall coherent structures in the former state acting as possible seeds for the aforementioned TS-wave analogs \footnote{Given that recent experimental evidence points to EIT and MDR states being one and the same, for low to moderate $Re$ values, it is worth mentioning here that the 2D TS-wave-analogs recently proposed to underlie EIT \citep{Shekar2019,Shekar_2020,Shekar_2021} stand in sharp contrast to an earlier interpretation that regarded the MDR regime as corresponding to a hibernating state of turbulence \citep{xigrahamPRL2010,Xi_Graham_2012} comprising 3D so-called edge-state solutions (lying on the basin boundary between the laminar fixed point and the turbulent attractor in an appropriate phase space). Such states already exist in Newtonian turbulence, and their frequency of occurrence is thought to be progressively enhanced with increasing polymer concentration (although, the hibernating periods have been found to be strongly box-size dependent \citep{Xi2019DRreview}). The relation between this earlier edge-state-based hypothesis, and the more recent TS-analog-based hypothesis, needs further investigation.}. However, the fact that there is no analog\,(linear or nonlinear) of the TS-mode in the Newtonian pipe-flow spectrum, and that the centermode remains the least stable one even in the weakly elastic regime \citep{chaudharyetal_2021}, suggests that the TS-analog-based subcritical mechanism may not be obviously applicable to pipe Poiseuille flow; more work is clearly required in this regard.

The recent subcritical continuation of the unstable center mode, in viscoelastic channel flow, to a nonlinear EIT structure \citep{page2020exact} implies that subcritical mechanisms based on the centermode might also be operative in certain regions of $Re$-$Wi$-$\beta$ space, and thus the relevance of the centermode might extend outside of the linearly unstable regions indicated; see the dashed line in Fig~\ref{fig:highbeta}. The very recent weakly nonlinear analyses of Buza \textit{et al.} \citep{buza_page_kerswell_2021} for channel flow and Wan \textit{et al.} \citep{wan2021subcritical} for pipe flow further confirm that the center-mode instability is likely subcritical in large parts of the parameter space. Despite these developments, it is relevant to point out that there still remain vast tracts of the viscoelastic parameter space where the mechanism of transition is not understood. As an example, Khalid \textit{et al.} \citep{khalid2021centermode} have shown that for $\beta = 0.97$, and for $0.02 < E < 0.5$, neither the center mode nor the wall mode is the least stable. Instead, it is the singular modes belonging to the continuous spectrum that are the least stable for these $E$ values, and that might therefore dictate the nature of the transition.While some work has been done on the role of continuous spectra in pattern formulation, in Hamiltonian systems  \citep{balmforth2013pattern}, more work is therefore required to clarify continuous-spectra-dominated transition mechanisms in viscoelastic shearing flows.

The above discussion of transition scenarios has been restricted to either new elastoinertial modal pathways, or the elastic modification of essentially Newtonian nonmodal nonlinear pathways\,(implicit in the examination of the effects of finite $Wi$ on the Newtonian ECS's, mentioned earlier). In the opposite limit of $Re \ll 1$, pipe and plane Poiseuille flows, as indeed all rectilinear shearing flows, are linearly stable for $Wi \sim O(1)$ \cite{Wilson1999,chaudharyetal_2021}, since a linear instability at such $Wi$'s requires a hoop-stress-based mechanism\,(see section \ref{sec:curvilinear}). The absence of a linear instability at moderate $Wi$'s has led to the exploration of novel nonmodal pathways due to elasticity alone \citep{jovanovic_kumar_2010,jovanovic_kumar_2011}, or due to a non-trivial interplay of elasticity and inertia \citep{Zaki2013,PageZaki2014}. While such efforts are discussed in detail in Sec.~\ref{sec:nonmodal} below, it is worth summarizing a few salient points that appear in Figs~\ref{fig:schematicpipe}, \ref{fig:lowbeta} and \ref{fig:highbeta}. The nonmodal pathways, in the inertialess limit in particular, point to the importance of spanwise varying disturbances (much like the Newtonian case) that are amplified by an elastic analog of the lift-up effect \citep{landahl_1980,roy_subramanian2014}, and by an amount that increases with increasing $Wi$. Page and Zaki \citep{pagezakJFM2015} have examined an elastoinertial nonmodal pathwayfor streamwise varying\,(2D) disturbances, termed the reverse-Orr mechanism, on account of the dominant algebraic growth occurring during the phase where the disturbance aligns with the ambient shear flow\,(in contrast to the Orr-mechanism-driven growth dynamics in the Newtonian case \citep{roy_subramanian2014}).

An alternate transition scenario, again relevant to the elasticity-dominant limit, is that of a subcritical 2D nonlinear instability \cite{morozov_saarloos2005,morozov_saarloos2007} based on the classical Stuart-Landau amplitude expansion, an approach originally developed to describe the Newtonian transition \citep{stuart_1960,Watson_1960}. This approach, described in more detail in Sec.~\ref{sec:nonlinear} below, has been demonstrated only for $Wi \sim O(1)$ and $\beta \rightarrow 0$. Both the elastic nonmodal and nonlinear (modal) pathways above are shown in Figs~\ref{fig:schematicpipe}, \ref{fig:lowbeta} and \ref{fig:highbeta}, in the vicinity of the $Wi$-axis, and are believed to trigger transition to an ET state (`ET1' in the figures). The existence of an additional linear instability for $Wi \sim O(1000)$  and $\beta \rightarrow 1$\cite{khalid_creepingflow_2021}, discussed in section \ref{subsec:zeroRe},  implies a possible bifurcation to a distinct elastic turbulent state. It is therefore possible to envisage (at least) two different ET states (ET1 and ET2 in Fig.~\ref{fig:highbeta}), in inertialess plane Poiseuille flow, depending on $Wi$.
There, however, remains a wide intermediate range of $\beta$ ($0 < \beta < \beta_c$) for which the nature of the subcritical transition is not fully understood.

\section{Interfacial instabilities in multilayer and shear-banded flows}
\label{subsec:Interfacial}

Next, we focus on interfacial instabilities which are a major concern in applications\,(coating, coextrusion and others) that involve multi-layer configurations, i.e.\, the flow of immiscible liquids as distinct layers in contact with one another. The objective in the said applications is to obtain composite materials with properties that are a desired combination of those of the individual layers. In order for these layered composites to have the desired properties, uniformity of the individual layers is crucial, in turn implying that instabilities must be avoided during processing; the formation of interfacial waves, for instance, can result in a significant deterioration of the product. Interfacial waves in such configurations are often driven by a stratification (i.e., a sufficiently rapid variation across the layers) of either material (static  or dynamic) or flow properties. For viscoelastic liquids, there  arises the specific scenario of a stratification in the elastic characteristics. 


Interfacial instabilities are well known to occur even in Newtonian fluids, where stratification can be due to the (rapid)\,variation of density, viscosity, and/or velocity across layers. A stratification in fluid density, with the heavier fluid lying above the lighter one, leads to the well-known Rayleigh-Taylor instability; for brevity, we will not discuss the role of density differences here. A difference in the velocities of two co-flowing fluid streams leads to the classic Kelvin-Helmholtz (`shear layer') instability, which has an essentially inviscid origin. 
Azaiez and Homsy \cite{azaiez_homsy_1994} used the Oldroyd-B model (in addition to Giesekus and co-rotational Jeffereys models) to show that elasticity has a stabilizing effect on the shear layer instability, with an increasing $E$ reducing both the growth rates and the unstable interval of wavenumbers. Even in the absence of a density and velocity stratification, a jump in viscosity across an interface can lead to an instability, as first demonstrated by Yih~\cite{Yih1967} for Newtonian fluids.  Yih analysed one of the simplest interfacial flows viz. wall-bounded two-layer plane Couette flow, and found that viscosity stratification can cause a long wave  instability for any non-zero $Re$; see Ref.~\cite{hinch_1984} for a discussion on the underlying physical mechanism. The lack of a threshold $Re$ for the onset of this instability should be contrasted with the Kelvin-Helmholtz instability above.
 An analogue of the viscosity stratification instability is also seen, for instance, in lubricated pipelining~\cite{Preziosi1989}, where a viscous core fluid (typically oil) is lubricated by a thin annulus of viscous fluid (water).

\subsection{Predicting interfacial instabilities using the Oldroyd-B model}

Moving beyond Newtonian fluids, there is the possibility of an elasticity mismatch between fluids having identical shear viscosities and densities. The first study of this scenario was carried out by Waters \& Keeley~\cite{Waters1987} for a two-layer plane Couette flow of Oldroyd-B fluids; however, the authors found no instability due to an error in the interfacial boundary condition.  Chen~\cite{Chen1991} carried out a rather similar calculation for a core-annular coextrusion flow of a pair of UCM fluids, corrected the error above, and discovered a new instability. The predictions were experimentally verified by Bonhomme \textit{et al.} \citep{Bonhomme_etal}. In the limit of long wavelengths, the instability arises due to a \textit{jump} in $N_1$ across the interface. However, the elastic stratification can be stabilizing or destabilizing, depending on the ratio of the volumetric fluxes. 
Hinch {\it et al.}~\cite{Hinch1992} gave a simple physical explanation to show which fluid arrangements would be stable or unstable to varicose \,(as studied by Chen) and sinuous modes, and showed that for cases where both modes are unstable, the sinuous modes were the more dangerous. Further analysis of interfacial instabilities, of UCM fluids in Couette flow, was carried out by Renardy~\cite{Renardy1988}; this effort identified five modes in the short-wave limit, of which only one is an interfacial mode.  The interfacial mode was again shown to become unstable in the said limit due a stratification in elasticity even when the viscosities of the two layers are identical.

These results were extended, still using the Oldroyd-B model, to all wavelengths in symmetric three-layer planar interfacial flows (essentially the 2D analogue of the axisymmetric coextrusion configuration) by Miller, Wilson \& Rallison~\cite{Miller2007A,Wilson1997,Miller2007B}, and to two-layer arrangements in plane Poiseuille flow by Su \& Khomami~\cite{Su1992}. More recent works have extended the above efforts in various directions: (i) to very high Weissenberg numbers~\cite{Miller2007A,Miller2007B}; (ii) to two-layer flows with moving boundaries (Couette-Poiseuille flow)~\cite{Chokshi2015}; (iii) by analyzing the effect of surfactants~\cite{Peng2011}, and (iv) by exploring the use of deformable solid boundaries as a way of suppression of interfacial instabilities \citep{Shankar2004,Shankar2005}.

\subsection{Shear banding and instabilities in the banded state}
\label{subsubsec:beyondOldB}
%

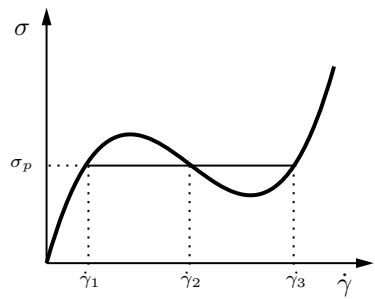
\begin{figure}[htb]
\centering
\tikzset{every picture/.style={line width=0.75pt}} 
\begin{tikzpicture}[x=0.75pt,y=0.75pt,yscale=-1,xscale=1]
\clip (80,0) rectangle (270,160); 
\draw (99.88,134.25) -- (99.88,8.25) ;
\draw [shift={(99.88,5.25)}, rotate = 450] [fill={rgb, 255:red, 0; green, 0; blue, 0 }][line width=0.08]  [draw opacity=0] (8.93,-2.5) -- (0,0) -- (8.93,2.5) -- cycle;
\draw (99.88,134.25) -- (261.88,134.25) ;
\draw [shift={(264.88,134.25)}, rotate = 180] [fill={rgb, 255:red, 0; green, 0; blue, 0 }][line width=0.08]  [draw opacity=0] (8.93,-2.5) -- (0,0) -- (8.93,2.5) -- cycle;
\draw  [line width=1.5]  (99.88,134.25) .. controls (148.22,-38.44) and (196.55,207.7) .. (244.88,35) ;
\draw [line width=0.75]  (119,85) -- (225.88,85) ;
\draw  [dash pattern={on 0.84pt off 2.51pt}]  (121,85) -- (121,135) ;
\draw  [dash pattern={on 0.84pt off 2.51pt}]  (172,85) -- (172,135) ;
\draw  [dash pattern={on 0.84pt off 2.51pt}]  (224.5,85) -- (224.5,135) ;
\draw  [dash pattern={on 0.84pt off 2.51pt}]  (100.88,85) -- (119,85) ;
\draw (82,12) node [anchor=north west][inner sep=0.75pt] {$\sigma $};
\draw (245,138) node [anchor=north west][inner sep=0.75pt]{$\dot{\gamma }$};
\draw (80,80) node [anchor=north west][inner sep=0.75pt][font=\scriptsize]{$\sigma _{p}$};
\draw (115,137) node [anchor=north west][inner sep=0.75pt][font=\scriptsize]{$\dot{\gamma }_{1}$};
\draw (166,137) node [anchor=north west][inner sep=0.75pt][font=\scriptsize]{$\dot{\gamma }_{2}$};
\draw (219,137) node [anchor=north west][inner sep=0.75pt][font=\scriptsize]{$\dot{\gamma _{3}}$};
\end{tikzpicture}
\caption{\label{fig:shearBanding0} Non-monotonic flow curve, as exhibited by the Johnson--Segalman model. If the applied stress is $\sigma = \sigma_p$, flow can coexist at two shear-rates $\dot{\gamma}_1$ (low-viscosity band) and $\dot{\gamma}_3$ (high-viscosity band). The intermediate shear-rate $\dot{\gamma}_2$ is thermodynamically unstable.}
\end{figure}

As discussed in another paper in this special issue~\cite{Hinch_Harlen2021}, Oldroyd introduced the Oldroyd-A model \citep{Oldroyd1950} as well as the more famous Oldroyd-B, implying the existence of everything between the two.  In other words, there exist valid constitutive equations with any of a one-parameter family of convected derivatives (collectively dubbed the `Gordon-Schowalter' (GS) derivative \citep{larson1988constitutive}) all of which are consistent with the principle of material frame indifference. The aforementioned parameter appears as a slip parameter, $a$, in the GS derivative. The Johnson-Segalman model replaces the upper-convected derivative in the UCM model with the GS derivative~\cite{Johnson1977}. With varying $a$, the convected derivative in this model transitions from a lower convected derivative\,(Oldroyd-A; $a = -1$) to an upper convected one\,(Oldroyd-B $a = 1$) via the corotational or Jaumann derivative $(a = 0)$. The resulting response in viscometric flows has a shear-thinning character for all $a$ values except those corresponding to the Oldroyd-B and Oldroyd-A limits. Further, for some values of $a$\,(and $\beta$), the shear thinning is intense enough that the shear stress exhibits a non-monotonic dependence on shear rate, as shown in Fig.~\ref{fig:shearBanding0}.

%

One of the most striking phenomena arising from the non-monotonicity of the flow curve is \textit{shear banding}, in which a simple shear flow of a complex fluid,  with a shear rate in the intermediate mechanically unstable portion,   spontaneously separates into high- and low-shear-rate bands \cite{yerushalmi1970stability,Cates2006}. There has been widespread interest in such banding flows since the phenomenon was first reported in the early 1990sin the context of worm-like micellar solutions \cite{Rehage91,Callaghan1999}.
Surprisingly, shear-banded flows of worm-like micellar solutions are themselves unstable \cite{Lerouge2010,Fardin2012b} and exhibit a variety of instabilities ranging from purely elastic instabilities localised in the more elastic band \cite{Fardin2010} to instabilities of the interface between the bands \cite{Lerouge2006,Lerouge2008,Nghe2010,Yamamoto2008}. Although the primary banding instability is beyond the scope of the Oldroyd-B model, significant understanding of the secondary instabilities of banded micellar systems can be gained from drawing analogies with the interfacial instabilities discussed above, and the purely elastic bulk instabilities discussed in detail in Sec.~\ref{sec:curvilinear} below, both based on the Oldroyd-B model. 

Thus, in shear-banded flows, one can capture the subsequent instabilities in the banded state by treating each band as a  distinct Oldroyd-B fluid. For instance, the interfacial instabilities seen in shear-banded Couette and plane Poiseuille flow
\citep{Cromer2011,Fielding2005,Wilson2006,Fielding2010a} can be explained, at least in the long-wave limit, by Chen's mechanism discussed above \citep{Chen1991}, adapted to allow for the fact that the two shear bands have mismatches of viscosity as well as of $N_1$. It has also been shown that the Pakdel-McKinley criterion, discussed below in Sec.~\ref{subsec:PakdelMcKinley},  can be adapted to describe (bulk) instabilities in shear-banded flows \cite{Fielding2010a}. 
  

%
%

In a very recent paper, Castillo \& Wilson~\cite{Castillo2020} found an interfacial instability while analyzing the stability of channel flow of a shear-banded thixotropic-viscoelasto-plastic fluid, which exhibits shear banding; thus, this instability shares some similarities with the instability of the shear banded state discussed above. For the aforesaid configuration, $N_1$ varies continuously across the interface, and there is only a jump in the viscosity; thus, the instability may be regarded as the elastic version of the inertial instability analyzed by Yih \cite{Yih1967}. The authors were able to reproduce the main points of the instability (seen in a highly shear-thinning fluid with many constitutive complications) by using two Oldroyd-B fluids, with the the interfacial value of $N_1$ matched, but having different shear viscosities. The explanatory power of the Oldroyd-B model is again seen to extend well beyond its expected realm of validity.

\section{Instabilities in curvilinear shearing flows}
\label{sec:curvilinear}

The term `purely elastic' instabilities has traditionally been used to refer to the instabilities observed in flows of viscoelastic fluids, in geometries with curved streamlines, including in particular viscometric configurations such as the Taylor-Couette, cone-and-plate and parallel-plate geometries.  This class of instabilities is present even when  inertial effects are not significant.  As already mentioned in the introduction, a precise prediction of the domain of existence of these instabilities is of immense importance to rheological characterization of polymeric liquids, since the inference of rheological properties presupposes the existence of viscometric flows in the aforementioned geometries; the occurrence of instabilities corrupts rheological measurements, precluding characterization. Further, the instabilities are of relevance to coating applications, and other polymer processing scenarios where flow configurations akin to the said viscometric flows occur. 
Excellent comprehensive reviews by the pioneers of this field, earlier ones by  Larson \citep{larson1992} and Shaqfeh \citep{Shaqfeh1996}, and the more recent one by Muller \citep{muller_review}, already exist in the literature; the goal of this section is to provide a self-contained, but more up-to-date summary of this important and novel class of instabilities. In fact, a discussion of these instabilities is all the more pertinent to the present review, because their prediction is one of the prominent success stories of the Oldroyd-B model. 


\subsection{Effect of viscoelasticity on the Newtonian Taylor-Couette instability}
\label{subsec:VEonNewtonianTC}
Purely azimuthal flow of a Newtonian fluid between concentric cylinders (the Taylor-Couette configuration), becomes unstable due to (inertial) centrifugal effects\citep{Taylor1923}. 
The instability is absent when only when both cylinders rotate in the same direction with the angular velocity of the outer cylinder exceeding that of the inner cylinder by the ratio $(R_{out}/R_{in})^2$, $R_{out}$ and $R_{in}$ being the radii of the outer and inner cylinders, respectively. 
This is consistent with the Rayleigh criterion for inviscid instability that requires the base-state angular momentum to monotonically decrease (in magnitude) with increasing radius \citep{Drazinreid}.
The domain of existence of both the primary linear instability and the various higher order transitions has been well characterized 
in a parameter plane consisting of the Reynolds numbers based on the radii and angular velocities of the inner and outer cylinders \citep{andereck_liu_swinney_1986}.  Excluding the case of strong counter-rotation, the unstable mode, at onset, is axisymmetric and stationary (i.e., with  a zero frequency). A similar centrifugal instability is also present in `Dean flow' entailing pressure-driven flow through a curved channel, and originally analyzed in the limit where the channel width is small compared to the radius of curvature  \citep{Dean1928}. When a combination of cylinder rotation and a streamwise pressure gradient drives the flow, the resulting centrifugal instability is dubbed the `Taylor-Dean' instability, and may be achieved experimentally by inserting a meridional obstruction in the original Taylor-Couette geometry\citep{joo_shaqfeh_1994}.

Early efforts by Ginn and Denn \citep{Ginn_Denn_1969} probed the role of weak viscoelasticity on the centrifugal instability using a second-order fluid model. The analysis showed that positive values of the first normal stress difference ($N_1$), corresponding to a tension along the base-state azimuthal streamlines, had a destabilizing effect. In contrast, a negative second normal stress difference ($N_2$), corresponding to a tension along the base-state axial vortex lines, had a stabilizing effect. The latter effect could be interpreted as being due to the resistance of the tensioned vortex lines to bending caused by axially modulated perturbations. This is somewhat analogous to the work of  Azaiez and Homsy \citep{azaiez_homsy_1994}  mentioned in the subsection above, where the bending resistance of tensioned streamlines acts to stabilize the viscoelastic shear layer. In the narrow-gap limit of $\epsilon \ll 1$ ($\epsilon$ being the ratio of gap width between the cylinders and the inner cylinder radius), the effect of an $N_2$ appears at a lower order in $\epsilon$ than $N_1$.
%
Hence, although $N_2$ is usually negative and much smaller in magnitude than $N_1$ (typically 10-30\% for polymer melts, and smaller for polymer solutions \citep{Maklad_Poole_2021}), the stabilizing or destabilizing action of viscoelasticity can nevertheless be expected to depend sensitively on $N_2$ for $\epsilon \ll 1$.

The second-order model used in  Ref.~\citep{Ginn_Denn_1969}, being restricted to weak flows in the quasi-steady limit, is only valid for $Wi \ll 1$ \citep{TannerBook}; experiments, particularly those examining instabilities, are most often  performed outside this regime, especially in the narrow-gap limit.
Using the more realistic UCM model,  Walters and coworkers \citep{thomas_walters_1964,thomas_walters_1964b,Beard1966} again found that $Re_c$ for the stationary Newtonian mode decreased with $E$ for small $E$, this decrease being consistent with the destabilizing role of a positive $N_1$ mentioned above. A new oscillatory `inertio-elastic' mode was found to become more unstable for higher $E$.
%
%
%
Note that this new oscillatory unstable mode is not connected to the Newtonian limit, and is therefore not captured by the second-order fluid model.  Beard \textit{et al.} \citep{Beard1966} found the $Re_c$ for this mode to also decrease monotonically with $E$ in the range $ 0 < E < 1$, although the authors did not extend their computations all the way down to the inertialess limit ($Re = 0$ or $E = \infty$).  The destabilizing effect of weak elasticity on the Newtonian centrifugal instability also holds for the Dean and Taylor-Dean configurations \citep{Joo_Shaqfeh_inertia_1992}.

\subsection{The purely elastic Taylor-Couette instability}
\label{subsec:purelyelasticTC}

\begin{figure}[htp]
 \centering
                \includegraphics[width=0.4\textwidth]{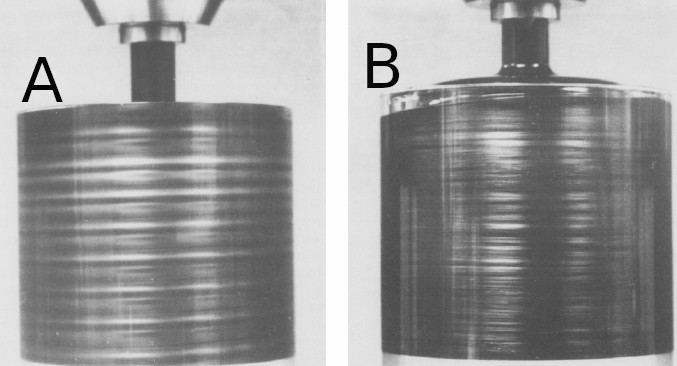}
                \caption{Visualization of Taylor-Couette flow using reflective mica flakes. Panel~A shows the patterns for flow of a Newtonian fluid with Taylor number $3.8 \times 10^3$, while panel~B pertains to viscoelastic Taylor-Couette flow with $Ta = O(10^{-8})$ and $De = 20$; here, $Ta = Re^2 \epsilon$. Figure reproduced with permission from Fig.~1 of Muller \textit{et al.}~\citep[][]{muller_etal_1989}}
                \label{fig:mullerfig}
\end{figure}

While Giesekus reported evidence for the onset of a cellular instability in Taylor-Couette flow of polymer solutions as early as 1966 \citep{Shaqfeh1996} at Reynolds numbers of $O(10^{-2})$, it is
the pioneering theoretical-cum-experimental efforts of Larson, Muller and Shaqfeh \citep{muller_etal_1989,larson1990purely,shaqfeh_muller_larson_1992} that led to the unequivocal establishment of an  inertialess instability in viscoelastic Taylor-Couette flow. 
In addition to carrying out 
a classical modal stability analysis, using the Oldroyd-B model in the inertialess limit,  the authors also characterized the transition experimentally using a Boger fluid\footnote{The term `Boger fluid' refers to a class of fluids prepared by dissolving small amounts of high-molecular weight polymer in a very viscous solvent \citep{James_AFM2009}, which leads to a high elasticity (owing to the long relaxation time) but negligible shear-thinning (in the viscosity); further, on account of an intermediate-shear  plateau in $\Psi_1$ \citep{Quinzani_1990}, these fluids have served as model systems reasonably well described by the Oldroyd-B model over a range of shear rates corresponding to the aforementioned plateau. 
Since the elasticity number $E$ scales directly with the solvent viscosity, and inversely with the square of the flow length scale, elastic effects can be enhanced, and inertial effects simultaneously suppressed, by use of Boger fluids in microscale flows};  figure~\ref{fig:mullerfig} shows the secondary\,(toroidal) recirculation patterns for both the Newtonian and purely elastic cases \citep{muller_etal_1989}.
The theoretical predictions, obtained for axisymmetric disturbances in the thin gap limit, were in qualitative agreement with experimental observations. Unlike the Newtonian case, the unstable mode existed with rotation of either cylinder, and was found to be oscillatory at onset, with the dominant measured frequency in good agreement with theory. 
 However, experiments showed the
vertical length scale of the cellular pattern at onset to correspond to an axial wavenumber smaller than the theoretical prediction.  
Interestingly, the cellular pattern in the experiments continued to evolve over times much longer than the nominal polymer relaxation time, with the cell height eventually shrinking to half its initial value. The authors attributed this discrepancy to the relatively flat neutral curve, implying the excitation of unstable modes across a broad spectrum of wavenumbers even in the immediate vicinity of the threshold, and the resulting nonlinear interactions then contributing to the aforementioned evolution\,(as discussed below, consideration of nonaxisymmetric disturbances leads to better agreement).
Further, 
the measured critical Weissenberg number was typically found to be  between $0.5$--$0.9$ times the predicted value. Plausible reasons behind this discrepancy are discussed in Sec.~\ref{subsec:exptsnonisothermal} below.
Analogous elastic instabilities have also been predicted in the Dean flow \citep{joo_shaqfeh_1991} and Taylor-Dean flow configurations \citep{joo_shaqfeh_1994}, with the unstable mode in the latter case changing from an oscillatory to a stationary one, as the pressure gradient becomes dominant in relation to cylinder rotation.

Both the elastic Taylor-Couette and Dean instabilities owe their origin to either the base-state or perturbation hoop stress fields that arise on account of tension along the curved base-state streamlines. For the former flow, Larson \textit{et al.} \citep{larson1990purely} proposed a physical mechanism based on a dumbbell model for a polymer molecule, consistent with the microscopic picture underlying the Oldroyd-B equation used for the stability analysis. The toroidal circulation associated with the axisymmetric unstable eigenmode leads to an extensional flow in the meridional plane that stretches the dumbbell in the radial direction.
This stretched dumbbell is now acted upon by the base-state azimuthal shear, which tilts it, leading to an increased separation between the beads along the azimuth (see Fig.~12 of Ref.~\citep{larson1990purely}). This drives a perturbation normal stress in the $\theta \theta$ direction\,(the hoop stress), which in turn produces a radial perturbation pressure gradient. The radial flow driven by this pressure gradient is out of phase with the original extensional flow, but for sufficiently high $Wi$, overwhelms the original radial perturbation,  leading to overstability, and a growing oscillatory response. 
Note that, on account of the underlying hoop stress, the mechanism above is relevant only for flows with curved streamlines, and and is absent, at linear order, in the rectilinear shearing flows discussed in Sec.~\ref{sec:rectilinear}. However, the mechanism can operate at a nonlinear order in rectilinear flows, where the streamline curvature itself arises on account of the perturbation, and this scenario is discussed in more detail in Sec.~\ref{sec:nonlinear}.

\subsection{Finite-gap effects and nonaxisymmetric disturbances}
\label{subsec:finitegap}

The early theoretical efforts \citep{muller_etal_1989,larson1990purely} were restricted to small gap-widths ($\epsilon \ll 1$) and axisymmetric disturbances. In order for the effects of curvature, in the Oldroyd-B model, to remain important for $\epsilon \ll 1$, one requires $Wi \sim \epsilon^{-1/2}$, and expectedly, the linear analysis \citep{larson1990purely} yields a threshold value of $Wi_c\,\epsilon^{\frac{1}{2}}$ for instability\,(note that, when inertial effects are negligible, the limit $Wi\,\epsilon^{1/2} \ll 1$ corresponds to plane Couette flow which, as already seen in section \ref{sec:rectilinear}, is linearly stable \citep{Gorodtsov1967,renardy1986linear,chaudhary_etal_2019}).
The above $Wi-\epsilon$ scaling, and similar scalings found for the other curvilinear flows (the parallel-plate and cone-and-plate geometries, with $\epsilon$ being replaced by the pertinent geometric factor), are discussed below in Sec. \ref{subsec:PakdelMcKinley}, in the context of the Pakdel-McKinley criterion.

The restriction to axisymmetric disturbances, for $\epsilon \ll 1$, is often justified based on the assumption of the azimuthal variation of the perturbation being on scales comparable to the cylinder radii; as a result, for $\epsilon \to 0$, both axisymmetric and nonaxisymmetric modes are governed by the same leading order eigenvalue problem. Working within the axisymmetric disturbance framework, Shaqfeh \textit{et al.} \citep{shaqfeh_muller_larson_1992} showed that, as $\epsilon$ increases, $Wi_c$ from both experiments and the finite-gap theory decreases much more slowly with $\epsilon$ than the aforementioned prediction\,($Wi_c \propto \epsilon^{-1/2}$) of small-gap theory. Significant (positive)\,deviations of the threshold are already predicted for $\epsilon = 0.05$, pointing to the restrictive range of validity of the small-gap assumption, with $Wi_c$ becoming nearly independent of $\epsilon$ in the range 0.1 to 0.25, this being consistent with experimental observations (Fig.~11 of Ref.~\citep{joo_shaqfeh_1994}).

As pointed out above, the effect of nonaxisymmetry is only expected to enter the analysis at a higher order for $\epsilon \ll 1$. Indeed, Joo and Shaqfeh \citep{joo_shaqfeh_1994} showed that the terms differentiating the various nonaxisymmetric modes are $O(\epsilon^2 Wi^3)$. Since $\epsilon Wi^2 \sim O(1)$ for curvature effects to remain finite, these additional terms are seen to be $O(\epsilon^{1/2})$, and therefore, asymptotically small compared to the leading order axisymmetric ones. However,
detailed calculations showed the thin gap assumption to again be very restrictive. Thus, Joo and Shaqfeh, even for $\epsilon \sim 10^{-4}$, found the additional terms\,(characterizing departure from axisymmetry) to be important; nonaxisymmetric modes, in fact, turned out to be more unstable than the axisymmetric mode.  
Joo and Shaqfeh \citep{joo_shaqfeh_1994} found the results obtained using nonaxisymmetric disturbances to be in good qualitative agreement with experimental observations of the variation of $Wi_c$ with $\beta$ and $\epsilon$. Further, the wavenumber of the axial vortices developing at onset, determined using an image analysis, was found to be in good agreement with the prediction for the most unstable nonaxisymmetric mode\,(about 5.5 inverse gap thicknesses). The results of Joo and Shaqfeh were also consistent with the earlier findings of Avgousti and Beris \citep{Avgousti1993}, who again found a nonaxisymmetric oscillatory mode in viscoelastic Taylor-Couette flow to be more unstable than the aforementioned axisymmetric mode, even for $\epsilon = 0.1$.

\subsection{The dominant instability in the $Wi-Re$ plane}
\label{subsec:inertialeffects}

The discussion above began with a brief description of the Newtonian Taylor-Couette instability\,($Wi=0$), and thereafter, shifted to an examination of the purely elastic instability\,($Re = 0$) arising in the same flow configuration. The initial efforts referred to above \citep{Ginn_Denn_1969,thomas_walters_1964,thomas_walters_1964b,Beard1966} emphasized the destabilizing role of weak elasticity on the Newtonian (centrifugal) instability\,(for $N_2/N_1 \rightarrow 0$). The destabilizing nature of weak elasticity on the Newtonian instability holds even for the Dean and Taylor-Dean configurations \citep{Joo_Shaqfeh_inertia_1992}.
A study of the opposite limit, that is, the role of weak inertia on the purely elastic instability \citep{larson1990purely}, was first carried out by Joo and Shaqfeh \citep{Joo_Shaqfeh_inertia_1992}. Expectedly, the rotation of the outer cylinder was found to have a stabilizing effect, and that of the inner cylinder a destabilizing one; note that this breaks the symmetry present in the purely elastic limit, for $\epsilon \ll 1$, where the instability is independent of the particular cylinder being rotated\footnote{It is perhaps worth noting that the analogous question for the other viscometric flow configurations viz.\ the cone-and-plate and parallel-plate geometries, is complicated by the fact that the purely azimuthal flow is no longer an exact solution for finite $Re$; see section \ref{subsec:coneplate}.}.

\begin{figure}
\includegraphics[scale=0.32]{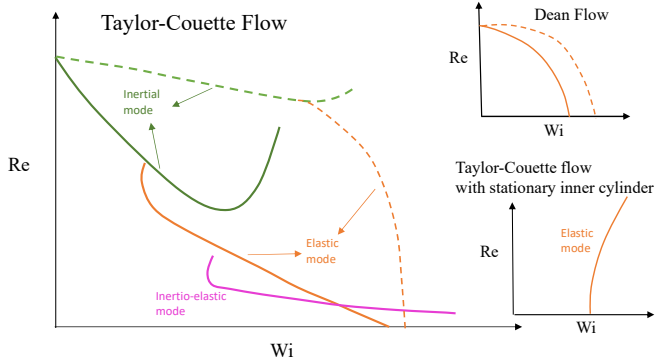} 
\caption{Dominant modes in the $Re$-$Wi$ plane for viscoelastic Taylor-Couette with outer cylinder stationary or rotating at half the angular velocity of the inner cylinder\,(main schematic). The continuous lines are for the UCM limit and the dashed lines are for $\beta = 0.8$. The two insets show the dominant mode(s) in Taylor-Couette flow with a stationary inner cylinder and in Dean flow. For the former configuration, the elastic mode is stabilized by inertia. 
In contrast to Taylor-Couette flow, a single unstable mode continuously spans the $Re$-$Wi$ plane for Dean flow.}
\label{fig:dominantmode}
\end{figure}

It is also of interest to examine the nature of the dominant unstable mode in the $Re-Wi$ plane as a whole (Fig.~\ref{fig:dominantmode}).  Recall from section \ref{sec:rectilinear} that, within the Oldroyd-B framework of a shear-independent viscosity and first normal stress coefficient, the elasticity number $E = Wi/Re$ is a convenient measure of the relative magnitudes of elasticity and inertia, with $E = 0$ and $\infty$ corresponding to the purely inertial and purely elastic limits, respectively. Avgousti and Beris \citep{Avgousti1993b}, using both linear stability analyses and symmetry arguments\,(that remain valid beyond the linear regime), reported the existence of three different eigenmodes in the $Re-Wi$ plane for Taylor-Couette flow of a UCM fluid subject to axisymmetric perturbations. For $E \ll 1$, the unstable mode is stationary and its structure analogous to the well-known Taylor vortices in the Newtonian limit; for $E \gg 1$, the unstable mode is oscillatory akin to the purely elastic mode analyzed by \citep{larson1990purely}. For finite values of $E$, however, a distinct inertio-elastic oscillatory mode becomes most unstable with a wavelength and flow structure intermediate between the purely elastic and inertial modes. Avgousti and Beris \cite{Avgousti1993} showed that the above trend, of new modes becoming dominant with increasing $E$, persisted for non-axisymmetric disturbances \citep{Avgousti1993b}, although this finding was restricted to the UCM limit ($\beta = 0$). Moreover, the unstable regions in the $Re$-$Wi$ plane were mapped out only for fixed wavenumbers, and for the specific case  $\Omega_{out}/\Omega_{in} = 1/2$.

In contrast, Joo and Shaqfeh \citep{Joo_Shaqfeh_inertia_1992}, in addition to the aforementioned angular velocity ratio, also considered the cases $\Omega_{out} =0, \Omega_{in} \neq 0$ (only inner cylinder rotating), and $\Omega_{out} \neq 0, \Omega_{in} = 0$ (only outer cylinder rotating), with the restriction of axisymmetric disturbances. 
For $\Omega_{out}/\Omega_{in} = 1/2$ and $\Omega_{out}/\Omega_{in} = 0$, the said authors presented results for the critical Reynolds number, minimized over the entire range of axial wavenumbers, as a function of $Wi$, and confirmed the existence of three distinct modes in the $Re$-$Wi$ plane in the UCM limit, similar to the results of Avgousti and Beris discussed above. However, the authors also showed that the nature of the dominant mode was sensitively dependent on $\beta$. Thus, while the inertio-elastic mode was the most unstable over a small, intermediate range of $Wi$ for $\beta = 0$, this range disappears entirely for $\beta$  slightly greater than zero, suggesting that the inertio-elastic mode may not be relevant for polymer solutions (see Fig.~\ref{fig:dominantmode}).  Thus, for $\beta = 0.8$ and $\Omega_{out}/\Omega_{in} = 0$, there are only two dominant modes in the $Re$-$Wi$ plane, viz., the continuation of the inertial mode from the Newtonian limit to finite $Wi$, and the continuation of the purely-elastic mode from the inertialess limit to finite $Re$ (see Fig.~\ref{fig:dominantmode}). For $\Omega_{in} = 0$, there is no instability in the Newtonian limit. Thus, the purely elastic instability,  present for $Re = 0$, is the only unstable mode in $Re$-$Wi$ plane, and the critical $Wi$ for this instability increases as $Re$ is increased from zero. In contrast to this rather complicated scenario in Taylor-Couette flow, wherein different modes become dominant as $E$ is increased, the same unstable axisymmetric mode continues all the way from $E = 0$ to $E = \infty$ for Dean flow \citep{Joo_Shaqfeh_inertia_1992}; moreover, this scenario holds for different values $\beta$ (Fig.~\ref{fig:dominantmode}). This simple picture may not, however, hold for nonaxisymmetric disturbances, although this remains to be investigated.

Interestingly, the nature of the dominant instability in the $Re-Wi$ plane in the limit $Re,Wi \rightarrow \infty$, with $E$ fixed, is of relevance to the astrophysical scenario. Note that this distinguished limiting scenario arises only for the Oldroyd-B model since shear thinning in the nonlinear constitutive models leads to elastic stresses scaling sub-quadratically, and thereby, becoming asymptotically small in relation to the inertial ones in the aforementioned limit. The limit may be termed the elastic Rayleigh limit since the equation governing linearized evolution is a second order ODE similar to the Rayleigh equation in the inviscid limit \citep{Drazinreid}, but with $E(1-\beta)$ as a parameter that measures the importance of elastic stresses\,(the reduction of order points clearly to the singular nature of this limit). It has been shown \citep{OgilvieProctor_2003} that the equations governing polymer solutions in the elastic Rayleigh limit are identical to the magnetohydrodynamics\,(MHD) equations in the relaxationless limit\, (that of infinite magnetic Reynolds number), with the dumbbell end-to-end vector field being analogous to the magnetic field. A consequence is that the instabilities of a polymer solution in this limit have corresponding astrophysical analogs; in particular, an inertio-elastic instability in this limit should map onto the so-called magnetorotational instability\,(MRI). The axisymmetric version of the latter instability was originally predicted by Vehlikov \citep{Velikhov_1959} and Chandrasekhar \citep{Chandra1961}, and rediscovered much later by Balbus and Hawley \cite{BalbusHawley_1991}. In the polymer solution literature, the elastic Rayleigh limit was first considered by Azaiez and Homsy \citep{azaiez_homsy_1994}, in the context of a viscoelastic shear layer, although elasticity had a stabilizing influence in this case, as already discussed in section \ref{subsec:Interfacial}. Rallison and Hinch \citep{rallison_hinch_1995} were the first to discover an inertio-elastic instability for a submerged jet configuration, arising from a novel mechanism involving a resonant interaction of elastic shear waves\,(the analogs of Alfven waves in the MHD context) in the aforementioned limit. More recently, an analogous shear-wave-driven instability has been shown to destabilize an elastic vortex column \citep{roy2021}; unlike the jet, this inertio-elastic instability exists in isolation for the vortex case, and is therefore of greater relevance. Importantly, the vortex column configuration bears an obvious relation to the Taylor-Couette geometry. There have indeed been experiments in the Taylor-Couette setup motivated by the above analogy \citep{Boldyrev_2009}, and also numerical stability calculations for relatively modest $Re$ and $Wi$ \citep{OgilviePotter_2008}. It is certainly of interest to theoretically investigate the instabilities of the Taylor-Couette configuration in the elastic Rayleigh limit, and in particular, examine the relation between any unstable modes arising from the aforementioned shear-wave-resonance mechanism, and the inertio-elastic mode, at finite $Re$ and $Wi$, that has been discussed in the elastic instability literature\citep{Avgousti1993,Joo_Shaqfeh_inertia_1992}.

\subsection{Purely elastic instability in cone-and-plate and parallel-plate flows}
\label{subsec:coneplate}

The mechanism underlying the purely elastic Taylor-Couette instability is suggestive of a similar instability in the two other rheometric flow configurations, viz., the cone-and-plate and parallel-plate geometries. Magda and Larson \cite{Magda_Larson_1988} observed an anomalous increase in the shear stress above a threshold rate of shear, in both these geometries, even for a vanishingly small $Re$.  Although similar observations had been made by earlier researchers \citep{Jackson_Walters_1984,Binnington_Boger}, the aforementioned authors were the first to ascribe the anomalous increase to a flow transition, rather than to the intrinsic rheological character of the fluid.  Subsequent visualization experiments by McKinley \textit{et al.} \citep{McKinley1991} showed the onset of a secondary flow at the critical shear rate, corroborating the inference of Magda and Larson. 

The first theoretical studies to demonstrate an instability of the flow of an Oldroyd-B fluid, in the above geometries, were due to Phan-Thien \citep{PhanThien1983,PhanThien1985}. Motivated by the form of the Newtonian flow in these geometries, the author assumed a self-similar so-called von-Karman ansatz for the base-state flow at finite $Re$ and $De$, with $De = \lambda\Omega$. Despite the base-state flow reducing to a purely azimuthal one for zero $Re$\,(regardless of $De$)\,\footnote{In the Newtonian case, the (inertial)\,centrifugal forces driving the meridional secondary flow scale with the velocity, which is larger near the rotating plate, and therefore give rise to a secondary flow. In contrast, the elastic stresses scale with the velocity gradient\,(the square of the gradient for Oldroyd-B), and are therefore uniform across the gap. The divergence of these stresses may be balanced with a radial pressure gradient, allowing the base-state flow to have only an azimuthal component.  This argument is applicable, however, only in the limit of small cone angles, which is when the velocity gradient across the gap is a constant.
 The existence of an elastic instability, of course, implies solution multiplicity, and the azimuthal flow is not the only possible axisymmetric flow beyond a threshold $De$.}, the author again invoked the Karman ansatz for the perturbation, and demonstrated the onset of an stationary and axisymmetric elastic linear instability beyond a critical $De$. The absence of a length scale in the assumed ansatz implied a disturbance flow field in the form of a single `roll', in the meridional plane, that `closes at infinity'. Thus, the only relevant dimensionless group is $De$ above, with the threshold criterion therefore coming out to be independent of the gap-angle\,($\theta$) for the cone-and-plate geometry, and of the non-dimensional plate spacing\,($H/R$, $H$ and $R$ being the inter-plate spacing and plate radii) for the parallel-plate geometry; note that $Wi = De/\theta$,  and $De\,(R/H)$ for the two geometries.

The aforementioned visualization experiments of McKinley \textit{et al.} \citep{McKinley1991}, for the parallel plate geometry, showed that the growing disturbance flow field was not of the similarity form assumed in \citep{PhanThien1983}; instead, the observed patterns were characterized by a radial scale of $O(H)$. The first  analysis to account for this scale was that of Oztekin and Brown \citep{Oztekin1993}, who used the Oldroyd-B model to conduct a local stability analysis of the parallel-plate flow in the vicinity of a particular radius\,($R^*$). While the analysis correctly showed the unstable modes to be time dependent non-axisymmetric spiraling patterns, the prediction of the threshold was qualitatively incorrect. The threshold $De$, obtained from the Oldroyd-B analysis, bore an inverse relation to $R^*$, so the torsional flow configuration was predicted to be unstable for any finite $De$ beyond a certain $R^*$\,(in contrast to the Phan-Thien analysis above). This, however, contradicted the later experiments of McKinley and coworkers \citep{byars1994}, who showed that the unsteady secondary motion, just beyond the threshold, was restricted to an annular region between a pair of critical radii. The restabilization of the flow beyond the second critical radius arises from effects of shear thinning which decrease the effective relaxation time at higher shear rates. This was confirmed by a linear stability analysis using the FENE-CR model \citep{byars1994}. The aforementioned restabilization also translates to the stability of the torsional flow configuration below a threshold inter-plate spacing; the resulting shape of the unstable region in parameter space is discussed in section \ref{subsec:PakdelMcKinley} below in the context of the Pakdel-Mckinley criterion. 

An analogous scenario was shown to hold for the cone-and-plate geometry \citep{McKinley1995}. Use of the Oldroyd-B model again reproduced the unstable spiral patterns in a qualitative sense; unlike the parallel-plate case, the homogeneity of the base-state shearing flow and the resulting absence of a characteristic length scale implies that the spirals in this case are no longer radially localized. Use of the FENE-CR model was needed to capture the correct nature of the threshold condition, which involved a stabilization of the flow for small enough cone angles; this is discussed in section \ref{subsec:PakdelMcKinley}. The use of a multimode Giesekus model, in fact, yielded quantitative agreement with experimental observations \citep{Oztekin1994}

\subsection{Experiment vs Theory (Non-isothermal effects)}
\label{subsec:exptsnonisothermal}
The comparison between theory and experiment, even for the primary transition in the viscoelastic case, is nowhere as quantitative as for the Newtonian case. The first and perhaps obvious reason is the uncertainty surrounding the constitutive equation used. The Oldroyd-B equation, used in the stability analyses, is only an approximation for the Boger fluids used in the experiments. Boger fluids have a nontrivial spectrum of relaxation times, and consistent modeling of both their steady state and transient rheology requires a nonlinear multimode constitutive equation \citep{Quinzani_1990}. The underlying relaxation spectrum manifests as a dependence of the apparent relaxation time on the method of measurement. The resulting arbitrariness in the choice of time scale, to be used in the Oldroyd-B model, leads one to expect an ambiguity in the theory-experiment comparison. A second reason has already been mentioned above in the context of the first experiments: the shallow nature of the viscoelastic neutral curve, in comparison to the Newtonian one (compare Figs.III-2 and III-11 of Ref.~\citep{larson1992}), has been speculated to lead to strong nonlinear interactions even close to onset, that then manifest as a slow drift of the observed pattern with the appearance of progressively smaller-scaled structure over times much longer than the nominal relaxation time \citep{larson1990purely,Northey1992}.
A second factor favoring the aforementioned importance of nonlinearity is that the bifurcation to the primary non-axisymmetric mode is likely a subcritical one \citep{sureshkumar_avgousti_beris}, in contrast to the Newtonian case. Thus, as briefly mentioned in Sec.~\ref{subsec:purelyelasticTC},  while the theoretical predictions based on axisymmetric disturbances over-predict the threshold for instability, consideration of nonaxisymmetric disturbances and the associated subcritical nature of the bifurcation leads to a narrowing of the gap between the theoretical and experimental thresholds.

\begin{figure}
  \centering
  \begin{subfigure}[htp]{0.35\textwidth}
    \includegraphics[width=\textwidth]{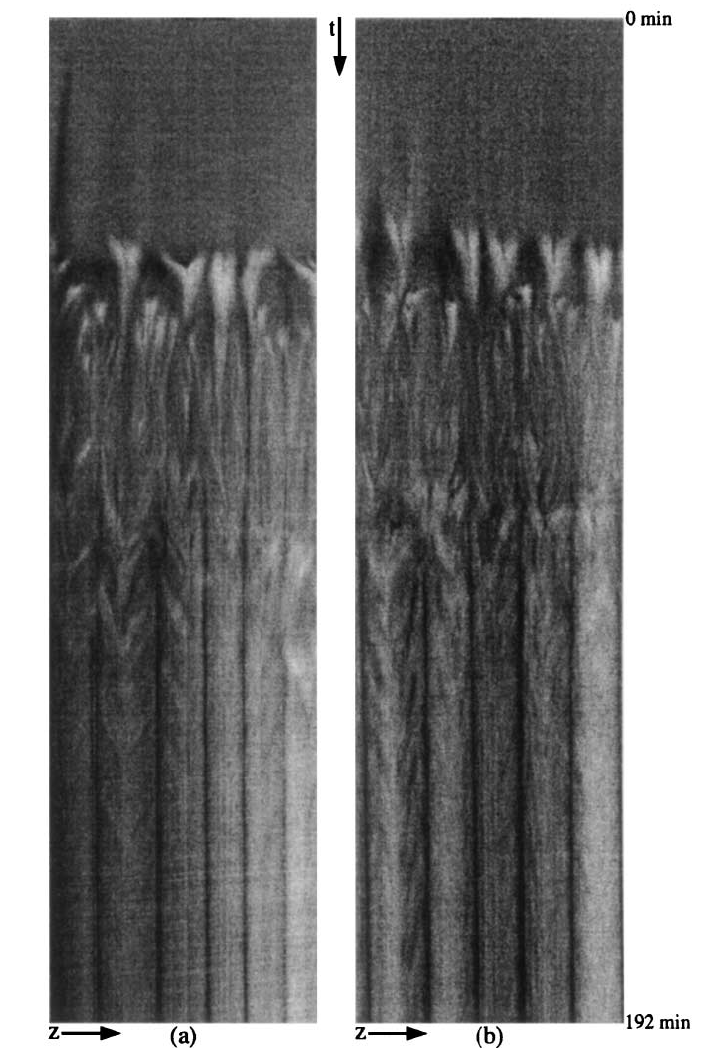}
    \caption{left panel: inner cylinder rotating; right panel: outer cylinder rotating}
    \label{fig:whitemullerfig4}
  \end{subfigure}
  \quad\quad
  \begin{subfigure}[htp]{0.35\textwidth}
    \includegraphics[width=\textwidth]{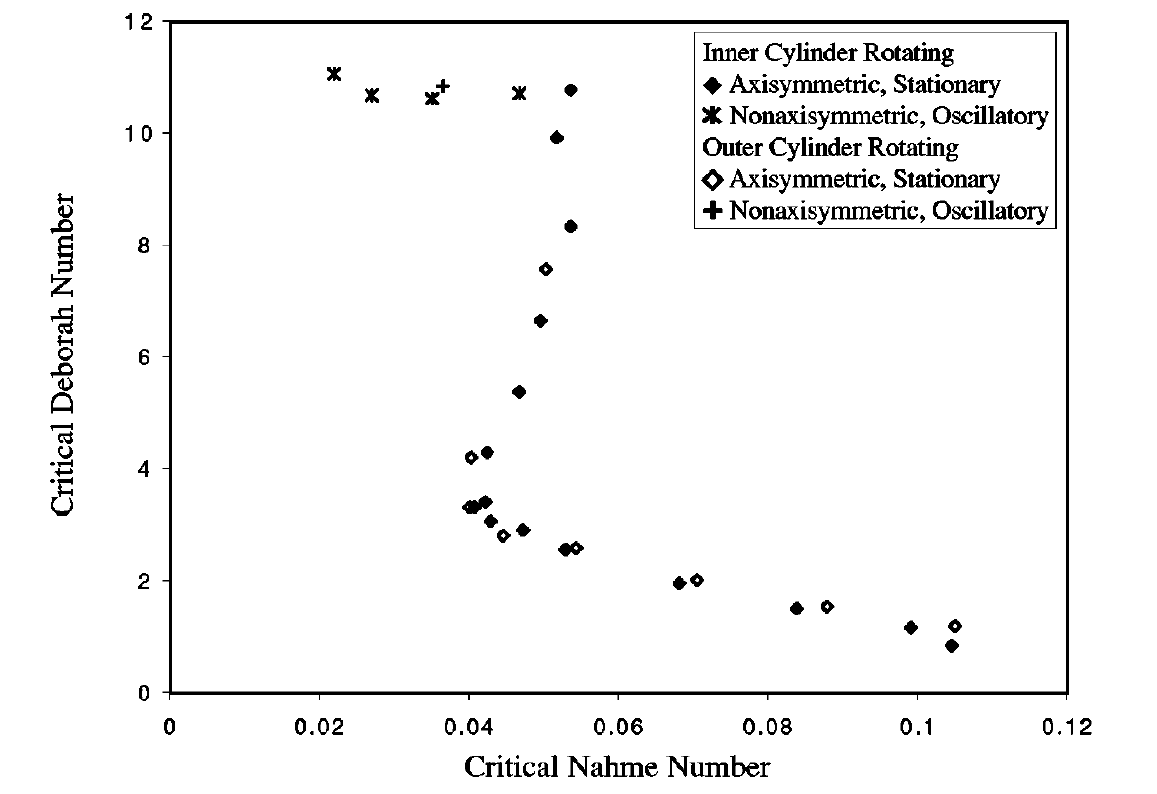}
    \caption{Critical $De$ vs Nahme number.}
    \label{fig:whitemullerfig19a}
  \end{subfigure}
  \caption{(a) Space-time plots of Taylor-Couette flow of a viscoelastic fluid at
$De = 2.7$, $Re = 2.7$, Brinkmann number $= 6.5 \times 10^{-3}$, and Prandtl number $\approx 24000$.
The extent of the z axis is 5.82 gap widths. Reproduced with permission from Ref.~\cite{White_Muller2000}. (b) $De_c$ vs. Nahme number, reproduced with permission from Ref.~\cite{White_Muller2003}.
These experiments differed from earlier efforts in adopting the so-called adiabatic\,(quasi-static) path, involving a slow ramp-up\,(as opposed to a step increase) to the rotation rate of interest.}
  \label{fig:whitemuller}
\end{figure}

A third and rather unexpected reason that has led to a stark difference between theory, and some of the experiments using Boger fluids, is the deviation from isothermal conditions arising due to viscous heating\,(Boger fluids have very high viscosities). Despite the critical parameters for instability onset being sensitive to the particular rheological model used, the elastic instability for Taylor-Couette flow, within the usual isothermal formulation, is predicted to be always non-axisymmetric and oscillatory at onset.  However, many of the experimental observations \citep{BaumertMuller_1997,Groisman_Steinberg1998b,White_Muller2000,White_Muller2003} have revealed a primary transition to a weak stationary axisymmetric mode on time scales much longer than the polymer relaxation time, and at $Wi$ much lower than the theoretical threshold. This discrepancy was addressed by Al-Mubaiyedh \textit{et al.} \citep{Al-Mubaiyedh1999,Al-Mubaiyedh2000a} who showed that the inclusion of viscous heating in the analysis, and the resulting prediction of a thermoelastic instability, leads to a good agreement between experiment and theory. That viscous heating leads to destabilization is somewhat counter-intuitive since both viscosity and  the relaxation time should decrease with an increase in temperature, and this ought to lead to a higher threshold rotation rate for a given $Wi_c$. This expected stabilizing effect has indeed been found for other curvilinear geometries such as the parallel-plate and cone-and-plate configurations \citep{Rothstein-McKinley-nonisothermal-2001,Olagunju2002}. The destabilizing mechanism \citep{Al-Mubaiyedh1999,Al-Mubaiyedh2000a} for Taylor-Couette flow is argued to arise due to the stratification of the hoop stress in the gap between the cylinders, which drives a radial secondary flow, and the convection of base-state temperature gradients by the radial perturbation velocity then leads to the instability. Figure~\ref{fig:whitemuller} shows the slow development of the stationary vortices on a space-time plot \citep{White_Muller2000}, with the adjoining plot showing the experimentally observed decrease in $De_c$\,(to be interpreted as $Wi_c$) with the Nahme number, the latter being a dimensionless measure of viscous heating \citep{White_Muller2003}.

\subsection{Secondary instabilities and elastic turbulence}
\label{subsec:secondaryinstab}

Transition from the laminar state, in the Newtonian case, occurs in contrasting fashions for the canonical\,(and non-inflectional) rectilinear shearing flows examined in section \ref{sec:rectilinear}, and for the curvilinear shearing flows examined here\,(the Taylor-Couette geometry in particular). In the former case, with increasing $Re$, turbulence appears abruptly and in its entire complexity, albeit in spatially localized forms \citep{mullin_2011}, while there is a gradual but global increase in the spatiotemporal complexity for the Taylor-Couette configuration. Thus, the topic of secondary or higher-order transitions following the primary instability, and that lead to the eventual turbulent state, is more pertinent to curvilinear shearing flows; indeed, the various flow regimes for Newtonian Taylor-Couette flow have been extensively studied and are well documented, as already mentioned in section \ref{subsec:VEonNewtonianTC}\,\citep{andereck_liu_swinney_1986,Grossmann_Lohse_Sun_2016,dutcher_muller_Newtonian_2009}.

At the other extreme of elasticity being dominant\,($E \gg 1$), a spatiotemporally disordered state has been shown to arise for sufficiently large $Wi$, and has appropriately been termed elastic turbulence\,(ET). The experiments of Groisman and Steinberg\,\citep{Groisman2000}  were the first to characterize ET in a wide-gap parallel-plate geometry\,($\epsilon = 0.263$ and  $0.526$), with the  shear stress in this state being up to $20$ times larger than the (hypothetical)\,laminar flow at the same rotation rate. The transition was found to be subcritical, although the detailed pathway and the associated rich sequence of intermediate states\,(more readily accessed in a narrow-gap setting) were characterized in later experiments, in the same geometry\,($\epsilon = 0.2$), by Schiamberg \textit{et al.} \citep{Schiamberg_2006}. Groisman and Steinberg\,\citep{Groisman2000} also examined the power spectrum of velocity fluctuations in the ET state via Doppler velocimetry; the plots decayed as $\omega^{-3.5}$ for large frequencies\,($\omega$), and thence, for sufficiently small length scales\,(via Taylor's frozen turbulence hypothesis). The rapid decay is in contrast to the Kolmogorov $(-5/3)$-law known for the inertial range in Newtonian turbulence, but is consistent with theoretical constraints obtained based on the Oldroyd-B model \citep{fouxon_lebedev}. The implication is that, in contrast to Newtonian turbulence, the velocity gradient is the maximum at the largest scales, and ET therefore corresponds to a stochastic but spatially smooth chaotic flow controlled by the large scales \citep{Steinberg2021}. The ET state has also been accessed in a curvilinear channel\,\citep{Groisman2001}, the analog of the Dean-flow configuration discussed earlier but one that does not conform to the infinite\,(large) depth assumption underlying the derivation of the theoretical stability threshold. The curvature ratio was large to facilitate transition, which was now found to be supercritical \citep{groisman2004elastic}, but with the power spectra again pointing to the large scales being dominant.

Groisman and Steinberg have also accessed an ET state in a wide-gap\,($\epsilon \approx 1$) Taylor-Couette geometry \citep{groisman2004elastic}, the transition being hysteretic similar to the parallel-plate configuration above. The velocity fluctuation power spectra, however, exhibited a large-$\omega$ decay exponent a little greater than $2$, thereby deviating from the scalings for the other  geometries above, in turn implying the possibly  non-universal geometry-dependent nature of the ET state. In earlier experiments involving a Taylor-Couette configuration with a narrower gap\,($\epsilon = 0.5$), Groisman and Steinberg have examined the higher-order transitions en-route to turbulence, as a function of $E$, although the aforementioned ET state itself was not identified\,\citep{Groisman_Steinberg1996,Groisman_Steinberg1997,Groisman_Steinberg1998,Groisman_Steinberg1998b}. For $E < 0.15$, the primary instability of the azimuthal flow gave rise to a Newtonian-like Taylor-Vortex flow\,(TVF) which, with increasing $Re$, became unstable to a new oscillatory state termed a `rotating standing wave'. At a higher $Re$, this state transitioned to yet another new oscillatory state termed `disordered oscillations', characterized by broad peaks in the frequency spectra and by the appearance of patches of both standing and travelling waves in space-time plots. For $E > 0.15$, the TVF directly transitioned to disordered oscillations, while for $E> 0.22$, disordered oscillations emerged directly from the purely azimuthal laminar base flow. At higher elasticities\,$(E \sim 20)$, the  transition to disordered oscillations became increasingly hysteretic. Thus, along a path of decreasing $Re$, these oscillations persisted down to an $Re$ much lower than that marking the onset of the forward transition, eventually giving way to stationary so-called `diwhirls'; solitary diwhirls have since been identified via fully nonlinear computations \citep{Kumar_Graham_2000}.  
Many of the aforementioned transitions, for small but finite $E$, have been reproduced in simulations\,(for $E = 1/3$) by Khomami and co-workers, using the Oldroyd-B and FENE-P models, with $L$ in the latter case being large enough for shear-thinning to be unimportant \citep{thomas_sureshkumar_khomami_2004}.

Detailed experiments, with the intent of mapping both the primary and higher-order transitions in the Taylor-Couette geometry, have also been conducted by Muller and co-workers. While the early experiments using high-viscosity Boger fluids were likely corrupted by the emergence of a thermoelastic instability\,(the later experiments of White and Muller pertaining to the thermoelastic instability are discussed in more detail in section \ref{subsec:exptsnonisothermal}), and possibly, its interaction with the elastic instability on account of adopting a non-quasistatic ramping protocol\,\citep{BaumertMuller_1997,baumert1999}, later experiments\,\citep{dutcher_muller_weakVE_2011,dutcher_muller_moderateVE_2013}, based on a quasi-static ramping protocol, have extended the known mapping of flow states in the $Re_{in}-Re_{out}$ plane\,(see section \ref{subsec:VEonNewtonianTC}) to non-zero $E$.  The general pattern is that of elasticity affecting the higher-order less symmetric flow states at a lower $E$, which is expected on account of the time scales of oscillation involved being comparable to the polymer relaxation time. For the highest $E$ examined\,($\approx 0.2$), the authors identify an elastically dominated turbulent state based on the emergence of a broadband frequency spectrum, although the relation with the ET-state, identified by Steinberg and coworkers above, is not clear\,\citep{dutcher_muller_moderateVE_2013}.

\subsection{Beyond Oldroyd-B}
\label{subsec:beyondOldB}

Here, we comment briefly on how features not captured by the Oldroyd-B model affect the prediction of the purely elastic instabilities in curvilinear flows. These include (i) a nonzero $N_2$,  (ii) the existence of a nontrivial relaxation time spectrum,  (iii) the shear-rate dependence of the viscosity and first normal stress difference, (iv) the use of a multi-scale simulation approach that avoids closure approximations, and (v) going beyond the dilute regime.

Most entangled polymer solutions have negative $N_2$'s \citep{Maklad_Poole_2021}, which are predicted to strongly stabilize the purely elastic instability in Taylor-Couette flow \citep{shaqfeh_muller_larson_1992}, especially
for small gap widths\,($\epsilon \ll 1$); as alluded to in section \ref{subsec:VEonNewtonianTC}, in the context of the second-order fluid model, the effects of $N_2$ enter at a lower order in $\epsilon$ (compared to $N_1$) in this limit. Indeed,
 Shaqfeh \textit{et al.} \citep{shaqfeh_muller_larson_1992} noted a significant discrepancy between the prediction from the Oldroyd-B model  ($Wi_c = 47$) and experiment ($Wi_c = 71$) for the smallest gap ratio ($\epsilon = 0.03$), where the small-gap theory should otherwise have been accurate.  To address this discrepancy, Shaqfeh \textit{et al.} used a modified Oldroyd-B model with an additional contribution to the stress tensor that, albeit of the second-order fluid form, gave rise to a nonzero $N_2$, and showed that this indeed had a stabilizing effect. Similarly, using the Giesekus model, Beris \textit{et al.} \citep{Beris1992} also concluded that negative second normal stress differences were strongly stabilizing. Indeed, for $- N_2 / N_1 > 0.1$ (typical for entangled polymer solutions), $Wi_c > 100$, and this could be one reason why the purely elastic Taylor-Couette instability is not reported for the flow of entangled polymeric solutions \citep{larson1992}.

Larson \textit{et al.} \citep{Larson1994} used the K-KBZ equation to incorporate both shear thinning and a relaxation time spectrum to analyze the elastic Taylor-Couette instability, and concluded that the critical conditions are a function of both the longest (the Oldroyd-B) and the average relaxation times. This immediately points to the arbitrariness inherent in the choice of a single-time-scale (either linear or nonlinear) model, as has already been pointed out in section \ref{subsec:exptsnonisothermal}. In a later effort pertaining to instability onset in the cone-and-plate and parallel-plate geometries, a multi-mode Giesekus model, with parameters tuned so as to best fit the rheological properties of the Boger fluid used in the experiments, has been shown to quantitatively predict the critical $Wi$ as a function of gap width or cone angle \citep{Oztekin1994}. Notwithstanding the subtle issue of stress-conformation hysteresis that might come into play for instabilities in non-viscometric flows\,(see discussion at the end of Sec.~\ref{subsec:flowcylinder}), multimode versions of nonlinear constitutive models appear necessary for a quantitative match with experiments.

Larson  \textit{et al.} \citep{Larson1994} found shear thinning to have a monotonically stabilizing effect on the instability, on account of the decrease in the first normal stress coefficient with the shear rate. As mentioned in Sec.~\ref{subsec:coneplate}, shear-thinning has, in fact, a profound effect on the neutral boundary demarcating the stable and unstable regions \citep{McKinley1991,McKinley1995,Oztekin1993,byars1994}, and this is discussed in more detail in the following subsection, in the context of the Pakdel-Mckinley criterion.

Although the commonly used constitutive equations like the Oldroyd-B, the FENE-P or the Giesekus models, have the obvious advantage of needing only modest computational resources, they are limited either by the simplicity of the microscopic picture\,(Oldroyd-B), by closure approximations\,(FENE-P) or by lack of a rigorous connection to an underlying kinetic theory framework\,(FENE-CR, Giesekus). There have been some efforts, in the context of hydrodynamic stability, that have attempted to go beyond the said limitations by adopting a `micro-macro' approach. For instance, Somasi and Khomami \cite{Somasi_Khomami_2000,Somasi_Khomami_2001} calculate the polymer stress field from Brownian dynamics simulations using Hookean or FENE dumbbells, which is subsequently coupled with a finite-element solution of the Cauchy momentum equations. 
Using this approach, the authors examined the linear stability of plane- and Taylor-Couette flows, finding good agreement with the results of a stability analysis carried out using closed-form constitutive equations. However, very large computational times were required (approximately 100 times that required for  a regular stability computation)  even for the simple elastic dumbbell-based models used.

Finally, the results obtained using the Oldroyd-B model are strictly valid only in the truly dilute regime. However, the polymer solutions used in experiments are not always in this regime, and it is important to include the concentration dependence of the fluid rheology in a  systematic manner. In this regard, as a first step, it seems appropriate to use the Giesekus model (discussed earlier in Sec.~\ref{sec:Intro}), with the anisotropic drag coefficient in the model acting as a proxy for polymer concentration. Recent efforts \citep{prabhakar2016} have attempted to account for, in a more rigorous manner, hydrodynamic interactions between chains that come into play close to and beyond the overlap concentration. Although the role of such inter-chain interactions has been examined in the context of the CABER technique, now commonly used to measure the relaxation time \citep{Bazilevsky,mckinley_tripathi2000}, in future, one could also envisage a stability analysis, based on the rigorous models above, that extends across the overlap concentration, spanning both the dilute and semi-dilute regimes.

\subsection{The Pakdel-McKinley criterion}
\label{subsec:PakdelMcKinley}

We now discuss a heuristic argument developed by Pakdel and McKinley \citep{Pakdel_McKinley,McKinley1996} which unifies the threshold criteria for instability onset in the different curvilinear shearing flows examined above. As will be seen below and in section \ref{sec:nonviscometric}, it allows one to arrive at sensible stability thresholds (to within a numerical factor of order unity) even when closed form results from linear stability analysis are not available\,(owing, for instance, to the geometrical complexity of the flow configuration).
The criterion incorporates the two key ingredients required for a purely elastic instability viz.\ streamline curvature in the base-state laminar flow as well as the magnitude of the streamwise (tensile) normal stresses, and expresses the threshold for elastic instability in terms a product of the Deborah and Weissenberg numbers. Recalling the definition $De=\lambda/T$, $T = {\cal R}/U$ now being the residence time with ${\cal R}$ the radius of curvature\,(for the curvilinear shearing flows in question), and writing $Wi$ in terms of the ratio of the normal and shear stresses as $Wi = N_1/2\tau$, the Pakdel-Mckinley criterion for elastic instability may be written as:
\begin{equation}
[(\lambda U/{\cal R}) (N_1/\tau)]_c > M^2, \label{PM_criterionGen}
\end{equation}
$M$ being an order unity number in the simplest cases (curvilinear viscometric flows), or a field\,(when the flow configuration is more complicated, and in particular, inhomogeneous, as is the case for the non-viscometric examples considered in section \ref{sec:nonviscometric}). Note that the ratio $\lambda U/{\cal R}$ may also be interpreted as the ratio of the  distance\,(along a streamline), over which disturbances relax elastically, to the characteristic radius of curvature. Replacing $\lambda U$ by an appropriate viscous boundary layer length scale recovers the known stability criteria associated with Newtonian instabilities driven by curvature; these include the classical Taylor-Couette instability itself, and the Gortler instability associated with laminar boundary layers on curved surfaces \citep{Saric_ARFM}. Note, however, that the  application of this criterion relies on the assumption that the instability mechanism is local and can be predictive only within this scenario.
  
We now obtain the explicit form for this criterion for the Taylor-Couette and cone-and-plate configurations. For the former, $U = \Omega R_{in}$, corresponding to the inner cylinder\,(say) rotating with angular velocity $\Omega$, and an obvious choice for the radius of curvature is the (inner)\,cylinder radius, viz., ${\cal R} = R_{in}$. The streamwise normal stress for the Oldroyd-B model is $\tau_{\theta \theta} = 2 \eta_p \lambda \dot{\gamma}^2$, and the total shear stress $\tau = \eta_t \dot{\gamma}$, $\eta_t$ being the solution viscosity. Substituting these expressions leads to
\begin{equation}
(\sqrt{De Wi})_c \geq \frac{M}{\sqrt{2 (1-\beta)} }\, ,
\end{equation}
for instability in the thin-gap limit, where $De = \lambda \Omega$, $Wi = \lambda \dot{\gamma}$, and $\eta_p/\eta_t =  (1-\beta)$. Using $\dot{\gamma} = \Omega R_{in}/d$ and the dimensionless gap width $\epsilon = d/R_{in}$ with $Wi = De/\epsilon$, the above equation reduces to 
\begin{equation}
Wi_c \epsilon^{\frac{1}{2}} \geq \frac{M}{\sqrt{2 (1-\beta)}},  \label{PM:Couette}
\end{equation}
in agreement with the original thin-gap analysis of Larson \textit{et al.} \citep{larson1990purely} (note that these authors refer to $Wi$ in (\ref{PM:Couette}) as $De$). In Eq.~\ref{PM:Couette}, $M$ is in general a complicated function of $\beta$, and asymptotes to a constant only in the limits $\beta \rightarrow 0$ and $1$. In the former limit, the stability analysis yields $M \approx 6$ for axisymmetric disturbances \citep{larson1990purely}, and for the latter case, the threshold $Wi$ diverges as the reciprocal square root of $(1-\beta)$ in  the limit of vanishing elasticity.

For the cone-and-plate configuration, the velocity at any radial position is $U = r \Omega$, and for $\theta_0 \ll 1$, the shear rate, which is now uniform across the gap, is $\Omega/(r \dot{\theta}) = \Omega/\theta_0$. Using ${\cal R} = r$, and the expression for the hoop stress mentioned for Couette flow above, one obtains:
\begin{equation}
Wi\, \theta_0^{\frac{1}{2}} \geq \frac{M}{\sqrt{2 (1-\beta)}}.
\end{equation}
The same comments as above apply with regard to the $\beta$-dependence. For $\beta = 0.5$, linear stability analysis yields  $M = 4.602$ \citep{McKinley1996}.

\begin{figure}
 \centering
 \includegraphics[width=0.9\linewidth]{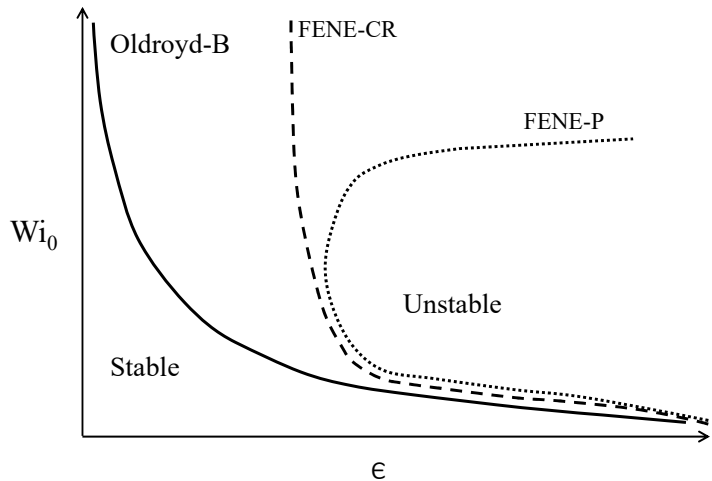}
                \caption{The variation of the threshold Weissenberg number $Wi_0$ (based on the zero-shear relaxation time) with the dimensionless parameter $\epsilon$, is shown qualitatively for the Oldroyd-B, FENE-CR and FENE-P models. Here, $\epsilon = d/R_{in}$ for Taylor-Couette flow, $\epsilon = \theta_0$ for the cone-and-plate geometry, and $\epsilon = H/R$ for the parallel plate geometry\,($H$ being the gap width between the plates, and $R$ being the radius of the circular plate).}
\label{fig:Pakdel}
\end{figure}

The real utility of the criterion above is evident when accounting for shear thinning effects which may be done by using the shear-rate-dependent analogs of the quantities that appear in Eq.~\ref{PM_criterionGen}; Thus, $Wi$ in  Eq.~\ref{PM_criterionGen} may be written as $[\Psi_1(\dot{\gamma})\dot{\gamma}^2]/2\eta(\dot{\gamma})$, in terms of the shear-dependent first normal stress coefficient and viscosity. This generalization is particularly important since, as already seen in section \ref{subsec:coneplate}, shear thinning associated with nonlinear constitutive models can lead to a qualitatively different character for the unstable mode, implying a corresponding difference in the shapes of the unstable regions in the relevant parameter space. Considering the FENE-CR model to begin with, one has a shear-rate-independent viscosity, but a shear-thinning first normal stress coefficient given by $\Psi_1(\dot{\gamma}) = \Psi_{10} \sqrt{\frac{L^2 -3}{2}} \frac{1}{\dot{\gamma} \lambda_0}$ for $Wi_0(=\!\dot{\gamma}\lambda_0) \gg 1$; here, the subscript `0' denotes the zero-shear-rate limit. Using these expressions, one obtains $\epsilon \frac{L^2 - 3}{2}  > \frac{M^2}{(1-\beta)}$, which points to a shear-rate-independent geometrical threshold for instability for $Wi_0 \gg 1$. Thus, for a given $\beta$ shear thinning eliminates the instability below a certain critical $\epsilon\,(= \frac{2M^2}{(1-\beta)(L^2-3)})$. Note that the above large-$Wi_0$ threshold applies to different configurations via an appropriate choice of $\epsilon$; for instance, $\epsilon = d/R_{in}$ for Taylor-Couette flow, $\epsilon = \theta_0$ for the cone-and-plate configuration, and $\epsilon = H/R$ for the parallel-plate configuration. Consideration of the more widely used FENE-P model, which predicts shear thinning of both $\Psi_1\,(\propto \dot{\gamma}^{-\frac{4}{3}}$) and $\eta\,(=\eta_p + \eta_s$ with $\eta_p \propto \dot{\gamma}^{-\frac{2}{3}})$\,\citep{Bird1980}, yields the Pakdel-McKinley criterion in the form $\epsilon^{\frac{3}{2}} \frac{L^3}{8 Wi_0}  > \frac{M^3}{(1-\beta_0)}$ for $Wi_0 \gg 1$, which points to a clear contradiction since $\epsilon \ll 1$; this implies that  shear thinning has a stronger role to play for the FENE-P case, with the unstable region turning over at a finite $Wi_0$.  A sketch of the boundaries demarcating the unstable region on the $\epsilon-Wi$ plane, obtained using the Oldroyd-B and nonlinear\,(FENE-CR and FENE-P) constitutive models, are shown in Fig.~\ref{fig:Pakdel}. While the Oldroyd-B captures the lower branch of this boundary, the upper branch arises solely due to effects of shear thinning, and as seen above, is dependent on the details of the nonlinear terms\,(the version of Fig.~\ref{fig:Pakdel}, on the $De_0-1/\epsilon$ plane appears in \citep{McKinley1996}, without the trend for the FENE-P model).

In summary, while the Oldroyd-B model and its refinements do provide a first-cut prediction of viscoelastic instabilities in flows with curvilinear streamlines, the comparison between experimental observations and theoretical predictions for instability in viscoelastic Taylor-Couette (and other viscometric) flows is not nearly as quantitative as their Newtonian counterparts for the following reasons:
(i) the sensitivity of the threshold conditions to details of fluid rheology such as shear thinning, spectrum of relaxation times, and a nonzero $N_2$, 
(ii) the relatively shallow neutral curve that results in multiple modes getting excited at the onset, and 
(iii) viscous heating effects\,(in certain instances).

\section{Instabilities in non-viscometric flows}
\label{sec:nonviscometric}

There is a wide variety of flow situations beyond the viscometric flows discussed in the previous two sections, which conform to a local simple-shear topology\,\footnote{Note that the local linear flow topology for Taylor-Couette flow, depending on the ratio of the cylinder angular velocities, may range over the entire one-parameter family of planar linear flows which include the hyperbolic and elliptic linear flows \citep{SubKoch_POF2006}; similar comments apply to the other curvilinear flows. However, the unsteadiness arising from the changing orientations of the principal axes leads to simple-shear-like kinematics in the neighborhood of a material point.}, and many of them are again susceptible to elastic instabilities. Poole \citep{poole_2019} constructed a purely elastic instability `flow map' which categorizes flows as being viscometric, shear- or extension-dominated, and those with mixed kinematics. By way of illustration, flows in the Taylor-Dean, Dean, and serpentine-channel configurations are predominantly shear-dominated, while the flow in a cross-slot geometry or a T-channel (in the vicinity of the stagnation points) would be largely extensional, with a fluid element experiencing elongation or compression in the streamwise direction. Flow past a cylinder, or a periodic array of cylinders, and that through a contraction-expansion geometry, would be characterized by mixed kinematics. In this section, we do not aim to be comprehensive in terms of coverage of the aforementioned non-viscometric flows, but instead provide a few canonical examples to showcase the range of real instabilities for which the Oldroyd-B model nevertheless provides an important foundation.

Before venturing into a discussion on stability, it is worth emphasizing that, for both the rectilinear and curvilinear viscometric flow configurations in sections \ref{sec:rectilinear} and \ref{sec:curvilinear}, the base-state is independent of $Wi$ for the Oldroyd-B model, being identical to that for a Newtonian fluid. This is no longer true for a non-viscometric flow. There is a strong dependence of the base-state flow itself on $Wi$ (or, $De$, as appropriate) for both the Oldroyd-B and nonlinear constitutive models, as also evident in experiments and computations \citep{RajMckinley1,RajMckinley2}. In fact, in contrast to the aforementioned viscometric cases, the dependence on $Wi$ is usually the strongest for the Oldroyd-B model, on account of the extensional stresses having the maximum magnitude\,(this in turn being due to the underlying divergence of the extensional viscosity, at an order unity $Wi$, in a homogeneous extensional flow, mentioned in the introduction). Thus, calculation of the base-state itself is often a non-trivial task. An example is the flow past a confined circular cylinder or a sphere, both of which have served as benchmark problems for viscoelastic fluid mechanics \citep{Hassager}; although, the degree of confinement specified has evolved over the years \citep{Brown_McKinley_Report}. Despite the absence of a geometrical singularity, and extensive research on improved numerical algorithms for this problem, the results for the drag coefficient for the cylinder, particularly for the Oldroyd-B model, are available only over a modest range of $Wi$\,(defined as $\lambda U/a$, $a$ being the cylinder or sphere radius, and being interchangeable with $De$ for weak confinement; see Refs.~\citep{Renardy2021,Alves2021}). The breakdown of the numerics at a $Wi$ of order unity, occurs in fact across a range of non-viscometric geometries that are dominated by an underlying extensional flow topology, and was originally thought to be on account of a possible physical singularity, therefore being dubbed the high-Weissenberg-number problem \citep{Keunings1986}.

For the aforementioned cylinder/sphere problem, the high-Weissenberg-number problem is now thought to arise due to the eventual inability of most numerical algorithms to resolve the increasingly steep stress boundary layer that develops in the neighborhood of the rear stagnation streamline for $Wi \geq O(1)$ (the large stresses arise from the slowly relaxing polymer molecules that were originally stretched in the vicinity of the rear stagnation point; see Ref.~\citep{Renardy2021}). In fact, the steady numerical solution breaks down in the immediate vicinity of the rear stagnation point at a $Wi$ even lower than the modest value affecting the drag coefficient; the region of large stresses above has been termed the `birefringent strand' on account of its appearance (as a bright line) in optical experiments \citep{HarlenRallison1}. As a result, nonlinear models, where the extensional thickening is alleviated by choosing
appropriately modest values of the finite extensibility parameter (the parameter $L$ in the FENE-CR model \citep{ChilcottRallison}, for instance), are often chosen for computational tractability. This allows one to obtain solutions for $Wi$ larger than the breakdown value for the Oldroyd-B model, although the steady solution still breaks down for modest values of $L$ in the immediate vicinity of the rear stagnation point; see Ref.~\cite{Oliviera2005}. Interestingly, there have been recent suggestions of an actual singularity (for the Oldroyd-B model), at the point corresponding to the maximum stress, along the rear stagnation streamline \citep{Bajaj2008,Hulsenetal}, harking back to the aforementioned high-$Wi$ problem. Given that the extensional flow in the vicinity of the rear stagnation point becomes arbitrarily weak\,(see the subsection below regarding stagnation points on a no-slip surface), so a given fluid element\,(polymer molecule) only experiences an extensional flow for a finite time\,(and thence, a finite strain) in travelling from the neighborhood of the stagnation point to any point on the rear stagnation streamline, the emergence of a singularity at a finite $Wi$ is not obvious. A definitive solution of the flow past a cylinder, based on the Oldroyd-B model for $Wi \geq O(1)$, remains an outstanding question; moreover, the solution would be relevant to interpreting the flow fields obtained from nonlinear constitutive models such as FENE-P or FENE-CR for large $L$'s\,(high moleular weight polymers).

%

%
%

\subsection{Cross-slot flow} \label{sec:CrossSlotInstabilties}

The instability that has garnered substantial attention in recent years is the family of flows seen in a symmetric planar cross-slot experiment; the geometry for this experiment may be regarded as a canonical one for an extensional flow topology. The classical cross-slot geometry consists of four bisecting rectangular channels, with pairs of opposing inlets and outlets, which result in a flow field with a free stagnation point nominally located at the center. The fluid velocity at this point is, of course, zero by definition, but the velocity gradient, characterizing the local linear extensional flow, remains finite. Similar free stagnation points occur, for instance, on the surface of a translating bubble, in 
the opposing-jet configuration, and in the flows generated in two-roll and four-roll mills. The flow transitions that occur in these settings share qualitative similarities in that they are confined to a small region near the stagnation point comprising highly stretched polymer molecules.

In light of the modified cross-slot geometry also discussed below, it is worth contrasting the nature of the aforementioned free stagnation points to those on a rigid particle. A stagnation point in the latter case is only one of a continuum of zero velocity points on the particle surface, and is distinguished by the presence of a local non-linear extensional flow in its neighborhood\,(with the streamwise velocity gradient going to zero right at the surface) in contrast to the simple shear flow in the vicinity of the other points. Similar to the case of the translating cylinder/sphere problem discussed above, a characteristic feature arising from the (linear or nonlinear) extensional flow associated with stagnation points is the formation of `birefringent strands' corresponding to highly localized regions of stretched polymers\,(associated with large stresses), extending downstream from the stagnation point. The early experiments of Gardner \textit{et al.} in 1982 \citep{Gardner_1982} in the cross-slot geometry showed the existence of a velocity minimum along the centerline of the outgoing channel, close to the stagnation point, and that  was attributed to the extended polymer molecules. Later simulations \citep{Feng_Leal1997}  in a four-roll mill geometry, using the FENE-CR and FENE-P models, demonstrated the existence of the birefringent strand, and a double-humped velocity profile along the outgoing flow axes for $Wi \sim O(1)$. The evolving structure of the birefringent strands with increasing $Wi$, and the associated singular stresses, in a flow that  still conforms to the symmetry of the cross-slot about the inlet channel centerline, has been studied in detail in other efforts \citep{HarlenRallison1,HarlenRallison2,Cruz2015,Becherer2009}. Clearly, even prior to the onset of the symmetry-breaking instability discussed below, the structure of the flow in these configurations is quite nontrivial (also see Ref.~\citep{Haward_OSCER_PRL_2012}). 

\begin{figure}[h]
  \centerline{\scalebox{1.3}{\includegraphics{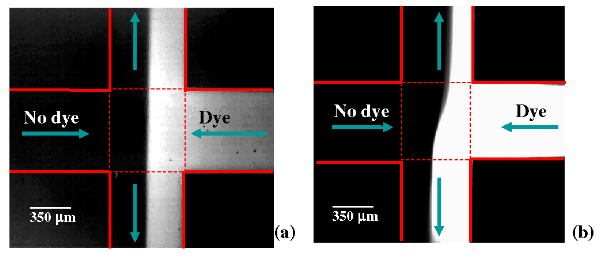}}}
  \caption{\label{fig:cross-slot} Dye advection patterns for a cross-channel flow with two inputs and two outputs at low $\Re$ ($<10^{-2}$) for (a) Newtonian fluid, and (b) PAA flexible polymer  solution  (strain  rate $\dot{\epsilon}=0.36\,\mathrm{s}^{-1}$, Weissenberg number $Wi=4.5$), where the interface between dyed and undyed fluid is deformed by an instability. Reproduced with permission from Ref~\cite{Arratia2006}.}
\end{figure}

While Muller, Odell and Tatham \citep{Muller_Odell_Tatham_1990} observed instabilities in viscoelastic flows of opposing jets, and  Broadbent, Pountney and Walters \citep{Broadbent_Walters}, and 
Ng and Leal \citep{Ng_Leal}, observed instabilities 
in stagnation point flows in two-roll and four-roll mills, respectively, it is the more recent experiments of Arratia and coworkers~\cite{Arratia2006}, using a high-molecular weight polyacrylamide solution in a microfluidic cross-slot configuration, that have unambiguously established the existence of an elastic instability; this instability limits the use of the cross-slot device as an extensional-flow rheometer. The instability observed is best understood visually; see figure~\ref{fig:cross-slot}. At low $Wi$, the incoming flow along the left and right channels in the figure) divides symmetrically into the ``up'' and ``down'' outflows, in the manner expected of a Newtonian fluid in the inertialess limit. As $Wi$ is increased, this symmetric flow becomes unstable and the system transitions to an alternate steady state. Additional unsteady instabilities also occur, as described by Arratia \textit{et al.}~\cite{Arratia2006} and, more recently, by Sousa {\it et al.}~\cite{Sousa2015}; but, perhaps because of its appealing simplicity, the steady transition has seen the most intense modeling activity. 

As mentioned in the Introduction, on account of the divergence of the extensional viscosity, the Oldroyd-B model is at its best in flows that are shear-dominated.
%
Thus, it is by no means obvious to choose the Oldroyd-B model to analyze the flow in the cross-slot geometry. Nevertheless, on account of the fewer parameters, Oldroyd-B remains a convenient first choice for modeling any new phenomenon, including ones that owe their origin to an underlying (transient) extensional flow. Indeed, the first simulations to replicate the observed cross-slot instability were carried out by Poole {\it et al.} in 2007~\cite{Poole2007} using the UCM model. The same flow was later simulated by Rocha \textit{et al.}\,\cite{Rocha2009}  using the FENE-P and FENE-CR models, to determine how the threshold $Wi$ varies with $\beta$ and $L$. The threshold $Wi$ was found to increase with decreasing $L$, and in addition, a slight smoothing of the cross-slot corners was found to have a minimal influence; both factors point to the apparent importance of extensional stresses close to the stagnation point. In contrast, for a strongly rounded corner, Rocha \textit{et al.} found that the threshold $Wi$ is significantly higher, indicative of the destabilizing role played by the sharp corners.
 The observation of the instability, and the above computations, sparked a renewed interest in the stability of pure extensional flows of the Oldroyd-B fluid \citep{Afonso_Alves_Pinho_2010,Haward_crossslot2012,Cruz_etal_2014,Wilson2012}.
In fact, it had been known since 1985 that unbounded planar linear flows (including planar extension) of an Oldroyd-B fluid were susceptible to a linear instability~\cite{Lagnado1985} involving 2D plane-wave  perturbations with a convected wavevector. This instability, however, has an essentially Newtonian origin~\cite{Lagnado1984}.
%
%

The aforementioned focus on the extensional flow topology is, in fact, not representative of the classical cross-slot geometry. While the fluid elements passing in the immediate vicinity of the cross-slot stagnation point do experience a predominantly extensional flow, those passing between the stagnation point and the re-entrant corners also 
experience shear; in fact, the flow topology in the immediate vicinity of the re-entrant corners is singular, and very different from that close to the stagnation point \citep{Hinch1993,Renardy1993}. In light of the heterogeneity in the local flow topologies, it is important to determine if the observed instability in Ref.~\citep{Arratia2006} is dominated by the extensional flow dynamics close to the stagnation point, or otherwise. The breakthrough in this regard came when Davoodi {\it et al.}~\cite{Davoodi2019} modified the cross-slot by placing a small solid cylinder, with the cylinder center at the point where the stagnation point would otherwise be located.  Based on the description of the flow topology above, in the vicinity of a no-slip surface, one no longer expects an extension-dominant character for the flow away from the corners. Nevertheless, and contrary to expectations, the presence of the cylinder (of sufficiently small radius)
made little difference to the instability, indicating that the stagnation point was not, in fact, the critical area of the flow. The authors went on to show that the onset of instability can be predicted rather well using the Pakdel-McKinley criterion discussed in section~\ref{subsec:PakdelMcKinley} above. While the value of $M$ required for instability, in the aforementioned criterion, is known from linear stability calculations for the viscometric flows in section \ref{sec:curvilinear}, for the cross-slot flow, it must be regarded as a scalar field,  either determined experimentally or from a detailed computation of the base-state; the maximum value of this `$M-$field' would then determine the threshold. 
Davoodi {\it et al.}~\cite{Davoodi2019} plotted contours of the $M$-field, and found it to be the largest near the four corners, both in the absence and presence of the cylindrical insert at the center, this largeness being attributed to the strong streamline curvature and high deformation rate prevalent near the corners. A further confirmation of the corners being responsible for this instability was the (numerical) prediction of delayed onset when the corners were strongly rounded\,\citep{Rocha2009}.
In hindsight, this explains why the Oldroyd model, even with its  problematic response to extension, turned out to reproduce experimental observations so well.

A closely related line of research is that of Haward, Alves, McKinley and collaborators \citep{Haward_OSCER_PRL_2012}, who have focused on a shape-optimized cross-slot flow geometry, in order to overcome the limitation of the aforementioned classical cross-slot geometry where the re-entrant corners likely decrease the neighborhood in which the flow is predominantly extensional, in turn leading to an instability that is (re-entrant) shear-dominated.
%
In this modified cross-slot, abbreviated as `OSCER', the sharp corners are replaced by smoothly varying contours, which result in a homogeneous elongation flow over a larger neighborhood of the stagnation point.
The OSCER configuration was found to be susceptible to a novel time dependent elastic instability at $Wi$'s lower than that corresponding to the steady asymmetric transition in the classical cross-slot configuration \citep{Haward_crossslot2012,Haward_McKinley_PoF2013,Haward_McKinley_Shen_SciRep_2016}.


To understand the origin of the aforementioned instability,
 Haward \textit{et al.}~\citep{Haward_McKinley_Shen_SciRep_2016} again plotted the `$M$-field' (of the Pakdel-McKinley criterion) for the OSCER geometry using observations from their experiments, and found that the maximum value of $M$ in the flow field, at the onset of the instability, was comparable to the critical value found numerically for planar stagnation flow \citep{Oztekin_McKinley_1997}, and to the experimental threshold for the onset of purely elastic instability in torsional shearing flows discussed in Sec.~\ref{subsec:PakdelMcKinley}. Crucially, the maximum $M$ occurred close to the stagnation point, in marked contrast to the classical cross-slot, suggesting an extensional origin for the elastic instability instead. In this regard, it is worth noting that the Pakdel-McKinley criterion was developed for viscometric\,(or nearly viscometric) shearing flows. That it works well in the prediction of elastic instabilities in  both the classical cross-slot and the OSCER configurations suggests that the observed instability relies only on the (local)\,coupling between the streamline curvature and streamwise tensile stresses, regardless of whether the underlying kinematics is shear or extension dominated. Nevertheless, the success of the criterion is not a substitute for the actual physical mechanism, and more work is needed in this regard that would likely benefit from the a focus on the dynamics in the vicinity of the birefringent strand \citep{Harris_Rallison1994,xi_graham_2009}.

Unlike the original experiments in the classical cross-slot geometry, the OSCER device has also been analyzed in regimes which allow effects of both inertia and elasticity to become important \citep{Haward_McKinley_PoF2013}, with the ratio of the two (as already seen) being characterized by the elasticity number $E$. The discussion in the preceding paragraphs pertain to the elasticity dominant limit. In the opposite limit of $E < 1$, the instability in the OSCER device occurred only beyond a critical $Re$, manifesting as an oscillatory motion of the birefringent strand alluded to above.
Owing to the importance of both elasticity and inertia, the authors have referred to this mode as an `inertio-elastic' mode. The experimental results have been summarized in the form of a stability diagram on the $Re-Wi$ plane, demarcating regions of occurrence of stable flows conforming to the cross-slot symmetry, and those corresponding to purely elastic and inertio-elastic instabilities.

\subsection{Contraction--expansion flow}
\label{sec:contraction-expansion}

\begin{figure}
\begin{center}
 \includegraphics[scale=0.35]{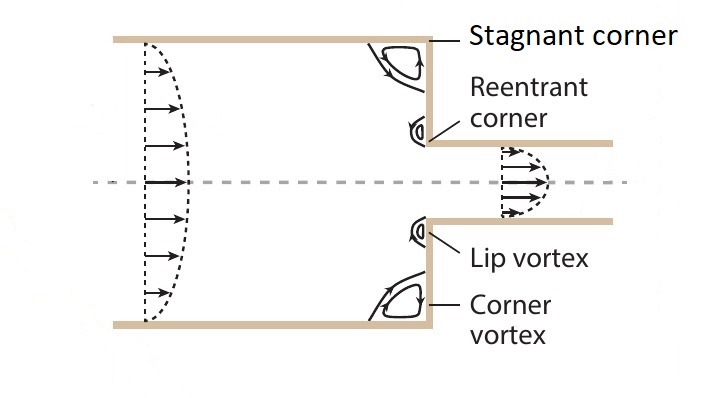}   
\end{center}
\caption{Schematic of the contraction flow geometry showing the key flow structures (i.e., the lip and corner vortices) and the locations of the corner and reentrant corner.}
\label{fig:schematic_contraction}
\end{figure}
One of the original benchmark problems, proposed in the context of numerical simulations of viscoelastic flows, was the 4:1:4 contraction-expansion flow (or, just the 4:1 contraction flow) of an Oldroyd-B fluid \citep{Hassager}, that captures some features of industrial flows like extrusion. As pointed out by Rothstein and McKinley \citep{Rothstein1999}, the geometry (see Fig.~\ref{fig:schematic_contraction}) allows for complex kinematics with a shearing deformation dominant near the walls, and a non-homogeneous extensional flow dominant along the centerline upstream of the contraction plane; one expects further complications in the vicinity of the geometric singularities\,(the reentrant corners). In the context of polymer extrusion, the aim is to optimize the entrance geometry, so as to minimize the pressure drop. While early efforts (e.g., Refs.\citep{Binding1991,Samsal1995})
focused on accurate pressure drop calculations, it became clear subsequently that
there exists a rich variety of flow patterns in this configuration, many of which owe their origin to elastic instabilities. As a result, there is now a much wider interest, from a fundamental perspective, in the different flow regimes and the relevant ranges of existence.

Previous studies have explored both axisymmetric \citep{mckinley_raiford_brown_armstrong_1991,Rothstein1999,Rothstein-McKinley-contraction-2001}  and planar \citep{Rodd2005,Rodd2007} contraction/extraction flows, although more attention has been given to the former case. In part, this difference in emphasis could be due to the much lower strains achieved in macroscopic planar contractions\,(for the same superficial velocity), leading to less pronounced viscoelastic effects; recent efforts (discussed below) by McKinley and coworkers \citep{Rodd2005,Rodd2007} have used micro-scale fabrication to partially overcome this limitation. An important question, in the present context, is as to how well the Oldroyd-B model captures the different flow transitions and instabilities.
%
%

%
%


We first briefly outline the salient observations concerning the sequence of flow patterns obtained with increasing flow rate.  The details of this sequence depend sensitively on the geometry (planar vs. axisymmetric), the contraction ratio, and on the nature of the polymer solution\,(as characterized by its transient extensional rheology). 
Boger's 1987 review \cite{Boger1987} provided a summary of the flow transitions for the axisymmetric case, for non shear-thinning\,(Boger) fluids at low $Re$, upstream of the contraction plane; in this limit, $Wi = 8 \lambda V_d/D_d$ can be used as a dimensionless measure of the flow rate, and is based on the velocity $V_d$ and diameter $D_d$ corresponding to the downstream section. For small $Wi$, the flow resembles Newtonian creeping-flow, with the appearance of a Moffatt eddy at the outer stagnant corners upstream. 
For higher $Wi$, two distinct sequences are observed depending on the nature of the polymer \citep{Boger1987}. For polyacrylamide/corn syrup Boger fluids, elastic effects lead to the corner vortex growing radially inward towards the re-entrant corner, and in the upstream direction, with increasing $Wi$; the flow remains largely steady in this regime. At higher $Wi$, the streamlines diverge away from the centerline upstream \citep{McKinley1991}. In contrast, for polyisobutylene/polybutene Boger fluids, the corner vortex decreases in size with increasing $Wi$. A separate lip vortex forms at the re-entrant corner (Fig.~\ref{fig:schematic_contraction}), the flow in the vicinity of the lip being unsteady and three-dimensional. Subsequent upstream vortex growth originates from the outward radial growth of the lip vortex \citep{mckinley_raiford_brown_armstrong_1991}. Despite the location of its inception being different for the two cases above, at very high $Wi$, the large upstream vortex becomes unstable to a global dynamical mode in both cases. The contraction ratio also determines as to which of the two aforementioned sequences occurs, with there being a transition from the lip-vortex sequence for lower contraction ratios, to the corner-vortex one for higher contraction ratios. Rounding the re-entrant corners leads to an increase in the $Wi$ required for flow transitions, but the overall structure of the flow field remains largely unaffected.

In addition to the novel flow structures engendered by elasticity, the pressure drop also shows nontrivial features compared to the Newtonian case. The pressure drop, normalized by the corresponding Newtonian value, is always greater than unity, and increases monotonically with $Wi$ before saturating; the initial increase with $Wi$ is not necessarily a signature of an elastic instability - for instance, the dimensionless pressure drop increases above unity at $Wi \sim 0.4$, but the onset of instability occurs only for $Wi > 2.6$. When instability does occur, it manifests initially as local small-amplitude fluctuations in the pressure measurements, and then, as global periodic oscillations in the pressure drop across the orifice \citep{Rothstein1999}. The additional pressure drop above,  even in the steady regime, is not captured by simulations that use nonlinear dumbbell models such as FENE-P \citep{Keiller1993} or FENE-CR \citep{Szabo1997}, with the simulations predicting a pressure drop that decreases with increasing $Wi$ for large\,(and realistic) values of $L$; an increase is predicted only for very modest $L$'s\citep{Szabo1997}; the implication for the Oldroyd-B model,  corresponding to the  infinite-extension limit, is that the pressure drop must decrease monotonically for all $Wi$.

\begin{figure}
\begin{center}
\includegraphics[scale=3.0]{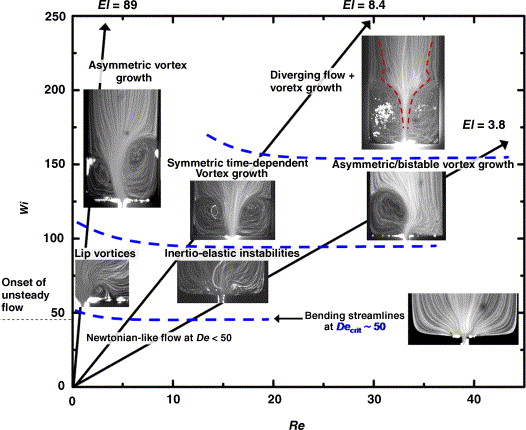} 
\end{center}
\caption{Summary of flow patterns in elastoinertial contraction flow in the $Re$-$Wi$ plane, for a contraction ratio of 16:1. Figure reproduced with permission from Rodd \textit{et al.} \cite{Rodd2005}.}
\label{fig:roddetal}
\end{figure}

For flow of Boger fluids through macroscopic planar contraction geometries,  as mentioned above, elastic effects are not as pronounced as in the axisymmetric case. 
In general, in going from axisymmetric to planar contraction flows, there is a reduction in the size, or even disappearance, of the corner vortex for the same contraction ratio.
However, Rodd and coworkers~\cite{Rodd2005,Rodd2007}, using micro-fabricated planar contraction-expansion geometries to access higher elasticity number regimes, as well as a larger fraction of the $Wi$--$Re$ parameter space, observed significant vortex growth upstream of the contraction plane for a contraction ratio of $16:1$; this growth being accompanied by a 200\% increase in pressure drop across the contraction. The  flow eventually becomes unstable and three dimensional.
%

The sequence of flow transitions, in the micro-scale planar contraction geometry above, is depicted in the $Re-Wi$ plane in Fig.~\ref{fig:roddetal}, and is seen to be sensitively dependent on $E$.
For high $E\,(\approx 90)$, onset of an elastic instability is marked by local velocity fluctuations at $W \approx 50$. This is followed 
by the development of coherent and stable lip vortices for
$50 < Wi < 100$, which subsequently develop into larger asymmetric
viscoelastic corner vortices that continue to grow upstream
for $Wi > 100$. Thus, for $E \approx 90$ and a contraction ratio of 16:1, the flow appears to take the lip-vortex route described above;  for the same contraction ratio, corner  (and not lip) vortices are observed for flow through an axisymmetric contraction \citep{Boger1987}\footnote{Note that the planar geometry used by \cite{Rodd2005,Rodd2007} has the depth dimension being much smaller than the characteristic in-plane length scales, in contrast to the usual planar limit corresponding to an infinite depth. The consequence is the existence of a strong shear in the depth coordinate, and the consequences, especially for viscoelastic flows, are not clear.}. For lower $E$, inertio-elastic
instabilities upstream of the contraction plane at $Wi \sim  50$
replace the lip vortices observed for higher elasticities.
For $100 < Wi < 150$, elastic vortices grow steadily upstream.
These vortices are essentially symmetric for $E = 9$ but become temporally unsteady, developing into spatially bistable structures for $E = 3.8$, as inertial effects become important.
Diverging streamlines eventually develop for  $ Wi > 150$, just
upstream of the elastic vortex structures. Upstream divergence of streamlines appears to, in fact, be a
common feature of non-Newtonian entry flows that are governed by the competing effects of inertia and fluid elasticity.

The dimensionless pressure drop across the micro-scale planar contraction  in general shows an increase with $Wi$, eventually saturating to a plateau. Rodd \textit{et al.} \citep{Rodd2005} attributed this saturation to the polymer molecules being fully extended at sufficiently high $Wi$, the solution behaving as an anisotropic viscous fluid in this limit, with the enhanced pressure drop being determined by the extensional viscosity corresponding to the aligned rod-like polymers. Interestingly, the pressure drop was higher for the $0.05$\% solution\,(of PEO) compared to the $0.3$\% solution. The authors attributed this to a departure from the dilute regime, with inter-chain interactions resulting in a overall decrease in polymer extensibility, in turn leading to a decreased pressure drop compared to the lower concentration solutions.

%
%
%

The recent review article by Alves \textit{et al.} \citep{Alves2021} provides a state-of-the-art summary of numerical simulations of viscoelastic flows, and includes a summary of contraction/expansion flows. 
Alves and collaborators have previously analyzed the sequence of flow  states  that emerge as a function of $Wi$, for both planar \citep{Alves2004} and axisymmetric \citep{Oliveira2007} contractions, using the Oldroyd-B and PTT models, and were able to qualitatively obtain the sequence of flow transitions and upstream vortex growth observed in experiments. The authors further
demarcated the occurrence of the various flow states discussed above, in the contraction ratio-$Wi$ plane. For the simple models used, the  nature of such phase diagrams were rather similar for both geometries, in sharp contrast to the experimental observations discussed above.  However, the computations have thus far focused on the steady regime and (unsteady) regimes where the vortices exhibit considerable oscillations $(> 1\%)$ in their size, but have not been used to predict the onset of elastic instabilities in this geometry. 

Given the discrepancies in the pressure drop between dumbbell-based simulations and experimental observations, discussed above, it remains to be seen if  use of the Oldroyd-B model would at all be useful in the prediction of elastic instabilities in this flow configuration. In this regard, it is worth mentioning one stability analysis of the contraction--expansion flow configuration~\cite{Hassell2008}, which found a primary downstream mode of instability -- but, this work pertained  to polymer melts, and accordingly used the Rolie-Poly model\,(described in Ref.\cite{LikhtmanGraham2003}); the mechanism was found to be critically dependent on chain stretch. Thus, it is not surprising that this downstream instability has neither been predicted in computations using the Oldroyd-B model nor observed in experiments using dilute solutions.  An accurate prediction of both the steady flow and the onset of instabilities in contraction flows, therefore, remains a challenge. The challenge here is two-fold: first, one needs a constitutive model that captures the stress response in a strong and transient extensional flow; second, one  needs accurate numerical schemes for implementing such a constitutive model.

%

\subsection{Viscoelastic flow past a cylinder}
\label{subsec:flowcylinder}

Flow around a circular cylinder is one of the most widely studied external flow configurations in fluid mechanics. As mentioned earlier, this configuration in a confined setting, with the cylinder diameter equaling half the separation between the confining boundaries (corresponding to the so-called blockage ratio of 0.5), was one of the original benchmark problems in viscoelastic fluid mechanics, having been extensively used to test how well constitutive models and numerical schemes are able to predict experimental data 
\citep{Hassager,Alves2021}.
With the advent of microfabrication and 3D printing technologies,  it is now possible to fabricate glass cylinders with radii of $O(10) \mu$m \citep{Haward_JNNFM_2018,Haward_SoftMatter_2019}. The fabrication of slender high-aspect-ratio cylinders, in a low-blockage-ratio ($\sim 0.1$) setting, allows experiments that mimic the theoretical limit of 2D flow past an infinite circular cylinder.
%
%
Further, for cylinders of such small radii, the typical Reynolds numbers will be significantly lower than unity, the Weissenberg numbers being very high at the same time, implying that elastic forces can be significantly enhanced ($Re < 10^{-4}$, $Wi \sim O(4000)$ and $E \sim O(10^8)$ in  Ref.~\citep{Haward_SoftMatter_2019}, with $Wi$  defined using the average velocity in the microfluidic channel)). Such flows provide ideal platforms for studying purely elastic instabilities in a canonical non-viscometric setting characterized by both shear (the gap between the cylinder and wall) and extensional (front and rear stagnation points) kinematics.

For lower $Wi$, the flow around the cylinder, although laterally symmetric, exhibits a fore-aft asymmetry. This asymmetry is on account of an elastic wake comprising the region of slow-moving fluid that develops along the rear stagnation streamline for $Wi \geq O(1)$.  The asymmetry increases with $Wi$, reflecting the increased length needed for the fluid to relax back to the ambient uniform flow at larger $Wi$; this behavior has been observed for both spheres  \citep{Fabris_sphere} and cylinders \citep{Haward_JNNFM_2018}, and the computational difficulties in capturing the elastic stresses around the rear stagnation streamline were already mentioned in the introduction to non-viscometric flows above. The experiments of Kenney \textit{et al.} \citep{Kenney_etal} demonstrated the existence of instabilities both downstream (for low blockage ratios) and upstream (for high blockage ratios) of the cylinder. In the latter case, the gap between the cylinder and the confining boundary becomes narrow compared to the cylinder radius, and hence the flow into the gap mimics that of a (smooth) planar contraction geometry; upstream instabilities are well-known to occur in this geometry, as discussed in Sec.~\ref{sec:contraction-expansion} above. 
%
Even prior to the advent of the microfluidic setting, flow past a cylinder has been shown to be susceptible to a purely elastic instability that leads to spanwise variations of the streamwise velocity in the elastic wake downstream \citep{McKinley1993}. The instability appears to have its origin in the extensional flow close to the rear stagnation point, and has been analyzed along these lines using the Oldroyd-B model \citep{Oztekin_McKinley_1997}; importantly, despite the extensional kinematics, the threshold has been successfully interpreted in terms of the Pakdel-Mckinley criterion.

\begin{figure}
\begin{center}
\includegraphics[scale=2.5]{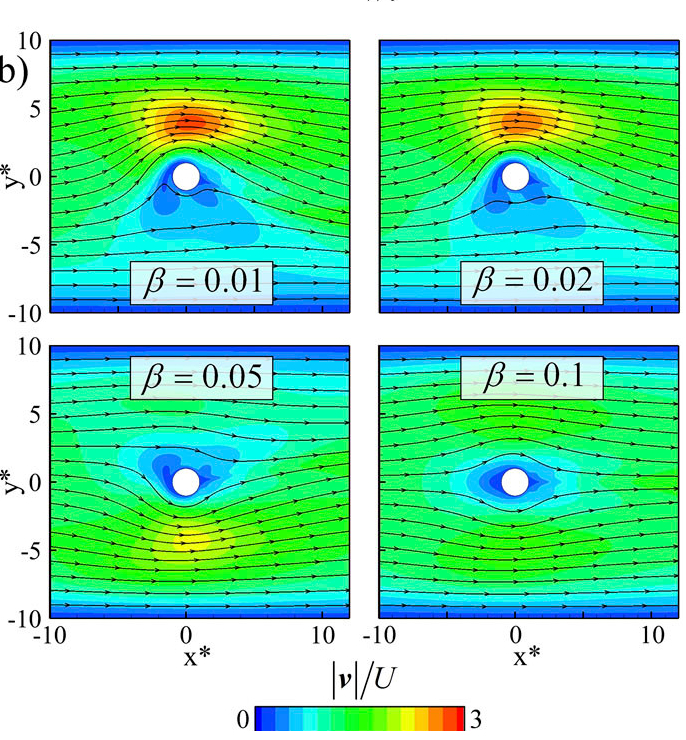} 
\end{center}
\caption{Effect of shear thinning on the emergence of flow asymmetry in flow past a cylinder as obtained using the l-PTT model. Note that shear thinning is more pronounced in this model as $\beta$ is decreased. Reproduced with permission from Haward \textit{et al.} \citep{Haward_PoF_2020}.}
\label{fig:patternshaward}
\end{figure}

Recent experiments have shown that the lateral symmetry of the flow is broken, beyond a threshold $Wi$, due to an instability that also leads to a distortion of the downstream elastic wake \citep{Haward_SoftMatter_2019,Haward_JNNFM_2020}.  The resulting asymmetric flow is characterized by the preferential passage of the fluid around one side of the cylinder. Theoretically speaking, the instability corresponds to a pitchfork bifurcation, with passage of the fluid around either side being possible, the preferred side being determined by experimental conditions (see Fig.~12 in Ref.\,\citep{Haward_JNNFM_2020}). Numerical simulations of flow past a cylinder, carried out \citep{Haward_PoF_2020} using the linear Phan-Thien and Tanner (l-PTT) model\,\footnote{The Phan-Thien and Tanner (PTT) model is a nonlinear generalization of the Johnson-Segalman model (Sec.\,\ref{subsubsec:beyondOldB}), with the relaxation time being a function of the trace of the stress tensor. Two functional forms have been proposed for the relaxation time: (i) a linear\,(resulting in the `l-PTT' model), and (ii) an exponential function, of the trace of the stress tensor. In addition to the slip parameter $a$ of the JS model, another parameter $\epsilon$ characterizes the strength of the nonlinearity, with $\epsilon = 0$  recovering the JS model. Similar to the JS model, the PTT model also predicts shear thinning, and could exhibit non-monotonicity of the constitutive curve in certain parametric regimes. The exponential version of the PTT model is suited for modelling extensional flows  and predicts strain softening of the extensional viscosity at high strain rates.}, showed that the threshold Weissenberg number, $Wi_c$, scales with blockage ratio, $B_R$,  as $Wi_c \sim 1/B_R$, consistent with the predictions obtained using the Pakdel-McKinley criterion \citep{McKinley1996}. Nevertheless, unlike the elastic-wake instability mentioned above \citep{McKinley1993}, the onset of a lateral asymmetry appears to require a combination of shear thinning and elasticity, with shear thinning playing an essential role (see Fig.~\ref{fig:patternshaward}). In the context of the l-PTT model shear thinning is mainly determined by the viscosity ratio parameter $\beta$; smaller $\beta$ indicates larger shear thinning.
%
%
%
%


%

\begin{figure}
\begin{center}
\includegraphics[scale=2.5]{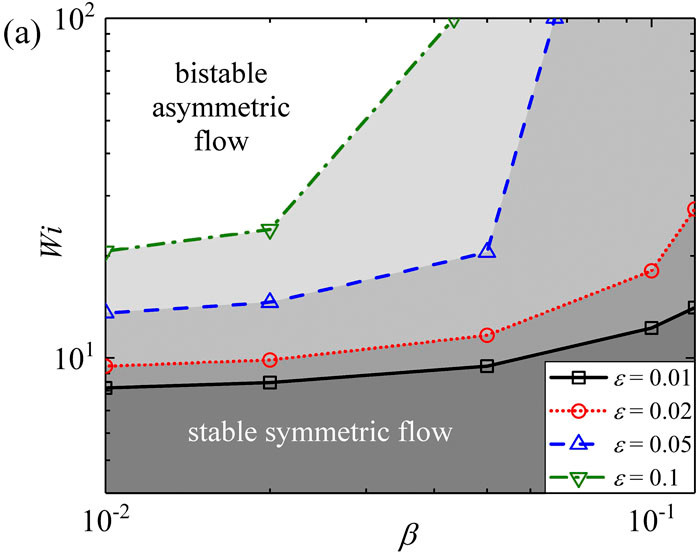} 
\end{center}
\caption{Critical conditions in the $\beta$-$Wi$ plane for the onset of asymmetric flow in flow past a cylinder as obtained using the l-PTT model. Reproduced with permission from Haward \textit{et al.} \citep{Haward_PoF_2020}.}
\label{fig:hawardetal}
\end{figure}

Haward \textit{et al.} \citep{Haward_PoF_2020} summarized their results in terms of a stability diagram in the $Wi$-$\beta$ plane, wherein the regions of stable symmetric flow and bistable asymmetric flow were demarcated (see Fig.~\ref{fig:hawardetal}).
Laterally asymmetric flows are observed only when the characteristic shear rate near the cylinder lies in the shear-thinning portion of the flow curve for the relevant constitutive model; thus,  for shear rates corresponding to the high-shear plateau, symmetric flows are recovered despite the large $Wi$. In light of this, the success of the Pakdel-McKinley criterion, developed for instabilities with a purely elastic origin (whose region of occurrence, in the relevant parameter space, could be modified by shear-thinning effects; see Figure~\ref{fig:Pakdel} in Sec.~\ref{subsec:PakdelMcKinley}) is a little surprising. Further, calculations based on the constant-viscosity FENE-CR model have only shown a localized oscillation/instability in the immediate neighborhood of the rear stagnation point, and it is unclear as to whether this time dependence is a physical or a numerical artifact (see discussion in the beginning of this section on the high-$Wi$ problem). For instance, it is unclear if the shear-thinning rheology amplifies this local breakdown of the steady symmetric solution into a global symmetry-breaking instability \citep{Haward_PoF_2020}. Thus, while the authors have constructed a stability diagram using $Wi$ and a parameter that characterizes the extent of shear thinning, a more comprehensive (and numerically difficult) investigation of the parameter space, particularly in the near-constant-viscosity limit, appears necessary to elucidate the physical mechanism.

We end this section by pointing out that the dumbbell (Oldroyd-B/FENE-CR/FENE-P) class of models have proved grossly inadequate in predicting the increase\,(with $Wi$) of either the pressure drop for contraction flows, or the drag coefficient for flow past a cylinder\,(or sphere). In both these configurations, the dynamics for $Wi$ of order unity or greater is dominated by fluid elements that experience a transient extensional flow\,(close to the centerline for the contraction flow, and in the neighborhood of the rear stagnation point for the sphere/cylinder). It is therefore thought that the phenomenon of stress-conformation hysteresis, that arises in transient extensional flows \citep{Doyle_Shaqfeh_1998,Doyle-Shaqfeh-McKinley1998}, and the resulting dissipative stresses, may play a key role in the aforementioned increase in frictional resistance \citep{Rothstein1999,Rothstein-McKinley-contraction-2001,McKinley2001}. Accounting for the hysteresis-induced increase in frictional resistance will require either a  fine-grained description of the polymer chains that accounts for internal degrees of freedom\,(a Kramers bead-rod chain was used in \citep{Doyle-Shaqfeh-McKinley1998}), or the incorporation of an additional dissipative stress contribution within the dumbbell framework \citep{Rallison1997,Verhoef1999}. However, even after using such a dumbbell model that accounts for dissipative stresses, Yang and Khomami \citep{Yang_Khomami_1999} concluded that it is not possible to predict experimental observations quantitatively. Much later, Khomami and coworkers have employed a multi-scale approach that adopted a Brownian configuration field approach to simulate the dynamics of a bead-spring chain, coupled with a finite-element simulation of the continuum field equations, for both the flow through the contraction-expansion geometry \citep{koppol_sureshkumar_abedijaberi_khomami_2009}, and the flow past a sphere \citep{abedijaberi_khomami_2012}. These studies have shown that it is indeed possible to capture the experimental trends exhibited by pressure drop or drag coefficient with $Wi$ using such an approach.  It is, however, an open question as to whether the prediction of instabilities in these configurations would also require an approach that accounts for the aforementioned subtleties.

\section{Nonmodal stability}
\label{sec:nonmodal}

Modal stability analysis has been successful at explaining experimental observations in large variety of fluid flows.  However, and perhaps surprisingly, it fails for rectilinear flows of Newtonian fluids.  Standard linear stability analysis predicts plane Couette flow and Hagen--Poiseuille (i.e., pipe) flow to be linearly stable for all Reynolds numbers.  Yet, instability is observed experimentally.  Although plane Poiseuille flow is predicted to be linearly unstable above a certain Reynolds number, in practice instability is observed at Reynolds numbers much lower than this critical value.

While these discrepancies may be attributed to the presence of finite-amplitude perturbations, which would cause the assumption of linearization to fail, it is now widely recognized that the discrepancies are likely due to other assumptions inherent in modal analysis \cite{Panton2013,Drazin2002,Charru2011,SchmidARFM2007}.  As briefly mentioned in Section~\ref{sec:Intro}, modal analysis provides predictions about {\it asymptotic stability}, i.e., whether perturbations grow or decay at long times.  It says nothing about behavior at short times.  However, two stable modes can interact in such a way that a disturbance grows at short times before decaying at long times.  This growth at short times could put the flow into a regime where nonlinear effects are no longer negligible, causing a transition to another flow state.  Thus, an initially small-amplitude disturbance can be amplified through a purely linear mechanism that is overlooked by modal analysis.

{\it Nonmodal analysis} provides information about this alternative type of perturbation growth \cite{SchmidARFM2007,MihailoARFM2021}.  It is worth mentioning that the original 
nonmodal analyses were restricted to infinitesimal disturbances \citep{Schimid-Henningson}. Within this assumption, the optimal disturbances corresponding to maximum
transient growth in Newtonian shear flows were identified in most cases as counter-rotating streamwise vortices
aligned along the spanwise direction, giving rise to growing streaks. The detailed
manner in which this growth would eventually be modified by nonlinear effects was
addressed much later \citep{Pringle_Kerswell2010,kerswell2018} for Newtonian flows. These efforts
have
obtained three-dimensional spatially localized structures, by accounting for the effects of
nonlinearity within a more general optimization approach. This approach has not yet been adopted for viscoelastic flows, and we restrict our discussion here to nonmodal growth in viscoelastic shear flows within a linearized framework.

In this section, we will provide an overview of some basic ideas, discuss their relevance to viscoelastic channel flows, highlight some recent results related to amplification of external disturbances, and identify some important open issues in the area.  Our discussion is not intended to be a comprehensive tutorial or review, but is instead aimed at providing non-expert readers a brief introduction to some fundamental concepts and selected results.

\subsection{Nonmodal amplification: Basic ideas}
\label{Sec5.1}

To illustrate the basic ideas of nonmodal amplification, we consider the 
 coupled pair of constant-coefficient linear ordinary differential equations (adapted from \cite{Grossmann2000}; see also \cite{liejovkumJFM13}):

\begin{eqnarray}
\label{eqn5.1-1SK}
\left(\begin{array}{c} \dot{x}_1 \\  \dot{x}_2 \\ \end{array} \right)
=\left(\begin{array}{cc}
\lambda_1 &  0 \\
R  & \lambda_2 \\
\end{array} \right)
\left(\begin{array}{c} x_1 \\  x_2 \\ \end{array} \right)
, 
\end{eqnarray}
where the dot denotes a time derivative and $\lambda_1$, $\lambda_2$, and $R$ are all real constants. We denote the matrix appearing in (\ref{eqn5.1-1SK}) as $\mathbf{A}$.  The solution to this equation system for $\lambda_1 \ne \lambda _2$ is
\begin{equation}
\label{eqn5.1-2SK}
x_1=x_1^0 e^{\lambda_1 t},
\end{equation}
\begin{equation}
\label{eqn5.1-3SK}
x_2=x_2^0 e^{\lambda_2 t}+\frac{R x_1^0}{\lambda_1-\lambda_2}\left(e^{\lambda_1 t}-e^{\lambda_2 t}
\right),
\end{equation}
where $x_1^0$ and $x_2^0$ are the initial conditions for $x_1$ and $x_2$.

With $\lambda_1, \lambda_2 < 0$, both $x_1$ and $x_2$ decay to zero as $t \rightarrow \infty$. However, if $R \ne 0$, $x_2$ can grow before decaying if $x_1^0 \ne 0$.  Note that the difference of the two decaying exponentials appearing in the expression for $x_2$ is zero at $t=0$ and approaches zero at long times, but is non-zero at intermediate times.  The influence of this term increases as the magnitude of $R$ does, and so does the amplification rate, since $dx_2/dt \sim Rx_1^0$ for $R \gg 1$.  Thus, there exist initial conditions such that $x_2$ can exhibit large {\it transient growth}.  Such growth is also referred to as {\it nonmodal amplification} because it would be missed by standard modal analysis, which focuses on responses triggered by individual eigenvalues, and thus long-time behavior.  We note that terms like $t e^{-t}$ that arise when $\lambda_1 = \lambda _2$ reflect a resonant interaction and also exhibit transient growth, but such terms are not required for transient growth as the above example illustrates.

Clearly, the parameter $R$ is causing this behavior.  If $R=0$, there is no transient growth and both $x_1$ and $x_2$ decay monotonically to zero.  When $R \ne 0$, the two stable modes interact, leading to nonmodal amplification.  Having $R \ne 0$ significantly changes the properties of the matrix $\mathbf{A}$ in  (\ref{eqn5.1-1SK}).  In particular, $\mathbf{A}$ no longer commutes with its adjoint, which in this case is simply the transpose. 

When a linear operator $\mathbf{L}$ commutes with its adjoint $\mathbf{L}^*$, then  $\mathbf{L} \mathbf{L}^* = 
\mathbf{L}^* \mathbf{L}$ and we refer to  
$\mathbf{L}$ as being a {\it normal} linear operator \cite{NaylorSell71,Strang88}.  Normal linear operators have orthogonal eigenfuctions.  However, 
if $\mathbf{L}$ does not commute with its adjoint, then  $\mathbf{L} \mathbf{L}^* \ne  
\mathbf{L}^* \mathbf{L}$, and $\mathbf{L}$ is {\it non-normal}.  Non-normal linear operators produce eigenfuctions that are non-orthogonal.

In the above example, the eigenvectors of the matrix $\mathbf{A}$ are given by 
\begin{eqnarray}
\label{eqn5.1-4SK}
\left(\begin{array}{c} 0 \\  1  \end{array} \right),
\frac{1}{\sqrt{1+R^2/(\lambda_1-\lambda_2)^2}}
\left(\begin{array}{c} 1 \\  \frac{R}{\lambda_1-\lambda_2}  \end{array} \right).
\end{eqnarray}
These eigenvectors are orthogonal only when $R=0$, and as $R \rightarrow \infty$, these eigenvectors become parallel.  In that situation, an initial condition that is nearly orthogonal to the eigenvectors would be ``misfit'', and the coefficients involved in the solution would have a very large magnitude 
(e.g., $\left(\,1\,\,\,0\,\right)^{T}=
-\epsilon^{-1}\left(\,0\,\,\,1\,\right)^{T}+
\epsilon^{-1}\left(\,\epsilon\,\,\,1\,\right)^{T}$, where $\epsilon \ll 1$).

If system (\ref{eqn5.1-1SK}) represents the linearization of a nonlinear system around a steady state, then simply focusing on the eigenvalues is misleading.  The steady state is asymptotically stable, but there may be large growth of perturbations at short times.  This would put the system into a regime where nonlinear terms are no longer negligible, and the system may transition to another state rather than returning to the steady state one started with.

Although the above discussion has focused on an unforced linear system with non-zero initial conditions, the same ideas apply to a forced linear system with zero initial conditions.  If the underlying linear operator is non-normal, then perturbations in the problem variables created by the forcing can have magnitudes considerably larger than the magnitude of the forcing (e.g., see \cite{SchmidARFM2007,MihailoARFM2021,
Mihailo2005}).

\subsection{Relevance to viscoelastic channel flows}
\label{Sec5.2}

The above example is directly relevant to channel flows of Newtonian and viscoelastic fluids \cite{jovanovic_kumar_2010}.  We consider channel flows driven by a constant pressure gradient (plane Poiseuille flow) or a constant boundary velocity (plane Couette flow).  The streamwise direction (direction of mean flow) is $x$, the wall-normal direction is $y$, the spanwise direction is $z$, and $t$ is time.  The equations are linearized around the base state and Fourier transforms are applied in the $x$- and $z$-directions to obtain a system of partial differential equations for the velocity, pressure, and stress fluctuations where $y$ and $t$ are the independent variables.  The Fourier transforms introduce the wavenumbers $k_x$ and $k_z$, characterizing variations in the streamwise and spanwise directions, respectively.    

For streamwise-constant disturbances ($k_x=0$), the linearized governing 
equations can be put into forms very similar to the example problem (\ref{eqn5.1-1SK}), allowing us to make powerful analogies \cite{jovanovic_kumar_2010}.  For channel flows of Newtonian fluids, the linearized governing equations can be written as
\begin{eqnarray}
\label{eqn5.2-1SK}
\left(\begin{array}{c} \dot{\psi} \\  \dot{u} \\ \end{array} \right)
=\left(\begin{array}{cc}
\bar{\mathbf{A}}_{11} &  0 \\
Re \bar{\mathbf{A}}_{21}  & \bar{\mathbf{A}}_{22} \\
\end{array} \right)
\left(\begin{array}{c} \psi \\  u \\ \end{array} \right)
, 
\end{eqnarray}
where $\psi$ is the streamfunction in the $yz$-plane, $u$ is the streamwise velocity component, and the dot now denotes a partial derivative with respect to time.  Here, $\bar{\mathbf{A}}_{11}=\Delta^{-1}\Delta^2$ is  the Orr-Sommerfeld operator,  and $\bar{\mathbf{A}}_{22}=\Delta$ is the Squire operator, and the operator $\bar{\mathbf{A}}_{21}=-ik_z\bar{U}'(y)$ is the vortex-tilting or lift-up operator \cite{Brandt2014,Ellingsen1975},  $\Delta=\partial_{yy}-k_z^2$, $\Delta^2=\partial_{yyyy}-2k_z^2\partial_{yy}+k_z^4$, $\bar{U}(y)$ is the base-state velocity, and the prime denotes a derivative.  System~(\ref{eqn5.2-1SK}) is characterized by the Reynolds number $Re = \rho U_0 L/\eta_s$, where $\rho$ is the density, $U_0$ is the maximum magnitude of the base-state velocity, $L$  is the channel half-height, and $\eta_s$ is the viscosity.  

Comparing (\ref{eqn5.2-1SK}) with (\ref{eqn5.1-1SK}), we see that the Reynolds number $Re$ in (\ref{eqn5.2-1SK}) plays the role of the parameter $R$ in (\ref{eqn5.1-1SK}).
When $Re$ is non-zero, the Orr-Sommerfeld and Squire modes become coupled, and the problem becomes increasingly non-normal as $Re$ increases.  Note that the analogy can be made more direct by recognizing that numerical discretization of the derivatives with respect to $y$  converts the operator in (\ref{eqn5.2-1SK}) into a standard matrix, and the example problem (\ref{eqn5.1-1SK}) can be generalized to higher dimensions.  The coupling term $\bar{\mathbf{A}}_{21}=-ik_z\bar{U}'(y)$ involves interaction between the mean shear, or vorticity, and three-dimensional velocity perturbations (This term is often referred to as vortex tilting, but it really involves the mean vorticity.)  It gives rise to alternating regions of high and low streamwise velocity, often referred to as streamwise streaks \cite{Brandt2014,Ellingsen1975}.

For inertialess channel flows of Oldroyd-B fluids, the linearized governing equations for the components of the polymer stress fluctuations $\tau_{ij}$ can be written as \cite{jovanovic_kumar_2010}
\begin{eqnarray}
\label{eqn5.2-2SK}
\left(\begin{array}{c} \dot{\mbox{\boldmath$\tau$}}_1 \\  \dot{\mbox{\boldmath$\tau$}}_2 \\ \end{array} \right)
=\left(\begin{array}{cc}
\mathbf{A}_{11} &  0 \\
Wi \mathbf{A}_{21}  &\mathbf{A}_{22} \\
\end{array} \right)
\left(\begin{array}{c} \mbox{\boldmath$\tau$}_1 \\  \mbox{\boldmath$\tau$}_2 \\ \end{array} \right)
, 
\end{eqnarray}
where $\mbox{\boldmath$\tau$}_1=\left(\,\tau_{22}\,\,\,\tau_{23}\,\,\,\tau_{33}\,\right)^{T}$ and $\mbox{\boldmath$\tau$}_2=\left(\,\tau_{12}\,\,\,\tau_{13}\,\right)^{T}$ with $(1,2,3)$ representing $(x,y,z)$.   The Weissenberg number is given by $Wi = \lambda U_0/L$, where $\lambda$ is the fluid relaxation time.

The operators $\mathbf{A}_{11}$, $\mathbf{A}_{22}$, and $\mathbf{A}_{21}$ are all independent of $Wi$ and their definitions can be found in \cite{jovanovic_kumar_2010}.  A static-in-time relationship \cite{jovanovic_kumar_2010} (not shown here) connects the polymer stress fluctuations to the velocity fluctuations.  The operators $\mathbf{A}_{11}$ and $\mathbf{A}_{22}$ involve the spanwise wavenumber $k_z$, derivatives with respect to $y$, and the viscosity ratio $\beta=\eta_s/(\eta_s+\eta_p)$, where $\eta_s$ and $\eta_p$ are, respectively, the solvent and polymer contributions to the viscosity.  The operator $\mathbf{A}_{21}$ shares these features as well, but it also involves the first and second derivatives of the base-state velocity.  This operator embeds the interaction between base-state velocity gradients and polymer stress fluctuations, and
the interaction between base-state polymer stress gradients and velocity fluctuations.  Physically, such interactions produce polymer stretching.

Comparing (\ref{eqn5.2-2SK}) with (\ref{eqn5.2-1SK}) reveals a remarkable analogy between creeping
flows of Oldroyd-B fluids and inertial flows of Newtonian
fluids for streamwise-constant disturbances \cite{jovanovic_kumar_2010}.  Polymer stretching and the Weissenberg number in elasticity-dominated flows of viscoelastic fluids play the role of vortex tilting and the Reynolds number in inertia-dominated flows of Newtonian fluids. When $Wi=0$, there is no coupling between $\mathbf{\tau}_1$ and $\mathbf{\tau}_2$ in Eq.~(\ref{eqn5.2-2SK}).  However, when $Wi \neq 0$, the evolution of 
$\mathbf{\tau}_2$ is influenced by $\mathbf{\tau}_1$ and system~(\ref{eqn5.2-2SK}) becomes increasingly nonnormal as $Wi$ increases. 

%

The discussion above lays bare the relevance of example problem (\ref{eqn5.1-1SK}) to nonmodal amplification in streamwise-constant channel flows of Newtonian and viscoelastic fluids.  In the case of Newtonian fluids, initially small-amplitude perturbations can become highly amplified through a nonmodal mechanism.  This puts the flow into a regime where nonlinear terms are no longer negligible, triggering a transition to turbulence even when no eigenvalue of the linearized problem has a positive real part \cite{Trefethen1993,Butler1992,Gustavsson1991,Henningson1993,Reddy1993}.  In the case of viscoelastic fluids, such disturbance amplification can potentially trigger a transition to elastic turbulence.  Further discussion of the linearized dynamics in viscoelastic fluids can be found in \cite{PageZaki2014}.  

In analogy with example problem (\ref{eqn5.1-1SK}), the discussion above implies that there exist initial conditions for systems (\ref{eqn5.2-1SK}) and (\ref{eqn5.2-2SK}) that lead to large transient growth of flow fluctuations.  This raises the question of how such initial conditions can be generated \cite{Bamieh2001,Farrell1993,Mihailo2005}, a topic we turn to next.

\subsection{Amplification of external disturbances}
\label{Sec5.3}

Fluid flows in practice are subject to disturbances that arise from sources such as vibrations and pressure fluctuations.  If these disturbance are amplified by the flow, they could produce the initial conditions that lead to significant transient growth, or could themselves induce a flow transition \cite{Bamieh2001,Farrell1993,Mihailo2005}.

While the exact form of disturbances that arise in  experimental settings may be unknown, it is still useful to consider how a flow model responds to various types of well-defined disturbances.   This is closely related to the topic of frequency response, which is often covered in undergraduate courses on control systems.  There, one typically considers systems of ordinary differential equations with disturbances that are time-periodic.  Of course, disturbances can be localized in time as well (e.g., an impulse), and with the partial differential equations that arise in flow models, the disturbances can be spatially varying.  Disturbance amplification can also provide insight into model robustness, or how sensitive a model is to neglected terms \cite{Bamieh2001,Mihailo2005,SchmidARFM2007,liejovkumJFM13}.  

Some of the most basic disturbances that can be considered in fluid flows are those that are random, e.g., white noise.  A simple way to account for these disturbances is to include them as body forces in the linearized equations.  Because disturbances are easily characterized by streamwise and spanwise wavenumbers, it is particularly useful to consider disturbances that are harmonic in those directions but stochastic in the wall-normal direction and time.  It then of interest to know how the disturbances to the base flow behave as a function of the streamwise and spanwise wavenumbers, and other parameters such as the Reynolds number, Weissenberg number, and viscosity ratio.

In control-systems courses, transfer functions relate input and output variables.  The same idea can be applied to fluid flows, with the input being the body force and the output measured in terms of scalar quantities like the kinetic energy of the velocity fluctuations \cite{MihailoARFM2021}.  Below, we discuss some important results concerning this topic, which draw heavily upon ideas and tools from linear systems theory and control theory.

\begin{figure}[ht]
\begin{center}
\includegraphics[width=5cm]{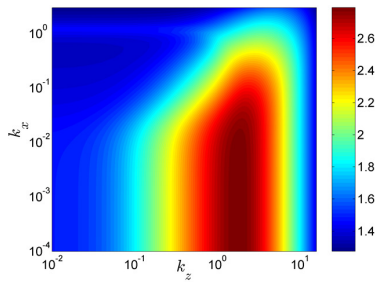}
\end{center}
\caption{\small Plot of (1/2)log$_{10}{\cal E}$ for plane Poiseuille of a Newtonian fluid at $Re=2000$; adapted (with permission) from Fig.~5 of \cite{Mihailo2005}.} 
\label{fig1-SK}
\end{figure}

We begin with inertial flows of Newtonian fluids subject to a body force that is harmonic in the streamwise and spanwise directions but stochastic (white noise) in the wall-normal direction and time \cite{Mihailo2005}.  Figure \ref{fig1-SK} shows a plot of the ensemble average energy density, ${\cal E}$.  This quantity, which we will refer to as the energy density for brevity, is simply the kinetic energy of the velocity fluctuations averaged over the wall-normal direction and time \cite{Bamieh2001, Mihailo2005}.

Figure \ref{fig1-SK} shows the energy density as a function of $k_x$ and $k_z$ for plane Poiseuille of a Newtonian fluid at $Re = 2000$.  The largest energy density occurs for $k_x \approx 0$ and $k_z \approx 1.78$, which corresponds to streamwise-constant disturbances.  In this limit, an analytical expression can be obtained for the energy density \cite{Bamieh2001,Mihailo2005}
\begin{equation}
\label{eqn5.3-1SK}
{\cal E}=Re f_N(k_z)+ Re^3 g_N(k_z).
\end{equation}
The function $f_N$ is a monotonically decreasing function of $k_z$, whereas the function $g_N$ has a peak at $k_z \approx 1.78$.  The function $f$ reflects the influence of viscous dissipation and the function $g$ reflects the influence of vortex tilting.  Thus, for $Re \gg 1$, the term involving $g$ dominates, and the energy density is largest for $O(1)$ values of $k_z$.  This highlights the importance of three-dimensional disturbances in inertial channel flows of Newtonian fluids.

\begin{figure}[ht]
\begin{center}
\begin{subfigure}[htp]{0.45\textwidth}
\includegraphics[width=0.9\textwidth]{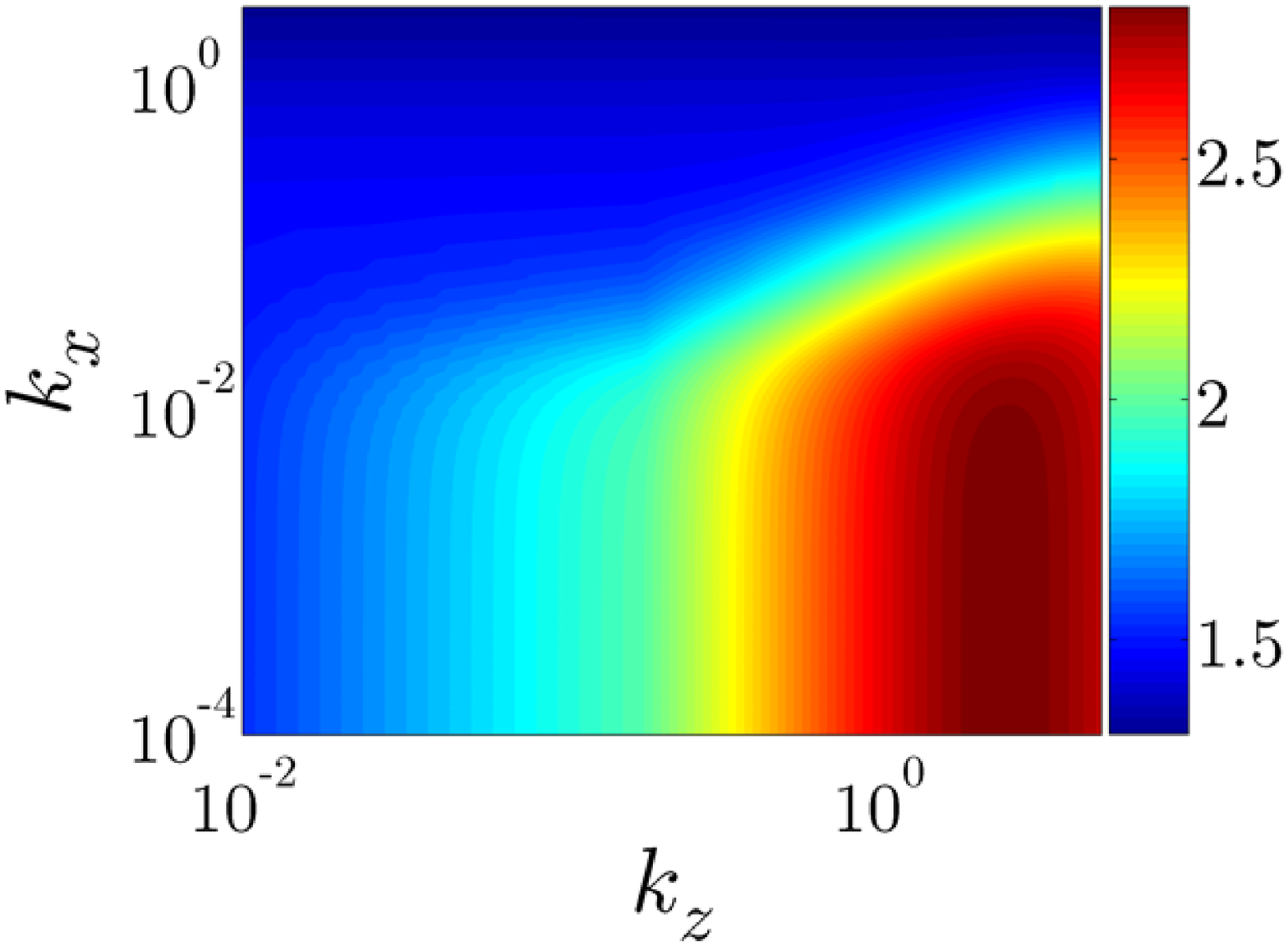}
\label{fig2a}
\end{subfigure}
\begin{subfigure}[htp]{0.45\textwidth}
\includegraphics[width=0.9\textwidth]{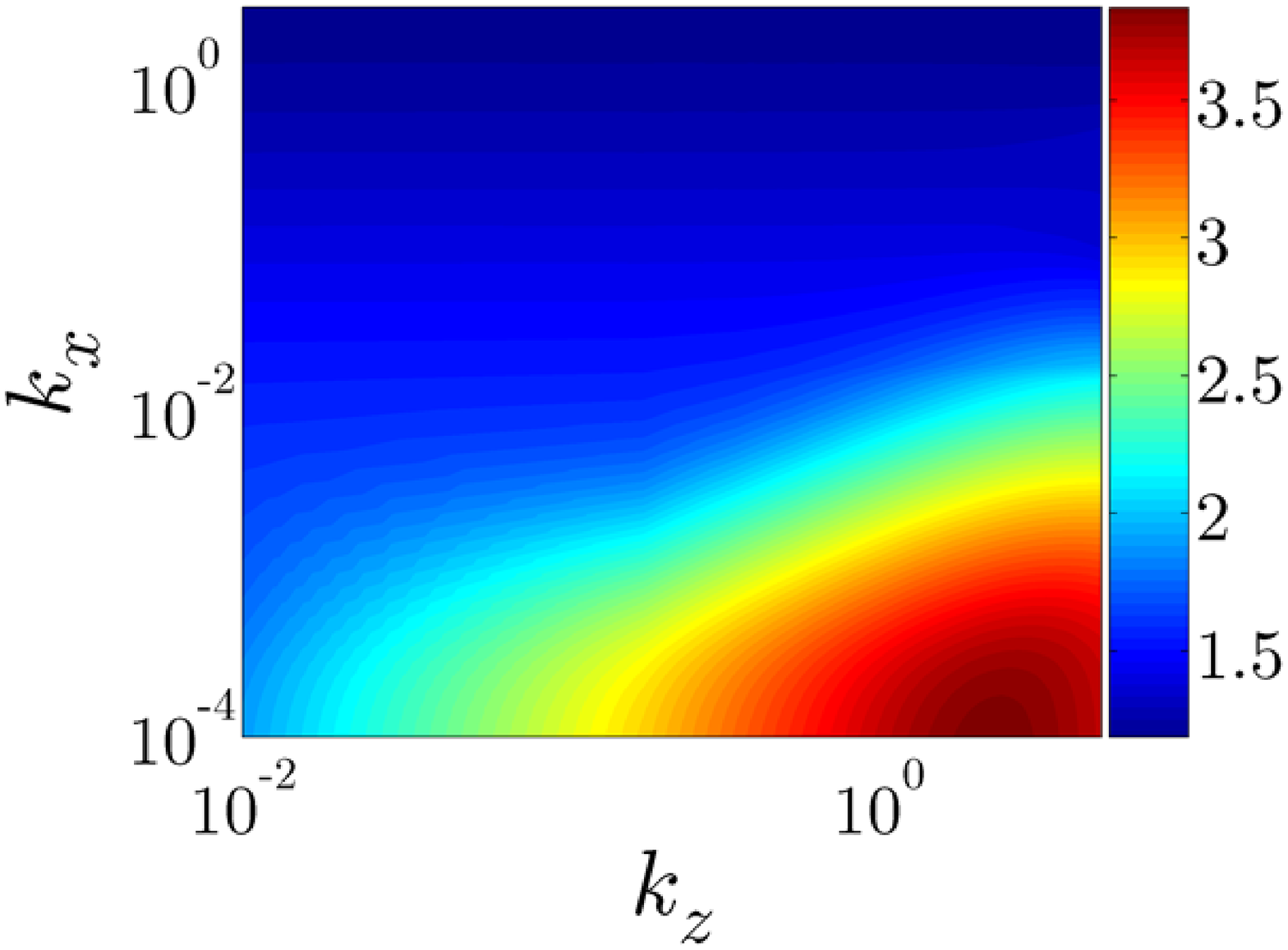}
\label{fig2b}
\end{subfigure}
\end{center}
\caption{\small Plots of (1/2)log$_{10}{\cal E}$ for plane Poiseuille of an Oldroyd-B fluid at $Re=1000$ and two different values of $E=Wi/Re$: (a) $E=0.1$ and (b) $E=10$; adapted (with permission) from Fig.~4 of \cite{Hoda2008}.}
\label{fig2-SK}
\end{figure}

Figure \ref{fig2-SK} shows the energy density as a function of $k_x$ and $k_z$ for plane Poiseuille of an Oldroyd-B fluid at $Re = 1000$, and two different elasticity numbers, $E= 0.1$ and $10$ \cite{Hoda2008}.
  It is seen that the energy density increases with increasing elasticity number.  In addition, the most amplified disturbances become increasingly streamwise constant as the elasticity number increases.  

For streamwise-constant disturbances at large elasticity numbers,  the energy density is found to obey the following expression \cite{Hoda2009}
\begin{equation}
\label{eqn5.3-2SK}
{\cal E} \approx Re f_{VE}(k_z, \beta)+ E Re^3 g_{VE}(k_z,\beta),
\end{equation}
As in the Newtonian case, the function $f_{VE}$ is a monotonically decreasing function of $k_z$, whereas the function $g_{VE}$ has a peak at $k_z = O(1)$.  The function $f$ again reflects the influence of viscous dissipation but the function $g$ now reflects the interaction of the polymer stresses with the velocity field.  When $Re \ll 1$, the term involving $g$  dominates if $E$ is large enough, and the energy density is largest at $O(1)$ values of $k_z$.  This highlights the importance of three-dimensional disturbances in 
strongly elastic channel flows of viscoelastic fluids. 

In the creeping-flow limit, the problem of energy amplification from stochastic forcing becomes ill-posed since the forcing affects the velocity and pressure directly without being ``filtered'' by the unsteady inertial term \cite{jovanovic_kumar_2011}.
This issue can be addressed by applying singular perturbation methods to treat the case of large elasticity numbers \cite{jovanovic_kumar_2011}.  Such an analysis shows that for inertialess streamwise-constant flows of Oldroyd-B fluids, the linearized dynamics of the wall-normal vorticity $\eta$ are governed by
\begin{equation}
\label{eqn5.3-3SK}
\partial_t \Delta \eta = -Wi (1/\beta-1)
(\partial_{yz}(U'(y)\tau_{22})+\partial_{zz}(U'(y)\tau_{23})) -(1/\beta)\Delta \eta
\end{equation}

In contrast, for streamwise-constant inertial flows of Newtonian fluids, the linearized dynamics of $\eta$ are governed by
\begin{equation}
\label{eqn5.3-4SK}
\partial_t \eta = -Re U'(y) \partial_z v + \Delta \eta,
\end{equation}  
where $v$ is the wall-normal velocity fluctuation.  The first term on the right-hand side of (\ref{eqn5.3-4SK}) corresponds to vortex-tilting and acts as a source term.  The spanwise vorticity of the
base flow, $U'(y)$ gets ``tilted'' in the wall-normal direction by spanwise changes in $v$.  This leads to the amplification of $\eta$ and thereby the streamwise velocity ($\eta=\partial_z u$), giving rise to streamwise streaks. This is the viscoelastic analog of the lift-up effect, which was briefly mentioned in Sec.~\ref{subsubsec:scenarios}.

We now consider the physical interpretation of (\ref{eqn5.3-3SK}); for additional details, see  \cite{jovanovic_kumar_2011} and Sec.~3.2 of Ref.~\cite{MihailoARFM2021}.  Here, the first term on the right-hand side represents spanwise variations in stretching of polymer stress fluctuations by the background shear.  This stretching, which acts as a source term, produces amplification of $\eta$ and, consequently, $u$, leading to streamwise streaks.  Thus, there is again a remarkable analogy between   
creeping flows of Oldroyd-B fluids and inertial flows of Newtonian fluids for streamwise-constant disturbances.  Polymer stretching and the Weissenberg number take the roles of vortex tilting and the Reynolds number.

Further analysis reveals that the energy associated with the velocity fluctuations is $O(Wi^2$) \cite{jovanovic_kumar_2011}.  One can also define an energy associated with the stress fluctuations, and this is $O(Wi^4)$ \cite{jovanovic_kumar_2011}.  The same scalings are obtained if one considers creeping flows with disturbances that are harmonic in time as well as in the streamwise and spanwise directions, and deterministic in the wall-normal direction \cite{liejovkumJFM13}.  Thus, in the creeping-flow limit, the more elastic the fluid is, the larger the disturbance amplification.
  
Although the above discussion focuses on work conducted over approximately the past decade, it must be pointed out that the potential importance of nonmodal amplification in channel flows of viscoelastic fluids was recognized much earlier \cite{Sureshkumar1999,Keunings2002,Kupferman2005,
Schumacher2006,Renardy2009}.  
However, that prior work mainly focused on two-dimensional (i.e., spanwise-constant) flows and did not consider amplification of external disturbances.


\subsection{Open issues}

The results highlighted above indicate that channel flows of viscoelastic fluids are exceedingly sensitive to external disturbances.  
Amplification of these disturbances could create the initial conditions needed for transient growth, or can itself put the flow into a regime where nonlinear effects are no longer negligible.  In this way, amplification of initially small-amplitude disturbances by a purely linear but nonmodal mechanism may eventually trigger a transition to a more complicated flow state such as elastic turbulence.

Despite this progress in our fundamental understanding of nonmodal amplification, there remain important open issues, some of which are 
identified and briefly discussed here.\\
\begin{figure}
\begin{center}
\includegraphics[width=8cm]{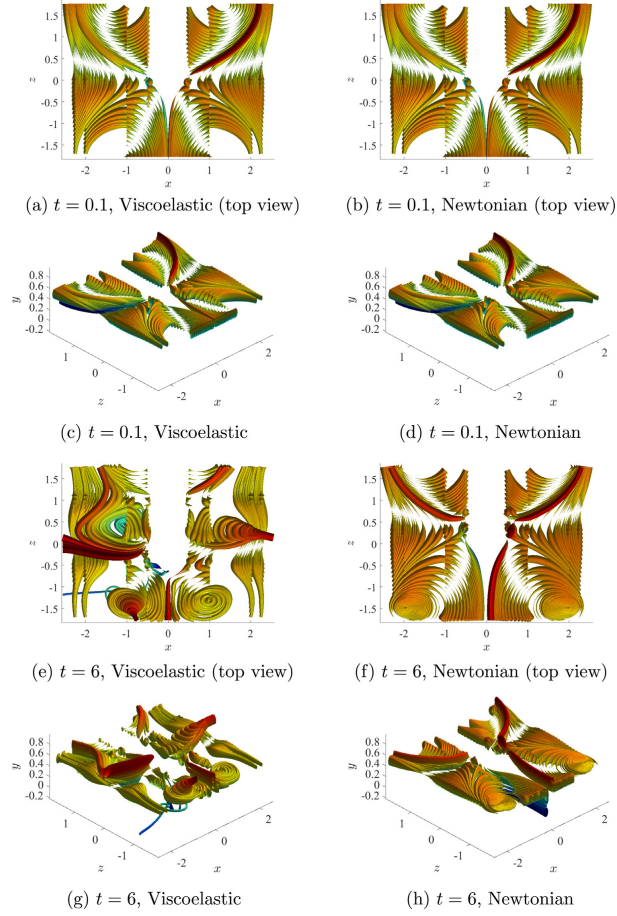}
\end{center}
\caption{\small Three-dimensional streamtubes of the velocity fluctuation vector arising from an impulsive disturbance in plane Poiseuille flow at $Re=50$. For the FENE-CR fluid, $Wi=50$, $\beta=0.5$, and $L=100$ (where $L$ is the maximum polymer extensibility). Reproduced with permission from Fig.~9 of \cite{Hariharan2018}.}
\label{fig4-SK}
\end{figure}

\noindent
{\it Localized disturbances}  

The discussion in \S\ref{Sec5.3} focused on disturbances in the form of body forces distributed throughout the flow domain.  However, this is unlikely to be the case in most experiments, where localized disturbances are easier to realize.  A simple way to introduce a localized disturbance in experiments is via obstacles, which exert a drag force on the fluid 
(e.g., \cite{Pan_2013_PRL,qin2017elastic,
nolan2016viscoelastic}).  Although these obstacles are of finite size, consideration in nonmodal analysis of disturbances that are localized at a point reveals rather rich behavior in the linearized dynamics.

An example of the dramatic influence viscoelasticity can have on the linearized dynamics of localized disturbances is shown in Fig. \ref{fig4-SK} \cite{Hariharan2018} (Other examples of the evolution of localized disturbances are discussed in \cite{agarwal2014linear,pagezakJFM2015}.)  Here, the disturbance occurs in plane Poiseuille flow and takes the form of an impulse in space and time.  The FENE-CR constitutive equation, which accounts for the finite extensibility of polymer molecules, is used for these calculations.  

It is seen from Fig. \ref{fig4-SK} that the presence of vortical structures is more pronounced in viscoelastic fluids than in Newtonian fluids.  The curved streamlines associated with these vortical structure could be susceptible to additional instabilities \cite{larson1992,Shaqfeh1996} if the amplitude of the velocity disturbance becomes sufficiently large.  Determining whether such a transition occurs will require nonlinear calculations, another open issue we discuss below.  Nonlinear calculations involving finite-sized obstacles similar to those used in experiments (e.g., \cite{Pan_2013_PRL,qin2017elastic,
nolan2016viscoelastic}) would also help bridge the gap between theory and experiment, as would additional experiments in which the disturbances are more localized (e.g., by using a small actuator).\\ 

As another example of the richness of the linearized dynamics, recent calculations using localized time-periodic disturbances in plane Poiseuille flow show that polymer-stress fluctuations can be amplified by an order of magnitude while there is only negligible amplification of velocity fluctuations \cite{gokul-jnnfm2}.  This appears consistent with experimental observations of elastic turbulence in microchannel flows of viscoelastic fluids, where the magnitude of velocity fluctuations decreases downstream before
increasing \cite{Pan_2013_PRL,qin2017elastic}.  Notably, the large stress amplification is highly localized in space, occurs for spanwise-constant disturbances, and was overlooked in prior studies that used square-integrated measures of disturbance amplification,
which are typically applied in nonmodal analysis \cite{gokul-jnnfm2}.  The large stress amplification can put the flow into a regime where
nonlinear terms are no longer negligible, and this could trigger a
transition to elastic turbulence.\\

\noindent
{\it Finite extensibility} 

As noted above, the calculations shown in Fig. \ref{fig4-SK} use the FENE-CR constitutive equation.  Accounting for the finite extensibility of polymer molecules is important for strengthening connections between theory and experiment.    Some progress has already been made on this front. In Ref.~\cite{liejovkumJFM13} the influence of harmonic body forces on creeping plane Couette flow of FENE-CR fluids was examined.  It was found that the velocity and polymer stress fluctuations are proportional to $\hat{L}^2$ and $\hat{L}^4$, respectively, as $Wi \rightarrow \infty$, where $\hat{L}$ is the maximum polymer extensibility.  In contrast, as $\hat{L} \rightarrow \infty$, the velocity and polymer stress fluctuations are bounded by $Wi^2$ and $Wi^4$, respectively (see also \S\ref{Sec5.3}).  Clearly, finite extensibility places bounds on the achievable level of nonmodal amplification.  Nonlinear calculations will thus be critical to ascertaining how important nonmodal amplification is in triggering flow transitions to complex states such as elastic turbulence.\\  

\noindent
{\it Nonlinear calculations} 

 Despite all the progress made in understanding the fundamentals of nonmodal amplification in channel flows of viscoelastic fluids, it is still unclear what role nonmodal amplification plays in transition to elastic turbulence.  For such a transition to take place, disturbances need to be of a large enough amplitude.  Whether this occurs through nonmodal amplification of an initially small-amplitude disturbance or through an externally imposed disturbance of finite amplitude \cite{Morozov2019} remains an open question.  Nonlinear calculations (both weakly nonlinear analysis and direct numerical simulations) are needed to resolve this issue.  
 
Two-dimensional numerical simulations conducted by Atalik and Keunings approximately two decades ago using a fully spectral method suggest that nonmodal amplification can lead to nonlinear flow states in plane Poiseuille flow of Oldroyd-B fluids \cite{Keunings2002}.  These flow states exhibit finite-amplitude periodic-like oscillations, and can occur at very low Reynolds numbers ($\sim$0.1) and sufficiently large Weissenberg numbers.  They are observed only when the ratio of the solvent viscosity to the total viscosity is sufficiently small, and are not observed in plane Couette flow.  Elastic turbulence was not observed, perhaps because the simulations were restricted to 2D.  The discussion in \S\ref{Sec5.2} and \S\ref{Sec5.3} highlights the importance of 3D---and in particular, spanwise-constant---perturbations.  
 
Experiments showing elastic turbulence in channel flows typically perturb the flow with obstacles such as cylinders located near the channel inlet  (e.g., \cite{Pan_2013_PRL,qin2017elastic,
nolan2016viscoelastic}).  Definitively unraveling what is actually happening in the experiments will likely involve careful interplay with detailed nonlinear calculations, particularly in the regime where the elasticity numbers ($E=Wi/Re$; see \S\ref{Sec5.3}) are large and steep polymer stress gradients may arise.  Such calculations remain an outstanding challenge that contains rich mathematical and computational issues (e.g., see \cite{Hameduddin2019,Biancofiore2017,gokul-jcp}).  The insights gained from nonmodal analysis are likely to be helpful for interpreting results from those calculations.

\section{Nonlinear stability of parallel shear flows}
\label{sec:nonlinear}

The analysis of viscoelastic flow instabilities presented in the previous Sections was largely restricted to the linear theory, either modal or non-modal.  Here, we demonstrate that the Oldroyd-B model can also be used to successfully describe strongly non-linear states that emerge beyond a linear instability or even in its absence. As an example, we treat the case of parallel shear flows of dilute polymer solutions, like pressure-driven flows in a channel or pipe. 

As discussed in Section~\ref{subsec:zeroRe}, such flows are mostly linearly stable. With the exception of the recently discovered linear instabilities in pressure-driven channel flow \citep{khalid_creepingflow_2021} at ultra low polymer concentrations, $1-\beta\ll 1$, and very high levels of elasticity, $\Wi >O(1000)$, there are no known linear instabilities in the majority of the parameter space. As a result, such flows have long been believed to exhibit no states that are not laminar. This assumption was challenged by Bonn, Morozov, van Saarloos, and collaborators \citep{Meulenbroek2003,meulenbroek2004weakly,bertola2003experimental}
  who proposed that, despite being linearly stable, parallel shear flows of viscoelastic fluids can exhibit non-linear instabilities: while infinitesimal perturbations to such flows always decay, a finite-amplitude perturbation may be sufficient to drive the flow unstable. Below, we review theoretical and experimental efforts to discover such instabilities, and discuss their implications for our understanding of elastic turbulence. While there have been very recent efforts \citep{wan2021subcritical,buza_page_kerswell_2021} that have carried out weakly nonlinear analyses for the elastoinertial center-mode instability  at finite $Re$ (see Sec.~\ref{subsubsec:scenarios}), the discussion in the present section will confine itself to the earlier body of work that has focused on inertialess flows.

\subsection{Weakly non-linear analysis}
\label{morozov_subsection_A}

To verify the suggestion made by Bonn, Morozov, van Saarloos, and collaborators in \cite{Meulenbroek2003,bertola2003experimental}, one has to demonstrate that viscoelastic parallel shear flows support other states rather than the laminar one. As there are no general methods of finding non-linear solutions of partial differential equations, one has to rely on approximate, often ad hoc, techniques. In the presence of a linear instability, a novel state bifurcating from the laminar profile is guaranteed to become arbitrarily weak as the distance to the linear instability decreases. This allows one to construct a non-linear solution perturbatively, using the distance to the linear instability threshold as a control parameter. Such techniques, that are often referred to as \textit{amplitude equations}, are the dynamical systems analogues of the Landau theory of second-order phase transitions, and have been successfully employed to study various pattern-forming systems \cite{cross1993,vanHecke1994}. In the absence of a linear instability, Morozov and van Saarloos formulated a novel version of the amplitude equation technique \cite{morozov_saarloos2005,Morozov2005,morozov_saarloos2007,Becherer2009}
that relies on the \textit{weakness} of the non-linear state rather than the distance to the instability threshold (which is formally infinite in this case). Below, we follow the latest version of the theory as formulated in \cite{Morozov2019}.

We consider a flow between two parallel infinite plates forced either by an applied constant pressure gradient along the plates (pressure-driven channel flow) or by the relative sliding of the plates (plane Couette flow). We select a Cartesian coordinate system with $(x,y,z)$ being along the streamwise, velocity gradient, and spanwise directions, respectively. As in the previous Sections, our starting point is the dimensionless version of the Oldroyd-B model. In the absence of inertia, it is given by
\begin{align}
&{\bm\tau} + Wi \left[ \frac{\partial {\bm\tau}}{\partial t} + {\bm v}\cdot\nabla{\bm\tau} - \left(\nabla {\bm v}\right)^T\cdot{\bm\tau} - {\bm\tau}\cdot\left(\nabla {\bm v}\right)\right] = \left(\nabla {\bm v}\right) + \left(\nabla {\bm v}\right)^T , \label{morozov_OB} \\
&\qquad\qquad -\nabla p + \beta \nabla^2{\bm v} + (1-\beta)\nabla\cdot{\bm \tau}  = 0, \label{morozov_Stokes} \\
&\qquad\qquad  \qquad\qquad  \nabla\cdot {\bm v} = 0, \label{morozov_incomp}
\end{align}
where $p$ is the pressure, $\bm \tau$ is the polymeric contribution to the total stress, and $\bm v$ is the velocity of the fluid assumed to satisfy the appropriate no-slip boundary conditions. These equations are rendered dimensionless by using the maximum value of the laminar fluid velocity $\mathcal{U}$ as a unit of velocity, half the distance between the plates $d$
as a unit of length, and their ratio $d/\mathcal{U}$ as a unit of time. This yields $\Wi=\lambda \mathcal{U} / d$ and $\beta = \eta_s/(\eta_s + \eta_p)$, where $\eta_s$ and $\eta_p$ are the solvent and polymer contributions to the total viscosity of the solution; $\lambda$ is the Maxwell relaxation time of the model. Equations \eqref{morozov_OB}-\eqref{morozov_incomp} can be written concisely as
\begin{equation}
\label{morozov_SymbolicEq}
\hat{\mathcal L}V + \hat{A}\frac{\partial V}{\partial t} = N\left(V,V\right),
\end{equation}
where the perturbation vector $V=\left(\bm{v}',{\bm \tau}',p'\right)^T$, comprises deviations of the velocity, stress, and pressure from their laminar values. Here, $\hat{\mathcal L}$ and $N$ are the linear operator and the quadratic non-linear operator, respectively, while the constant diagonal matrix $\hat{A}$ encodes the fact that only some equations in \eqref{morozov_OB}-\eqref{morozov_incomp} contain time-derivatives. Their explicit expressions can be found in \cite{Morozov2019}.

In the most general terms, a solution of Eq.\eqref{morozov_SymbolicEq} can be written as a Fourier series in the streamwise and spanwise directions, i.e.
\begin{align}
\label{morozov_arbibrary_solution}
V(x,y,z,t) = \sum_{n, m=-\infty}^{\infty} U_{n,m}(y,t) e^{i n k_x x} e^{i m k_z z },
\end{align}
where the wavenumbers $k_x$ and $k_z$ set the dominant length-scales in the corresponding directions. The weakly non-linear analysis of Morozov and van Saarloos \cite{Morozov2019} approximates this expression with
\begin{align}
\label{morozov_solution_form}
& V(x,y,z,t) = \Phi(t)e^{i (k_x x + k_z z)}V_{lin}(y) + \Phi^{*}(t)e^{-i (k_x x + k_z z)}V_{lin}^{*}(y) \nonumber \\
& \, + U_0(y,t) + \sum_{n=2}^{\infty}\left[ U_n(y,t)e^{i n (k_x x + k_z z)} + U_n^{*}(y,t)e^{-i n (k_x x + k_z z)}\right],
\end{align}
where ''$*$'' denotes complex conjugation. This expansion is built around the eigenfunction $e^{i\left(k_x x + k_z z\right)}V_{lin}(y)$ of the linear operator,
\begin{equation}
\label{linear_eigenmode}
\hat{\mathcal L} e^{i\left(k_x x + k_z z \right)}V_{lin}(y)=-\chi \hat{A}e^{i\left(k_x x + k_z z \right)}  V_{lin}(y),
\end{equation}
where $\chi$ is an eigenvalue. In what follows, we use the least stable eigenvalues as the basis for our analysis. In plane Couette flow, these are given by the extension to the Oldroyd-B model of the Gorodtsov-Leonov modes \cite{Gorodtsov1967} discussed in Section~\ref{subsec:zeroRe}, while in pressure-driven channel flows, we use the least stable of the three leading eigenvalues. (e.g. see  Fig.~6 of \citep{Wilson1999}).

The higher Fourier harmonics $U_n$ are assumed to be forced by the dynamics of the time-dependent amplitude of the first Fourier mode  $\Phi(t)$ and are thus produced by the recursive non-linear interaction of the first Fourier mode with itself and other modes produced in the process. This yields
\begin{align}
&U_0(y,t)=|\Phi(t)|^2 u_0^{(2)}(y) + |\Phi(t)|^4 u_0^{(4)}(y) + \cdots , \nonumber \\
&U_2(y,t)= \Phi^2(t) u_2^{(2)}(y) + \Phi^2(t)|\Phi(t)|^2 u_2^{(4)}(y) + \cdots ,\label{Un} \\
&U_3(y,t)= \Phi^3(t) u_3^{(3)}(y) + \cdots , \nonumber \\
&\cdots, \nonumber
\end{align} 
where the unknown functions $u_n^{(m)}(y)$ are to be determined from the analysis. One can view Eq.\eqref{morozov_solution_form} as a version of the Fourier expansion, Eq.\eqref{morozov_arbibrary_solution}, with an extra assumption about the form and interrelation between the coefficients. Unlike the usual amplitude equation technique \cite{cross1993,vanHecke1994}, there is no guarantee that this procedure yields a converging, meaningful solution; this has to be checked \textit{a posteriori}. 

The goal of the theory is thus to determine the time-evolution of the amplitude $\Phi(t)$, which is obtained by projecting the dynamics of Eq.\eqref{morozov_SymbolicEq} on to the slow manifold with the help of the adjoint operator; see \cite{Morozov2019} for details. Its main result is the derivation of the following equation for the amplitude:
\begin{align}
\label{morozov_amplitude_equation}
&\frac{d\Phi}{d t} = \chi \Phi + C_3 \Phi |\Phi|^2 + C_5 \Phi |\Phi|^4 + C_7 \Phi |\Phi|^6 + C_9 \Phi |\Phi|^8 + C_{11} \Phi |\Phi|^{10} \cdots,
\end{align}
where the complex coefficients $C$'s are functions of $k_x$, $k_z$, $Wi$, $\beta$, and the particular eigenmode selected for the analysis. For sufficiently small values of $\Phi(t)$, the amplitude equation reduces to the long-time decay predicted by the linear stability analysis, i.e. $\Phi(t)\sim e^{\chi t}$, while for larger values of $\Phi(t)$ it can exhibit non-trivial behaviour. Although the type of solutions it can support varies from steady states and periodic orbits to chaotic dynamics, below we focus on travelling waves in the form $\Phi(t) = \Psi\,e^{i\,\Omega\,t}$, where $\Psi$ and $\Omega$ are real numbers. 

The asymptotic nature of Eq.\eqref{morozov_amplitude_equation} implies that only converging series can represent a physical solution. To lie within the radius of convergence, defined by the coefficients $C$'s, the solution amplitude $\Psi$  has to be sufficiently small. In turn, this implies that the solutions that can be found by this method have to be sufficiently close to the original eigenmode used as the starting point of the theory. Convergence of the series for $\Psi$ can be assessed by studying the travelling wave solutions of Eq.\eqref{morozov_amplitude_equation} with a progressively increasing number of terms, i.e.
\begin{align}
&0=Re \left (\chi\right) + Re\left(C_3\right) \Psi_1^2,\\
&0=Re \left (\chi\right) + Re\left(C_3\right) \Psi_2^2 + Re\left(C_5\right) \Psi_2^4,\\
&\cdots
\end{align}

\begin{figure}[htp]
\centering
\includegraphics[width=0.48\textwidth]{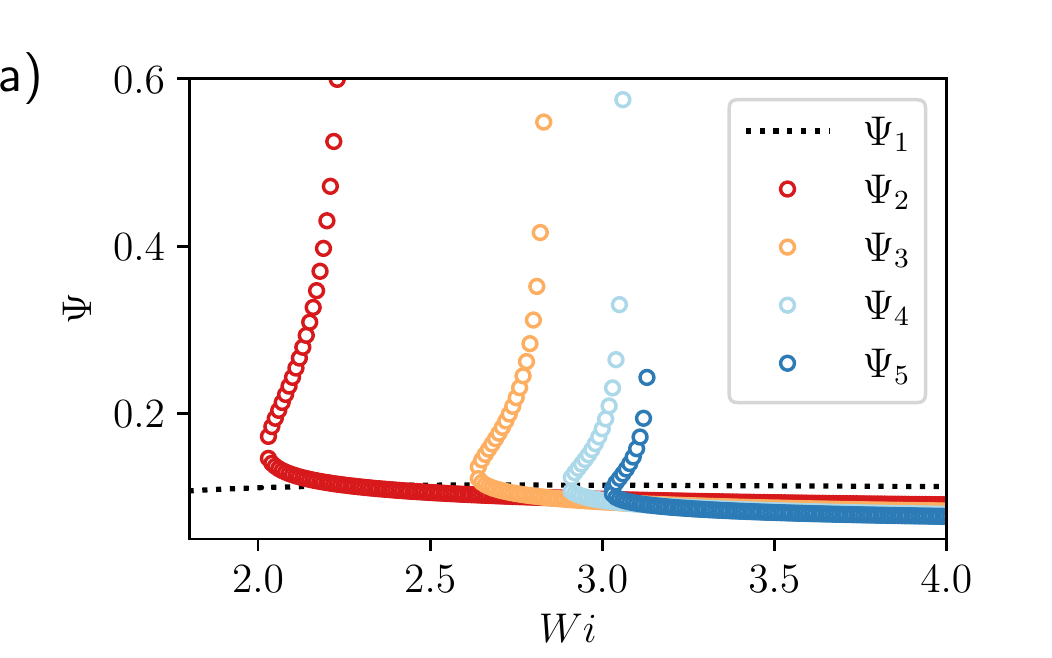}
\includegraphics[width=0.48\textwidth]{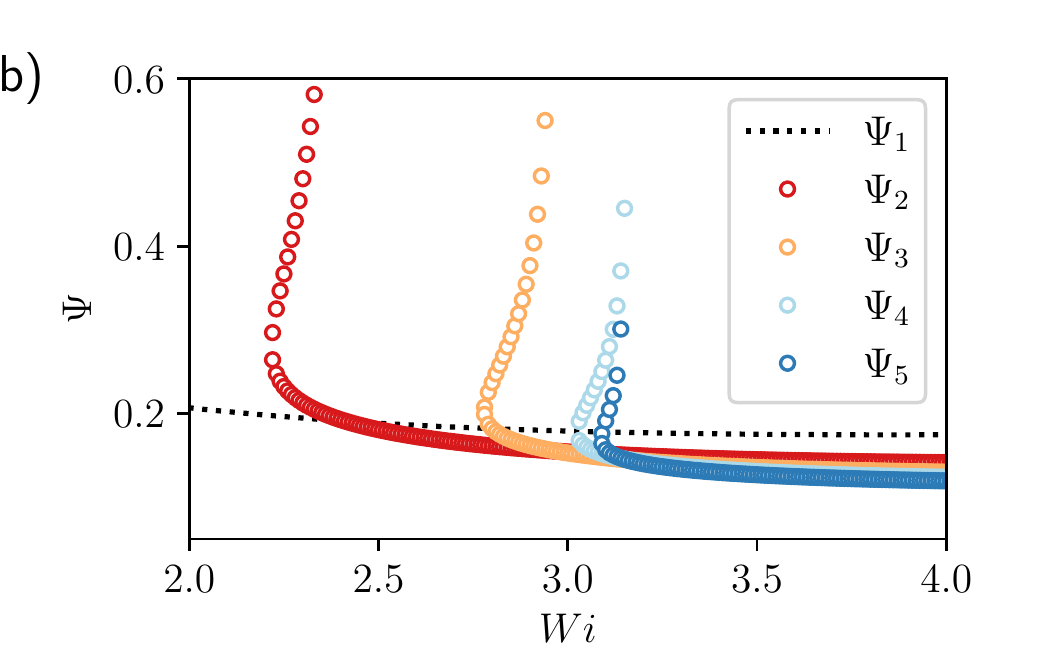}
\caption{Steady-state amplitude $\Psi$ of the travelling-wave solution as a function of $\Wi$. a) Plane Couette flow of a UCM fluid ($\beta=0$) with $k_x=k_z=1$. Calculations were performed in the presence of small amounts of inertia with the Reynolds number $Re=10^{-3}\Wi$.  b) Plane channel flow of an Oldroyd-B model with $\beta=0.05$, $k_x=1$ and $k_z=2$, in the absence of inertia, $Re=0$. Replotted under Creative Commons CC licence from \cite{Morozov2019}.
}
\label{Fig:morozov_bifurcation}
\end{figure}

\begin{figure}[htp]
\centering
\includegraphics[width=0.5\textwidth]{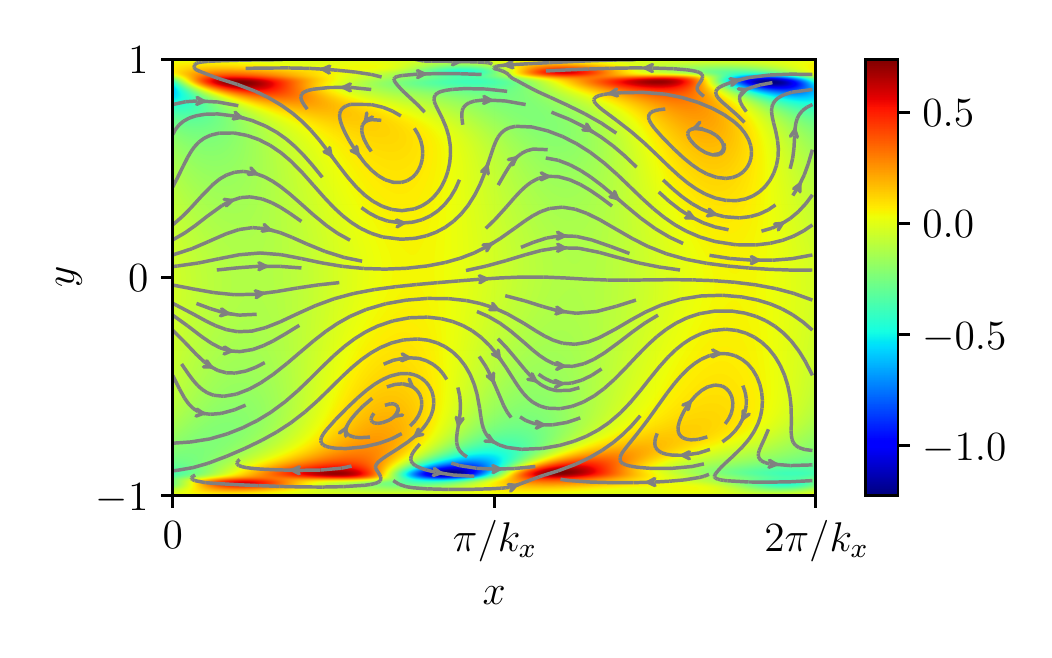}
\caption{The velocity profile at $z=0$ in the $xy$-plane for plane channel flow of an Oldroyd-B model with $\beta=0.05$, $k_x=1$ and $k_z=2$, as in Fig.\ref{Fig:morozov_bifurcation}; arrows denote the in-plane components of the deviation of the velocity from its laminar profile, while the colour gives the spanwise velocity. Reproduced under Creative Commons CC licence from \cite{Morozov2019}.}
\label{Fig:morozov_SpatialState}
\end{figure}

In Fig.\ref{Fig:morozov_bifurcation} we plot the consecutive approximation to the travelling wave amplitude $\Psi$ using this procedure for plane Couette and channel flows. Although we had no reason to expect this \textit{a priori}, the low-branch amplitude values and the position of the saddle-node (the point where the lower branch turns vertically upwards)
appear to converge; see \cite{Morozov2019} for the details of convergence tests. As can be seen from Fig.\ref{Fig:morozov_bifurcation}, the upper branches, which set the saturated amplitude of the travelling wave solutions, diverge rapidly close to the saddle-node of the bifurcation, indicating that the corresponding values of $\Psi$ lie outside the radius of convergence of the asymptotic series in Eq.\eqref{morozov_amplitude_equation}. Nevertheless, the highest-order upper-branch amplitude values of the amplitude $\Psi$ in Fig.\ref{Fig:morozov_bifurcation} allow us to study the spatial structure of the solution predicted by the theory. As an illustration, in Fig.\ref{Fig:morozov_SpatialState} we plot the velocity profile at $z=0$ in the $xy$-plane, where arrows trace the in-plane components of the deviation of the velocity from its laminar profile, while the colour gives the spanwise velocity.
Additional plots of the stress and velocity fields can be found in \cite{Morozov2019}.

The main conclusion to be drawn from these results is that both plane Couette and channel flows of model Oldroyd-B fluids exhibit travelling-wave solutions above the saddle-node value of the Weissenberg number $\Wi_{sn}\approx 3$. We now summarise how these results compare against recent studies of perturbed viscoelastic channel flows.

\subsection{Experimental evidence}

First indication that parallel shear flows of viscoelastic fluids can exhibit sub-critical instabilities can already be seen in the \textit{early turbulence} experiments \citep{Samanta2013}  already mentioned in Section~\ref{subsec:finiteRe}. There, it was observed that addition of polymers to Newtonian pipe flows before the onset of Newtonian turbulence often leads to drag-enhancement, betraying a transition that is different in nature. These novel instabilities can be observed as long as the pipe diameter is small enough (typically a few millimetres), yielding high values of $\Wi$, in which case they exist at Reynolds numbers significantly smaller than the onset of Newtonian turbulence. It is natural to suggest that these instabilities have a purely elastic origin, as was proposed by Samanta \textit{et al.} \cite{Samanta2013},
with their region of existence spreading from $Re=0$ to the values observed in early turbulence experiments \citep{Choueiri2018,Bidhan2018,chandra_shankar_das_2020,choueiri2021experimental}. 

A more convincing, yet still indirect, evidence for the existence of sub-critical instabilities in parallel shear flows comes from the phenomenon of melt fracture (mentioned earlier in Sec.~), observed when concentrated polymer solutions or melts are extruded from a thin capillary \cite{bird1987dynamics}. Above a critical extrusion speed, the extrudate develops unwanted long-wave undulations of its surface and even breaks entirely. The origin of melt fracture, which limits virtually every industrial process that involves extrusion, is hotly debated, with possible explanations ranging from bulk phenomena to the stick-slip behaviour at the capillary exit \cite{Denn2001}. By studying a wide range of viscoelastic fluids, Bertola \textit{et al.} \citep{bertola2003experimental} have shown that 
extrudate undulations appear through a sub-critical instability with the saddle-node Weissenberg number being around $Wi=5$. Supported by an early version of the weakly non-linear analysis developed by Meulenbroek \textit{et al.} \cite{Meulenbroek2003} that predicted the same saddle-node value of the Weissenberg number, Bertola \textit{et al.} \cite{bertola2003experimental} concluded that their observations are consistent with a sub-critical instability originating inside the capillary and being advected downstream by the flow.

To demonstrate the bulk origin of such instabilities, Bonn \textit{et al.} \cite{Bonn2011} conducted a series of experiments with dilute and semi-dilute polymer solutions that were fed from a thin capillary into a capillary of a larger radius. The resulting sudden-expansion flow at the entry to the large capillary provided a high level of flow fluctuations capable of inducing the instability.  For low $\Wi$, Bonn \textit{et al.} \cite{Bonn2011} observed that the inlet perturbations decayed along the large capillary, while for $\Wi\ge 4$ perturbations persisted far downstream. While being consistent with the scenario proposed here, that study has not demonstrated the sub-critical nature of the instability.

The issue was finally settled by Arratia and co-workers \citep{Pan_2013_PRL,qin2017elastic,qin2019elastic}
who simultaneously demonstrated the existence of large three-dimensional flow fluctuations inside a microfludic channel flow of dilute polymer solutions and the sub-critical nature of the transition. Their experimental setup consisted of a long microfluidic channel partially blocked at the entrance by a row of cylindrical obstacles along the flow direction. Flows of polymer solutions around cylinders have been extensively studied, both experimentally and numerically, and are known to exhibit a linear instability above a critical $\Wi$; we refer to \citep{Pan_2013_PRL,qin2017elastic,qin2019elastic} for relevant references. This instability was used by Arratia \textit{et al.} to generate flow perturbations and they observed their development far downstream of the cylinders. Above $\Wi\approx 5-6$, they reported that the inlet perturbations stayed at a constant level far downstream from the cylinders, indicating a sustained non-linear state. By reducing the flow rate from that state, Arratia and co-workers observed that the fluctuations disappeared at lower values of $\Wi$ than at their onset, thus confirming their sub-critical nature. The instability around the cylinders has been shown to be a supercritical bifurcation and, thus, is not responsible for the phenomenon observed. Later work by Qin \textit{et al} \cite{qin2017elastic,qin2019elastic} has demonstrated the three-dimensional nature of the non-linear flow state thus created. An interesting feature of the experiments by Bonn \textit{et al.} \cite{Bonn2011} and by Arratia and co-workers \cite{Pan_2013_PRL,qin2017elastic,qin2019elastic} is that a large inlet perturbation is required to drive the transition.
The saddle-node value of the Weissenberg number reported by Arratia \textit{et al.} is somewhat higher than the one predicted in Fig.\ref{Fig:morozov_bifurcation}, which can be due to a different value of $\beta$ for the solutions used in the experiments, their shear-thinning nature not accounted for in the Oldroyd-B model, and the approximate nature of the weakly non-linear analysis presented above. Nevertheless, these experiments convincingly demonstrate the existence of non-trivial flow states in channel flows of viscoelastic fluids, in line with the original suggestion of Bonn, Morozov, van Saarloos, and collaborators 
\citep{Meulenbroek2003,bertola2003experimental,meulenbroek2004weakly,morozov_saarloos2007}
; see the summary in Table \ref{Table:morozov_summary}.

\begin{table*}
\caption{Saddle-node value of $\Wi$ for various flow geometries. Here $\beta = 0$ for plane Couette and pipe Poiseuille flows, while $\beta = 0.05$ for plane Poiseuille flow. In the experiments of Pan \textit{et al.} \citep{Pan_2013_PRL}, $\beta \approx 0.5$. \\\hspace{\textwidth} 
\tiny{$^{\rm{a}}$Bonn \textit{et al.} \cite{Bonn2011},
$^{\rm{b}}$Bertola \textit{et al.} \cite{bertola2003experimental},
$^{\rm{c}}$Pan \textit{et al.} \cite{Pan_2013_PRL},
$^{\rm{d}}$Morozov and van Saarloos \cite{morozov_saarloos2005},
$^{\rm{e}}$Meulenbroek \textit{et al.} \cite{Meulenbroek2003},
$^{\rm{f}}$Morozov and van Saarloos  \cite{Morozov2019}}}
\centering
\begin{tabular}{c c c c c c}
\hline
 & Plane Couette Flow & & Pipe Flow & & Channel Flow \\ [0.5ex]
\hline
Experiment & $-$ && $4^{\rm{a}}-5^{\rm{b}}$ && $5-6^{\rm{c}}$\\
Weakly non-linear analysis & $3^{\rm{d}}$ && $5^{\rm{e}}$ &&  $3^{\rm{f}}$\\ [1ex]
\hline
\end{tabular}
\label{Table:morozov_summary}
\end{table*}


It is important to stress that the secondary flow structures predicted above might be difficult to detect experimentally. In the purely elastic regime, or at low $Re$, the dynamical variable is the polymeric stress, as can be seen from Eqs. \eqref{morozov_OB}-\eqref{morozov_incomp}, while the velocity field adiabatically adjusts to its evolution. As the weakly non-linear analysis presented above suggests, even weak velocity fields can develop sufficient gradients to re-inforce non-linear dynamics of the stress. In the absence of reliable techniques to measure three-dimensional profiles of polymeric stresses, this might lead to a situation where a strongly non-linear state is only weakly manifested through the available observables, i.e. the mean-velocity fluctuations as used in \cite{Bonn2011,Pan_2013_PRL}. The weakly non-linear analysis further suggests that the streamwise vortices and streaks in a plane perpendicular to the flow direction might be the best candidates for experimental detection \cite{Morozov2019}. Such structures are a hallmark of Newtonian coherent structures \cite{Graham2021}; they also feature prominently in the non-normal growth analysis by Kumar and Jovanovi\'{c} \cite{jovanovic_kumar_2010,jovanovic_kumar_2011}; see also Section~\ref{sec:nonmodal}. Recent experiments by Qin \textit{et al.} \cite{qin2017elastic} and Jha and Steinberg \cite{jha2020universal} present preliminary evidence for the existence of such coherent structures.

In view of the potential difficulties in resolving three-dimensional velocity fields associated with purely elastic instabilities in straight channels, it would be natural to address this question in direct numerical simulations. Unfortunately, such calculations are made notoriously difficult by the so-called High-Weissenberg Number Problem \cite{OwensPhillips}, that renders simulations unphysical at sufficiently high values of $\Wi$. In the absence of shear-thinning, the Oldroyd-B model often suffers from the High-Weissenberg Number Problem even at very low $\Wi\sim 1-2$ \cite{OwensPhillips}. Atalik and Keunings \cite{Keunings2002} performed numerical simulations of two-dimensional parallel shear flows of various constitutive models and reported large fluctuations of vorticity at low $Re$ and $\Wi\sim 0.5$. Unfortunately, that study did not provide the spatial profiles of the associated velocity field. Also the low value of $\Wi$ needed to generate oscillations and the low numerical resolutions used in those simulations bring in question whether they are indeed related to the instabilities discussed here. Sadanandan and Sureshkumar performed similar simulations of an Oldroyd-B fluid in channel flows and observed large velocity fluctuations above $\Wi\sim3$ (private communication). Unfortunately, those simulations ultimately suffered from the High-Weissenberg Number Problem and did not yield a turbulent-like steady-state. In the past years, there emerged a class of numerical techniques to ensure positive-definiteness of the conformation tensor (absence thereof was implicated as a cause of the High-Weissenberg Number Problem) \cite{Fattal2005,Pimenta2017,Alves2021}. Their use in unsteady parallel shear flows should provide the ultimate argument for the existence of sub-critical instabilities in such flows.
 
\subsection{Elastic turbulence in parallel shear flows}

The sub-critical transition in viscoelastic fluids presented above echos the instability scenario in parallel shear flows of Newtonian fluids \cite{Eckhardt2008,Barkley2016}. Recently, the transition to Newtonian turbulence has been understood to be organised by the exact solutions of the Navier-Stokes equations. These three-dimensional coherent structures in the form of travelling waves or periodic orbits comprise streamwise vortices, streaks, and three-dimensional flows connecting them dynamically \cite{Nagata1990,Hamilton1995,waleffe_1997,Hof2004,waleffe_2001,stone_graham2002,wedin_kerswell_2004}, and appear through a sub-critical bifurcation from infinity. Importantly, they are linearly unstable forming saddle-like structures in phase space: A turbulent trajectory passing by in the vicinity of such a structure is attracted towards it only to be eventually repelled along one of the few unstable directions \cite{Eckhardt2008,Barkley2016}. In the presence of a sufficient number of such solutions in the phase space, after overcoming a critical threshold, a turbulent trajectory performs a random walk in the phase space, being trapped among a large number of such structures. This scenario, coupled with the process of spatial splitting and merging of the localised coherent structures, was recently shown to place the transition to Newtonian turbulence within the directed percolation universality class \cite{Barkley2020}.

The experimental results by Bonn \textit{et al.} \cite{Bonn2011} and by Arratia and co-workers \cite{Pan_2013_PRL,qin2017elastic,qin2019elastic} demonstrate that while parallel shear flows of dilute polymer solutions are linearly stable, they exhibit sub-critical transitions that lead directly to a chaotic state related to purely elastic turbulence previously only observed in shear flows with curved streamlines \cite{Groisman2000,groisman2004elastic,Burghelea2007}. The results of the weakly non-linear analysis suggest that that transition might also be guided by unstable coherent structures: While the non-linear states described in Section~\ref{morozov_subsection_A} might be linearly unstable, and thus inherently  unobservable, they could organise the phase space dynamics in a way similar to their Newtonian counterparts. Their spatial structure, however, might be fundamentally different from the Newtonian coherent structures. Indeed, the weakly non-linear analysis predicts that viscoelastic solutions could equally exist as two- and three-dimensional structures, while the recent work on elasto-inertial turbulence
\citep{Samanta2013,Dubief2013,Sid_2018_PRF,Shekar2019,page2020exact,dubief2020coherent}
  points towards the importance of (quasi) two-dimensional coherent structures. Future work is needed to uncover the exact nature of purely elastic turbulence in parallel shear flows and its similarities and differences to the Newtonian transition scenario.

We conclude by noting that the recent discovery of a linear instability in purely elastic channel flows by Khalid et al \citep{khalid_creepingflow_2021} also provides a useful suggestion regarding the spatial structure of flow fields associated with purely elastic coherent structures. Although that instability exists only for $1-\beta < O(10^{-2})$, the corresponding nonlinear state has been shown to be sub-critical \citep{buza_page_kerswell_2021}, similar to its elastoinertial counterpart \citep{page2020exact}. It is possible that, while no longer being associated with a linear instability, that solution persists to the experimentally relevant values of $\beta$ as a bifurcation from infinity.
If that is indeed the case, it could potentially connect to the structures suggested in Refs.\citep{morozov_saarloos2005,morozov_saarloos2007,morozov_saarloos2019} \, as they are related to the same eigenmode of the linear operator.

\section{Conclusions and Outlook}
\label{sec:Concl}

The present review, written on the occasion of the birth centenary of James Oldroyd, has provided a summary of various instabilities encountered in viscoelastic shearing flows. 
Despite not accounting for a shear-dependent viscosity and first normal stress coefficient, the Oldroyd-B model has been quite versatile in its ability to predict, at least qualitatively, a host of instabilities in rectilinear and curvilinear viscometric flow configurations. Likewise, despite predicting a divergent stress response in steady extensional flows beyond a threshold $Wi$, the Oldroyd-B does provide a qualitative explanation even for instabilities in non-viscometric flows, provided the base-state stresses remain finite; artifacts might arise, however, in parameter intervals where the base-state stresses diverge \cite{Lagnado1985}. The model has also proved useful in interpreting secondary interfacial instabilities in multilayered shearing flows that might arise as the saturated state of a primary shear-banding instability.

While the Oldroyd-B model was originally intended for dilute polymer solutions, the pioneering experimental efforts that led to the discovery of EIT in pipe flow  (Samanta \textit{et al.} \citep{Samanta2013}), and subsequent efforts that further elucidated this transition \citep{Bidhan2018}, use a range of polymer solutions with polymer concentrations extending to the overlap value and beyond, and therefore, may not be deemed truly dilute. Thus, although the Oldroyd-B model has been used to predict a linear elastoinertial instability in pipe flow \citep{Piyush_2018,chaudharyetal_2021} that coincides with the observed onset of EIT, an accurate prediction of the elasto-inertial instability beyond the dilute regime will require a more detailed constitutive model that accounts for inter-chain interactions. Prabhakar and co-workers have developed a constitutive equation that accounts for such interactions, although this model is yet to be used within the framework of a stability analysis \citep{Prabhakar2006,prabhakar2016}. Further, it is well known that polymer solutions that are dilute, based on the equilibrium coil dimension\,(which determines the overlap concentration), may no longer be so at high shear rates, since the volume fraction that controls hydrodynamic interactions is determined by the longest chain dimension. While this chain-extension-induced departure from the dilute regime may not be relevant to a linear stability analysis, it might be important in analyses that account for the nonlinear feedback from finitely extended polymer molecules; turbulent drag reduction being one example.

Curvilinear viscometric shearing flows undoubtedly remain the bedrock on which the success of the Oldroyd-B model has been built, with the model providing a first-cut prediction of purely elastic instabilities in all of the viscometric flow configurations including the Taylor-Couette, parallel-plate and cone-and-plate geometries. The inclusion of multiple relaxation modes and shear thinning does lead to a more accurate prediction of the threshold criterion. Shear thinning, in particular, has an important effect in terms of leading to a restabilization at sufficiently high shear rates. The rigorous thresholds obtained from linear stability analyses of the aforementioned viscometric flows has led to the development of the heuristic Pakdel-McKinley criterion that has since been applied over a wider range of shearing flows, with the incorporation of additional physics; for instance, the shear-thinning-induced stabilization above can be explained by generalizing the said criterion to account for shear-rate-dependent rheological properties. Interestingly, and perhaps unexpectedly, Oldroyd-B-based stability analyses have had some success even for non-viscometric flows that include elastic wake instabilities \citep{Oztekin_McKinley_1997} and the instabilities observed in a cross-slot configuration \citep{Poole2007}. Again, perhaps a bit unexpectedly, the Pakdel-Mckinley criterion has been used successfully to rationalize the onset of many non-viscometric flow instabilities \citep{Haward_JNNFM_2018, Haward_PoF_2020}.

It must, however, be said that our understanding of non-viscometric flow instabilities is not at par with that of the viscometric ones. The original identification of the elastic instabilities in viscometric flows, based on linear stability analyses, were accompanied by physical arguments involving the interaction of the polymer molecules\,(dumbbells) with the imposed perturbation, that clearly pointed to a  positive feedback mechanism, in turn implying an exponential growth\,\citep{Shaqfeh1996,larson1990purely}. There exist no analogs of these detailed microscopic arguments for the non-viscometric instabilities discussed in Sec.\,\ref{sec:nonviscometric}. For instance, with regard to  Sec.\,\ref{sec:CrossSlotInstabilties} concerning the cross-slot configuration, the current understanding points to the existence of two apparently distinct elastic instabilities with one correlating to the sheared regions between the center and the re-entrant corners regions \citep{Rocha2009}, and the other originating at the stagnation point at the center of the slot \citep{Haward_McKinley_Shen_SciRep_2016}. The underlying flow configuration in both cases is a cross-slot with rounded corners\,(one of them being the OSCER \citep{Haward_crossslot2012}), and the rationalization of the instability origins is based solely on the Pakdel-Mckinley criterion, specifically, the differing $M$-field contours for the two cross-slot configurations. Microscopic physical arguments, along the lines of those developed for the viscometric flow instabilities, may provide further clarification. For the other widely studied non-viscometric flow configuration, that of a contraction-flow geometry, an accurate prediction of both the novel flow states and elastic instabilities observed in experiments, with the associated macroscopic signatures\,(pressure drop) remains an open challenge in viscoelastic fluid mechanics.

With regard to rectilinear shearing flows, predictions based on recent linear stability analyses \citep{Piyush_2018,chaudharyetal_2021}, using the Oldroyd-B model, appear to correlate well to observations of transition in viscoelastic pipe flow \citep{choueiri2021experimental}, in sharp contrast to the well-known disconnect in the Newtonian case. Important questions regarding the nature of post-transition scenarios remain,  however. For instance, what is the nature of the bifurcation at onset? What is(are) the pathway(s) to the EIT state from the initially growing elastoinertial centermode  eigenfunction? Is there a generic connection between the EIT and ET states, as has been shown for the case of viscoelastic channel flow\,\citep{khalid_creepingflow_2021,buza_page_kerswell_2021}  ? The answers to these questions will likely differ depending on the particular region in the $Re\!-\!Wi\!-\!\beta$ space. In this regard, there is a need for direct numerical simulations and weakly nonlinear analyses of channel and pipe flows. Recent efforts have taken a step towards answering these questions by showing that the elastoinertial center-mode instability is subcritical, especially for dilute solutions ($\beta > 0.8$), both in channel \citep{page2020exact,buza_page_kerswell_2021} and pipe \citep{wan2021subcritical} flow, in turn showing that the instability could be relevant even in parameter regimes where the flow is linearly stable; see dashed curve in Fig.~\ref{fig:channelscenarios}.  It is also of interest to go beyond the Oldroyd-B model and examine, for instance, the role of finite extensibility both in the initial instability and the subsequent nonlinear transition process. Interestingly, and in contrast to what is known for curvilinear shearing flows \citep{McKinley1996}, the preliminary results obtained by Buza \textit{et al.} \citep{buza_page_kerswell_2021}, using the FENE-P model, point to a potential destabilizing role of finite extensibility, with the center-mode instability now present in a larger region of parameter space.

It is worth commenting briefly on the topic of exact nonlinear solutions of the governing equations of motion - the so-called exact coherent states. In the Newtonian case, these ECSs have been a valuable aid in an understanding of the self-sustaining process that underlies transitional Newtonian turbulence, and form the basis of a dynamical systems interpretation of transition. A large number of Newtonian ECSs have now been found, across the canonical shearing flows, and a number of review articles have appeared on this topic \citep{kerswell_2005,eckhardt_etal_2007,Graham2021}. In comparison, there exists a paucity of elastic and elastoinertial coherent structures that might help unravel the dynamical processes underlying ET or EIT\,(an exception in this regard is the `diwhirl' solution found by Kumar and Graham \citep{Kumar_Graham_2000}, an effort that was motivated by the Groismann-Steinberg experiments \citep{Groisman_Steinberg1997} on the pronouncedly hysteretic transitions in viscoelastic Taylor-Couette flow). While there exist both non-modal and nonlinear analyses of rectilinear shearing flows, in the purely elastic limit, as described in sections\,\ref{sec:nonmodal} and \ref{sec:nonlinear}, the elements in these efforts have not, as yet, come together in the discovery of an elastic ECS. Very recent experiments in microfluidic channels \citep{jha2020universal} have taken a step towards exploring coherent structures underlying the purely elastic transition, although the interpretation of the results\,(especially, the attribution of the dynamics to Alfen-like elastic waves \citep{Varshney_2019_Alfen}) requires further investigation. In this regard, the recent linear elastoinertial \citep{Piyush_2018} and elastic \citep{khalid_creepingflow_2021} eigenfunctions, underlying the respective center-mode instabilities, may serve as useful initial guesses for numerical continuation to exact nonlinear states; as mentioned above, the results of Page \textit{et al.} \citep{page2020exact} exemplify such an effort.

We end with the mention of a few instabilities that cannot be captured by the Oldroyd-B model. As already discussed in Sec.~\ref{subsubsec:beyondOldB}, the Oldroyd-B model cannot capture `constitutive instabilities', exhibited by a range of complex fluids particularly worm-like micellar solutions, since they arise due to the non-monotonic nature of the flow curve \citep{yerushalmi1970stability}. Next, there have been both experimental \citep{bodiguel-et-al-2015,poole-2016,wen-et-al-2017,picaut-et-al-2017,Bidhan_PRF} and theoretical \citep{wilson-rallison-1999,wilson-loridan-2015,castillo-wilson-2017} efforts that have studied instabilities, at very low $Re$, in rectilinear shearing flows of concentrated, and thence, strongly shear thinning, polymer solutions. Such instabilities are now understood to be driven by the combined action of fluid elasticity and shear thinning, their prediction again being outside the purview of the Oldroyd-B model; the (phenomenological) White-Metzner model\,\citep{larson1988constitutive}, where the degree of shear thinning can be independently specified, has been used to examine such instabilities. The Oldroyd-B model also cannot predict instabilities that rely essentially on a non-zero second normal stress difference \citep{Maklad_Poole_2021}. This includes `edge fracture' which arises when a viscoelastic fluid is sheared, for example, in the cone-and-plate geometry. The free surface at the rim gets destabilized, leading to a complicated edge profile resembling fracture in elastic solids. The Johnson-Segalman and Giesekus models, both of which predict a negative $N_2$, have been used \cite{Hemingway_Fielding2020} to predict edge fracture; the interpretation of a negative $N_2$ as a tension along the radial vortex lines in the said geometry is suggestive of its destabilizing action in the above geometry. Another $N_2$-driven phenomenon is the spanwise instability of a two-layer sheared suspension flow, driven by a jump in $N_2$ across the interface \citep{Carpen_Brady2019}. Finally, the Oldroyd-B model cannot predict phenomena where coupling between the flow and polymer concentration plays a central role, since the model assumes a uniform (and dilute) polymer concentration. This coupling has been shown to lead to an instability in plane Couette flow, even when the constitutive curve of the fluid is monotonic \citep{fielding2003early,fielding2003flow,cromer2013shear,cromer2014study,Eggers2014,Peterson2019,larson_concentration_1992}.

To conclude, one may regard the Oldroyd-B model as a  sufficiently simple, yet realistic, model that allows for detailed mathematical analysis and numerical computations.  It can, therefore, be argued that, even after seven decades since it was proposed by James Oldroyd, the Oldroyd-B model remains the `go-to' model when one is faced with the prediction of a novel instability in viscoelastic flows.

\section*{Acknowledgements}
SK thanks Tamer Zaki for helpful comments concerning nonmodal analysis.  SK and MRJ thank the National Science foundation for support under Grant no. CBET-1510654, and the Minnesota Supercomputing
Institute (MSI) at the University of Minnesota for providing computing resources.
HACS and HJW acknowledge the support of the National Council of Science and Technology of Mexico (CONACyT Grant number: 299629/411301).


\bibliography{References}

\end{document}